\documentclass[aip,graphicx]{revtex4-1}
\usepackage{bm}
\usepackage{graphicx}
\usepackage{amsmath}
\usepackage{amsfonts}
\usepackage{color}
\usepackage{epstopdf}
\usepackage{bigdelim,multirow}
\usepackage{subfigure}
\draft % marks overfull lines with a black rule on the right

\begin{document}

%\preprint{}

\title{Extended-Kalman-filter-based dynamic mode decomposition for simultaneous system identification and denoising}


\author{Taku Nonomura}
\affiliation{Department of Aerospace Engineering, Graduate School of Engineering, Tohoku University}
\affiliation{Presto, JST}
\author{Hisaichi Shibata}
%\email{shibata@flab.isas.jaxa.jp}
\affiliation{Institute of Space and Astronautical Science, Japan Aerospace Exploration Agency}
\author{Ryoji Takaki}
\affiliation{Institute of Space and Astronautical Science, Japan Aerospace Exploration Agency}

\date{\today}

\begin{abstract}
A new dynamic mode decomposition (DMD) method is introduced for simultaneous online system identification and denoising in conjunction with the adoption of an extended Kalman filter algorithm\color{black}. The present paper explains the extended-Kalman-filter-based DMD (EKFDMD) algorithm and illustrates that EKFDMD requires significant numerical resources for many-degree-of-freedom (many-DoF) problems and that the combination with truncated proper orthogonal decomposition (trPOD) helps us to apply the EKFDMD algorithm to many-DoF problems. The numerical experiments of the present study illustrate that EKFDMD can estimate eigenvalues from a noisy dataset with a few DoFs better than or as well as the existing algorithms, whereas EKFDMD can also denoise the original dataset online. In particular, EKFDMD performs better than existing algorithms for the case in which system noise is present. The EKFDMD with trPOD can be successfully applied to many-DoF problems, including a fluid-problem example, and the results reveal the superior performance of system identification and denoising. Note that these superior results are obtained despite being an online procedure. 
\end{abstract}


\maketitle %\maketitle must follow title, authors, abstract and \pacs

\section{Introduction}
Recently, modal decomposition\cite{Taira2017} for fluid dynamics has attracted attention from the viewpoints of data reduction, data analysis, and reduced-order modeling of complex dataset. This is one method for data-driven science in fluid dynamics. The most conventional method of modal decomposition is a proper orthogonal decomposition (POD),\cite{Rowley2004,Berkooz1993} which is also called principal component analysis (PCA) and Karhunen-Lo\'{e}ve expansion. The standard POD can be computed by singular value decomposition (SVD), and this fact explains that the obtained modes are orthogonal with respect to each other. Proper orthogonal decomposition modes can be computed by snapshots of fluid data and can be used for both numerical and experimental approaches. Based on POD modes, a reduced-order model can be constructed with the Galerkin projection method for instance, although only a numerical approach can be used for reduced-order modeling in this way. 

Another conventional method is global linear stability analysis (GLSA),\cite{Theofilis2011, Shibata2015, Ohmichi2016} which shows that the eigenmodes of the system of linearized governing equations (i.e., the Navier-Stokes equations for most of the fluid problems) around the steady state of nonlinear dynamics. Here, GLSA shows the most unstable eigenmodes and judges whether the steady-state solution is stable. The modes obtained by GLSA are a solution of the original linearized equations, although this method always requires numerically complex approaches and cannot be applied to experimental data. Unlike POD modes, the modes obtained by GLSA are not orthogonal unless otherwise the system is written with an Hermite operator.  

In recent decades, a new method, dynamic mode decomposition (DMD),\cite{Schmid2010} has been proposed and developed as a data-driven science method and has been applied to numerous  fluid problems.\cite{Tu2014,Wang2017,Priebe2016,Ohmichi2017a} Here, DMD has characteristics of both POD and GLSA, whereas DMD can be computed only by a time-series of snapshots of numerical and experimental data. 
This method processes snapshots of sequential unsteady nonlinear flow fields and yields eigenvalues and corresponding eigenmodes for the case in which the dataset is assumed to be explained by a linear system $\bm{x}_{k+1}=A\bm{x}_k$, where ${x}_k$ is the $k$th snapshot of sequential data and $A$ is a system matrix. These dynamic modes are generally nonorthogonal\color{black}, and each mode possesses a single-frequency response with amplification or damping as a natural characteristic of a linear system expression, which leads to a more intrinsic understanding of the role of each mode. Thus far, there are several methods by which to compute the dynamic modes: standard DMD\cite{Schmid2010}, exact DMD, noise-cancelling DMD (ncDMD),\cite{Dawson2016} forward-backward DMD, (fbDMD),\cite{Dawson2016} total least-squares DMD (tlsDMD),\cite{Hemati2017,Dawson2016} online DMD,\cite{Zhang2017} and  Kalman-filter-based DMD (KFDMD),\cite{Nonomura2018} where ncDMD, fbDMD, tlsDMD and KFDMD focus on the noisy dataset.  The standard DMD and the exact DMD adopt SVD and a Moore-Penrose pseudo-inverse matrix for low-rank approximation of matrix $A$, respectively\color{black}. This implies that these algorithms compute dynamic modes as a kind of least-squares problem. A robust method for a noisy dataset, tlsDMD, adopts a truncated POD for pair data and successfully increases the accuracy of obtained dynamic modes. A recent KFDMD is written in the form of system identification using the Kalman filter algorithm\cite{Kalman1960} and can be optimized based on the prior knowledge of the noise superimposed on the data. This is different from the usage of Kalman filter in Reference\cite{Surana2016,Surana2017} in which the Kalman filter is used for data reconstruction and prediction\color{black}.

However, the application of DMD to noisy data and the denoising process are still limited. For example, tlsDMD has been developed for accurately estimating the dynamic modes and corresponding eigenvalues, but a method by which to reconstruct the data has rarely been shown except for the data reconstruction using first snapshot,\cite{Kutz2016} which is conventionally adopted. If we adopt the conventional simple estimation of initial amplitudes to reconstruct the data, then the data is greatly affected by the noise on the initial data, as expected.  One of a few advanced data reconstruction methods is use of Kalman filter for linear system that is corresponding to the Koopman operator after the linear system is estimated\color{black}.\cite{Surana2016,Surana2017}

Optimized DMD (optDMD),\cite{Askham2018} and the combination of tlsDMD\cite{Hemati2016,Dawson2016} and the sparsity-promoting DMD (spDMD)\cite{Jovanovic2014} could be used for the denoising of noisy data. Here, optDMD gives us the dynamic modes and eigenvalues and corresponding initial values that best fit the noisy time series data under the assumption of no system noise. On the other hand, spDMD\cite{Jovanovic2014} selects finite-number modes for the reconstruction of flow fields considering the $L_0$ or $L_1$ norm of regularization terms, as is often used in sparse modeling and compressed sensing. These methods are very useful for reconstructing flow fields, but the reconstructed data are governed by the initial value of the strength of each mode and possibly cannot handle the change in phase of dynamic modes in long-time data due to the system noise including modeling error, nonlinear processes, or unexpected events in the experiments. Furthermore, the optDMD requires fitting of all of the data, and the combination procedure of tlsDMD and spDMD requires two-step computation. At present, an online method for simultaneous system identification and denoising using the DMD framework has not yet been proposed. 

In the present paper, a new online method for simultaneous system identification and denoising using the DMD framework is proposed using the extended Kalman filter. In addition to the system identification of the previously proposed KFDMD,\cite{Nonomura2018} the observed data are simultaneously filtered online. The present paper first explains the algorithm of the proposed extended-Kalman-filter-based DMD (EKFDMD). The drawback of the computational costs of EKFDMD is addressed, and combination with a truncated POD (trPOD) is proposed for reduction of the computational cost. Finally, the proposed method is applied to various problems and its performance is illustrated. 

%\section{Formulations}

%\subsection{Standard DMD}
%

\section{Previous Methods Compared in the Present Study}
\subsection{Problem settings}
\label{sec:pre}
Here, the previous algorithms compared in the present study are briefly explained. For the extension in the next subsection, the linear system model is assumed for the time series dataset as follows:
\begin{eqnarray}
\bm{x}_{k+1} &=& A \bm{x}_k +\bm{v}_k,\label{eq:system1}\\
\bm{y}_{k  } &=&   \bm{x}_k +\bm{w}_k.\label{eq:system2}
\end{eqnarray}
Here, $A$, $\bm{x}$, $\bm{y}$, and $n$ are the system matrix, the state variable vector, the observation vector, and the dimension of the state and observed variables, respectively. Moreover, $\bm{x}_k$ is assumed to be the true value. Usually, we can only access $\bm{y}$ in the present paper, though $\bm{x}$ has been used as the observation vector in the previous DMD study. Therefore, the reader should take care when considering the notation used herein. 
First, three methods, DMD, tlsDMD, and KFDMD are briefly explained in Subsections \ref{sec:DMD}, \ref{sec:tlsDMD}, and \ref{sec:KFDMD}, respectively, and a conventional data reconstruction method for these algorithms is introduced in Subsection \ref{sec:DR}. Finally, optDMD, which is a state-of-art offline algorithm for both estimating the dynamic modes and reconstructing data, is explained in Subsection \ref{sec:optDMD}.  

\subsection{DMD}
\label{sec:DMD}
The $m$-sample observation data matrix including observation noise is defined as follows:
\begin{eqnarray}
Y_{1:m} &=& \left( \bm{y}_{m}, \bm{y}_{m-1}, \cdots , \bm{y}_{2}, \bm{y}_{1}\right),
\end{eqnarray}
whereas $\bm{y}_k=\bm{x}_k$ if the observation noise is absent. 
The original DMD is performed with SVD for $Y_{1:m-1}$ as follows: 
\begin{eqnarray}
Y_{1:m-1} &\sim& U_{1:m-1} \Sigma_{1:m-1}V_{1:m-1}^\text{T}.
\end{eqnarray}
Here, $U$, $\Sigma$, and $V$ are a left singular matrix, a diagonal matrix with singular values, and a right singular matrix, respectively. As described in the original DMD paper, a truncated POD (SVD) is used to filter the noise. Therefore, the rank $r$ approximation of the observation data matrix is obtained as follows:
\begin{eqnarray}
Y_{1:m-1} &=& \tilde U_{1:m-1} \tilde \Sigma_{1:m-1} \tilde V_{1:m-1}^\text{T}.
\end{eqnarray}
In this case, the projected $r \times r$ matrix $\tilde A$ of matrix $A$ onto the low-dimensional space can be obtained as follows:
\begin{eqnarray}
\tilde A=\tilde U_{1:m-1}^{\text{T}} Y_{2:m} \tilde V_{1:m-1} \tilde \Sigma^{-1}_{1:m-1}.
\end{eqnarray}
Then, the eigendecomposition is carried out:
\begin{eqnarray}
\tilde A W_\text{DMD}= W_\text{DMD} \Lambda_\text{DMD}.\label{eq:AWWL}
\end{eqnarray}
Here, $W_\text{DMD}$ are the eigenvectors, and $\Lambda_\text{DMD}$ is the diagonal matrix with the eigenvalues. Using $W_\text{DMD}$, the dynamic mode matrix in the original space is recovered:
\begin{eqnarray}
\Phi = \tilde U_{1:m-1} W_\text{DMD}. 
\end{eqnarray}
Here, $\Phi$ contains the dynamic mode vectors as follows:
\begin{eqnarray}
\Phi = [\bm{\phi}_1 \quad \bm{\phi}_2 \quad \dots \quad \bm{\phi}_r].\label{eq:PHI}
\end{eqnarray}

\subsection{tlsDMD}
\label{sec:tlsDMD}
For total least-squares DMD, the pair snapshot is considered. In this case, trPOD data or raw data can be used.\cite{Dawson2016,Hemati2017} In the present study, raw data are directly used as in the original code\color{black}.\cite{Rowley2017} The procedure for the time series data are as follows. First, define a pair data matrix: 
\begin{eqnarray}
Z=\left[\begin{array}{c}
Y_{1:m-1}\\Y_{2:m}
\end{array}\right]
\end{eqnarray}
and POD is applied to the pair data matrix above:
\begin{eqnarray}
Z=\left[\begin{array}{c}U_{1:m-1}\\U_{2:m}\end{array}\right]\Sigma_{Z}V^{\text{T}}_{Z},
\end{eqnarray}
Then, we obtain an $r$-rank truncated pair POD, as follows:
\begin{eqnarray}
\widehat{Z}=\left[\begin{array}{c}\widehat{U}_{1:m-1}\\\widehat{U}_{2:m}\end{array}\right]\widehat{\Sigma}_{Z}\widehat{V}^{\text{T}}_{Z},
\end{eqnarray}
Here, we obtain a snapshot pair of POD projections $\widehat{X}$ and $\widehat{Y}$ of $X$ and $Y$ as follows:
\begin{eqnarray}
\widehat{Y}_{1:m-1}&=&Y_{1:m-1}\widehat{V}_{Z}\label{eq:xbhat}\\
\widehat{Y}_{2:m} &=&Y_{2:m}\widehat{V}_{Z}\label{eq:ybhat}.
\end{eqnarray}
Using these matrices, $\tilde A$ is computed by SVD of $\hat X$:
\begin{eqnarray}
 \hat X&=&U_{\hat{Y}_{1:m-1}} \Sigma_{\hat{Y}_{1:m-1}} V_{\hat{Y}_{1:m-1}}\\
\tilde A&=&\hat U_{1:m-1} \hat{Y}_{2:m} \hat V_{1:m-1} \hat \Sigma^{-1}_{1:m-1}.
\end{eqnarray}
The dynamic mode and eigenvalue estimations are exactly the same as DMD in Eqs~\ref{eq:AWWL} to \ref{eq:PHI}.

\subsection{KFDMD}
\label{sec:KFDMD}
The components of matrix $A$ are considered to be state variables of the Kalman filter. The state variable vector $\bm{\theta}$ are written as follows:
\begin{eqnarray}
\bm{\theta}^\text{KF}=\text{vec}(A^\text{T})
\end{eqnarray}
Using the state variable vector described above, the system and observation equations can be written as follows:
\begin{eqnarray}
\bm{\theta}^\text{KF}_{k+1}&=&\bm{\theta}^\text{KF}_{k+1}+\bm{v}_k,\\
\bm{y}_{k+1}&=&H_k^\text{KF} \bm{\theta}_{k+1}^\text{KF}+\bm{w}_k,\\
\end{eqnarray}
where $H_k^\text{KF}$ is the following observation matrix defined as follows:
\begin{eqnarray}
H_k^\text{KF} &=& \overset{\text{$n^2$ dimensions}}{\overbrace{\left.\left[
		\begin{array}{ccccc}
		\bm{y}_{k-1}^{\text{T}} & \bm{0} & \cdots & \cdots & \bm{0} \\
		\bm{0} & \bm{y}_{k-1}^{\text{T}} & \bm{0} & \cdots & \bm{0} \\
		\bm{0} & \bm{0} & \ddots & \bm{0} & \bm{0} \\
		\bm{0} & \cdots & \bm{0} &\bm{y}_{k-1}^{\text{T}} & \bm{0} \\
		\bm{0} & \cdots & \bm{0} & \bm{0} & \bm{y}_{k-1}^{\text{T}} \\
		\end{array}
		\right] \right\}}} \text{\small{$n$ dimensions}}
\end{eqnarray}
Note that we have the following relationship:
\begin{eqnarray}
H_k^\text{KF} \bm{\theta}^\text{KF}_k = A_k \bm{y}_k, 
\end{eqnarray}
where $A_k$ represents the estimation of $A$ in the $k$th time step.
Here, $\bm{v}_k$ and $\bm{w}_k$ are system and observation noises, respectively.
Using the state equation given above, the linear Kalman filter is constructed with the fast algorithm shown in Reference. \cite{Nonomura2018} After obtaining matrix $A$, the dynamic mode and corresponding eigenvalues are obtained through the eigendecomposition of matrix $A$.

\subsection{Data reconstruction using DMD, tlsDMD, and KFDMD}
\label{sec:DR}
The DMD, tlsDMD, and KFDMD methods only estimate matrix $A$ and do not estimate the reconstructed time series data using dynamic modes. In a conventional method\cite{Kutz2016} of reconstruction, we assume that the data can be reconstructed as follows:
\begin{eqnarray}
X_\text{reconst}=\Phi B_0 V_\text{and},
\end{eqnarray}
Here, $X_\text{reconst}$ is the reconstructed data matrix, $B_0$ is a diagonal matrix of the initial amplitudes $b_i$ of dynamic modes $\Phi_i$, where 
\begin{eqnarray}
B_0=\left[
\begin{array}{cccc}
b_{1} & {0} & 0 & 0 \\
{0} & b_{2} & {0} & 0  \\
{0} & {0} & \ddots & \vdots \\
{0} &  {0} &\dots & b_{r} \\
\end{array}
\right], 
\end{eqnarray}
and $V_\text{and}$ is a Vandermonde matrix representing the temporal behaviors of dynamic modes while assuming the system noise to be absent:
\begin{eqnarray}
V_\text{and}=\left[
\begin{array}{ccccc}
{1}    & \lambda_1 & \lambda_1^2 & \dots & \lambda_1^m \\
{1}    & \lambda_2 & \lambda_2^2 & \dots & \lambda_2^m \\
\vdots & \vdots    & \vdots      & \dots & \vdots \\
{1}    & \lambda_r & \lambda_r^2 & \dots & \lambda_r^m \\
\end{array}
\right].
\end{eqnarray}
The initial value vector $\bm{b}_0$, which is defined as
\begin{eqnarray}
\bm{b}_0=[b_1 \quad b_2 \quad \dots \quad b_r]^\text{T}, 
\end{eqnarray}
can be obtained using the pseudoinverse of $\Phi$, as follows:
\begin{eqnarray}
\bm{b}_0=\Phi^+ \bm{y}_0,
\end{eqnarray}
where the plus symbol superscript denotes the Moore-Penrose pseudoinverse matrix. As discussed later herein, $\bm{y}_0$ includes the observation noise superimposed on the initial snapshot, and this reconstruction does not work well due to this noise, even if the eigenvalues are well estimated. 

\subsection{optDMD}
\label{sec:optDMD}
In the optimized DMD,\cite{Askham2018} the following problem is solved:
\begin{eqnarray}
[\Phi \quad B_0 \quad \Lambda]
&=&\mathop{\text{argmin}}_{\Phi, B, \Lambda}\| Y-X_{\text{reconst}} \|_F^2\\
&=&\mathop{\text{argmin}}_{\Phi, B, \Lambda}\| Y-\Phi B_0 V_\text{and}|_F^2
\end{eqnarray}
Although there are several ways to solve this nonlinear problem above, the variable projection method is adopted in the present study. In this case, the best-fit reconstructed data matrix is obtained under the assumption that system noise is absent. In the case of spDMD, 
$\Phi$ and $\Lambda$ are fixed using another DMD method, and optimum sparse $\bm{b}_0$ is solved while adding the $L_1$ or $L_0$ regularization term of $\bm{b}_0$.
 The original code\cite{Askham2017} is employed in the present study\color{black}. 
\section{Extended Kalman Filter DMD}

\subsection{Algorithm}
\label{sec:algorithm}
As introduced in the section above, we consider the system expressed by Eqs. \ref{eq:system1} and \ref{eq:system2}.
For simplicity, we introduce the tensor expressions for Eqs. \ref{eq:system1} and \ref{eq:system2}, as follows:
\begin{eqnarray}
x_{i,k+1} &=& a_{ij} {x}_{j,k}+v_{i,k}\\
y_{i,k+1} &=&  {x}_{i,k}+w_{i,k}
\end{eqnarray}
where $A=(a_{ij})$ and $\bm{x}={x_i}$.

Then, the Kalman filter algorithm is considered. In this problem, we would like to simultaneously conduct the online system identification and denoising of the observed variable. Therefore, the observed variables and elements of matrix $A$ are chosen as state variables of the considered system. The state variable vector $\bm{\theta}$ is defined as follows:
\begin{eqnarray}
\bm{\theta}_k = \left( 
\begin{array}{*{20}{c}}
\bm{x}_k \\\mathrm{vec}\left(A^{\text{T}}\right)
\end{array} \right) 
=\left.\left(
\begin{array}{*{20}{c}}
{{x_{1,k}}}\\{{x_{2,k}}}\\\dots\\{{x_{n,k}}}\\
{{a_{11}}}\\{{a_{12}}}\\ \vdots \\{{a_{1n}}}\\{{a_{21}}}\\{{a_{22}}}\\ \vdots \\{{a_{nn}}}
\end{array} \right) \right\}\text{$n+n^2$ dimensions}.
\end{eqnarray}
Using these state variables, the system transient can be written as follows:
\begin{eqnarray}
	\bm{\theta}_{k+1} &=&\left( 
\begin{array}{*{20}{c}}
\bm{x}_{k+1} \\\mathrm{vec}\left(A^{\text{T}}\right)
\end{array} \right) 
 = \bm{f} \left(\bm{\theta}_{k}\right)  = \bm{f} \left(\bm{x}_{k}, A\right) +\bm{v}_k \\
{{\bm{y}}_{n + 1}} &=& H \bm{\theta}_k+\bm{w}_k
\end{eqnarray}
where the $\bm{v}_k$ and $\bm{w}_k$ are the system and observation noise, respectively, and the nonlinear function $\bm{f}$ and the observation matrix are expressed as follows:
\begin{eqnarray}
\bm{f}&=& \left( 
    \begin{array}{*{20}{c}}
    A \bm{x}_k \\\mathrm{vec}\left(A^{\text{T}}\right)
    \end{array} \right) 
=\left(
\begin{array}{*{20}{c}}
{{a_{1j}x_{j,k}}}\\{{a_{2j}x_{j,k}}}\\\dots\\{{a_{nj}x_{j,k}}}\\
{{a_{11}}}\\{{a_{12}}}\\ \vdots \\{{a_{1n}}}\\{{a_{21}}}\\{{a_{22}}}\\ \vdots \\{{a_{nn}}}
\end{array} \right), \quad \quad
H= \overset{\text{$n+ n^2$ dimensions}}{\overbrace{\left.\left( I \quad \bm{0} \right)\right\} }}\text{\tiny{$n$ dimensions}} \label{eq:f}\\
\end{eqnarray}
The upper half of the system is written as the multiplication of state variables $x_{j}$ and $a_{ij}$, and, as such, the system is considered to be nonlinear. The lower half of the system corresponds to the constant or slowly varying system coefficients to be identified and does not change explicitly. For the construction of the extended Kalman filter, the linearization is required. The Jacobian matrix $F$ of a nonlinear function $\bm{f}$ of the state variables $\bm{\theta}$ is calculated as follows:
\begin{eqnarray}
F_k=\frac{\partial \bm{f}}{\partial \theta_{\bm{k}}} 
&=&  \left( {\begin{array}{*{20}{c}}
{\frac{\partial {A\bm{x}_k}}{\partial \bm{x}_k}}& {\frac{\partial {A\bm{x}_k}}{\partial \mathrm{vec}\left({A^{\text{T}}}\right)}}\\
{\frac{\partial \mathrm{vec}\left(A^{\text{T}}\right)}{\partial \bm{x}_k}}& {\frac{\partial {\mathrm{vec}\left(A^{\text{T}}\right)}}{\partial \mathrm{vec}\left({A^{\text{T}}}\right)}}\\
\end{array}} \right)
=
\overset{\text{$n+n^2$ dimensions}}{\overbrace{\left.
\left( {\begin{array}{*{20}{c}}
{{A}}&{\begin{array}{*{20}{c}}
{{\bm{x}}_{k}^{\text{T}}}&{}&{}&{\bf{0}}\\
{}&{{\bm{x}}_{k}^{\text{T}}}&{}&{}\\
{}&{}& \ddots &{}\\
{\bm{0}}&{}&{}&{{\bm{x}}_{k}^{\text{T}}}
\end{array}}\\
{\bm{0}}&{\bm{I}}
\end{array}} \right).  \right\}}} \text{\tiny{$n+n^2$ dimensions}}
\label{eqn:Fmatrix}
\end{eqnarray}
Using matrices $F_k$ and $H$, the extended Kalman filter can be constructed for the nonlinear system. Note that $F_k$ is a time-varying matrix. 

Following the theory of a Kalman filter, a priori prediction of a state variable vector $\bm{\theta}_k$ and a covariance matrix $P_{k|k-1}$ can be achieved using the state variable vector $\bm{\theta}_k$ and covariance matrix $P_{k-1|k-1}$ from the previous time step, 
\begin{eqnarray}
    \bm{\theta}_{k|k-1} &=& \bm{f}( \bm{\theta}_{k-1|k-1} ) \label{eq:predicttheta}\\
	P_{k|k-1} &=& F_k P_{k-1|k-1} F_k^{\text{T}} + Q_{k},\label{eq:predictP}
\end{eqnarray}
where the system matrix $F_k$ is expressed by Eq.~\ref{eqn:Fmatrix}, and $Q$ is a covariance matrix of the system noise.  

When a new observation is available, the state variables and covariance matrix are updated using the Kalman gain, which is computed as 
\begin{eqnarray}
K_k = P_{k|k-1} {H}^{\text{T}} S_k^{-1},\label{eq:updatekg}
\end{eqnarray}
where $S_k$ is a noise covariance matrix and is expressed as follows: 
\begin{eqnarray}
	S_k = R_k + {H} P_{k|k-1} {H}^{\text{T}}.\label{eq:updates}
\end{eqnarray}
Here, $R_k$ is a covariance matrix of observation noise $\bm{w}_k$. 

A modification vector for state variables $\bm{\theta}$ is computed as follows:
\begin{eqnarray}
	\label{eq:deltatheta}
	\delta \bm{\theta}_{k|k}& &       = K_k \left( \bm{y}_k - H \bm{\theta}_{k|k-1} \right)\\
                            & &\left( = K_k \left( \bm{y}_k - A_{k|k-1} \bm{x}_{k-1|k-1} \right) \right).
\end{eqnarray}

Finally, the state variable vector and the covariance matrix after the observation are updated as follows:
\begin{eqnarray}
	\bm{\theta}_{k|k} &=& \bm{\theta}_{k|k-1} + \delta \bm{\theta}_{k|k}.\label{eq:updatetheta}\\
	P_{k|k} &=& (I - K_k {H}) P_{k|k-1}.	\label{eq:updateP}
\end{eqnarray}

This extended Kalman filter requires the multiplication of the large matrix of dimension of $(n^2+n) \times (n^2+n)$, as discussed in Section \ref{sec:Comp}. This is a clear drawback of this formulation for many-degree-of-freedom (many-DoF) problems, and using this algorithm together with trPOD is recommended, as explained in Section \ref{sec:trPOD}. This drawback of EKFDMD is the same as that of KFDMD designed for only the system identification, though the drawback of KFDMD is somehow relaxed owing to the fast algorithm proposed in the previous study,\cite{Nonomura2018} in which the large matrix is assumed to be decomposed into several identical block matrices. Although we attempt to use a concept similar to the previous KFDMD,\cite{Nonomura2018} we could not find a similar method for EKFDMD in the present state. Therefore, the computational cost for EKFDMD is severer than that for KFDMD designed for only system identification, and the use of the present algorithm together with trPOD is strongly recommended for many-DoF problems. 

It should be noted that, in the early implementation of EKFDMD, we employed the several initial time steps for only the estimation of $A$ without filtering of $\bm{x}$, but they are found to just degrade the results. In the present implementation, the simultaneous estimation is impulsively started from the first step. \color{black}
\newpage
\subsection{Combination with a truncated POD}
\label{sec:trPOD}
As discussed in the previous section, the computational cost of the present algorithm is high, and, therefore, a truncated POD (truncated SVD) should be used for the reduction in the number of DoFs of the dataset of the observed variables. Similar to a previous study on KFDMD for only system identification, the obtained data are processed as follows:
\begin{enumerate}
    \item the batch POD is applied, 
    \item a proposed Kalman filter is then applied to the amplitude of each POD mode, and
    \item the mode shape of a fluid system is finally recovered by multiplying the spatial POD modes.
\end{enumerate}

As the first step (step 1), POD is applied to an observed data matrix and an observed data matrix is expressed in SVD form as follows:
\begin{eqnarray}
Y_{1:m}=U_{1:m}D_{1:m}V^{\text{T}}_{1:m}. 
\end{eqnarray}
Here, $U$ and $V$ are matrices consisting of the spatial and temporal POD modes, respectively. The $r$-rank approximation of the observed data matrix is calculated as follows:
\begin{eqnarray}
 Y_{1:m}  \sim\tilde U_{1:m} \tilde D_{1:m} \tilde V^{\text{T}}_{1:m},
\end{eqnarray}
where quantities with tildes indicate $r$-rank approximations. Here, the $r$-dimension matrix of $\tilde D$ consists of $r$-largest singular values of $D$. In addition, the row vectors of $\tilde U$ and $\tilde V$ are the same as the corresponding first $r$ row vectors of $U$ and $V$. Using these matrices,\\
 reduced-order $\tilde Y$, which represents mode strength, is constructed as follows:
\begin{eqnarray}
\tilde Y_{1:m}  = \tilde D_{1:m} \tilde V^{\text{T}}_{1:m}.\label{eq:trun}
\end{eqnarray}
 In the second step (step 2), $\tilde Y$ and $\tilde y_k$ are treated in a manner similar to $Y$ and $y_k$ in the proposed EKFDMD procedures, and $\bm{x}_k$ and $A$ are simultaneously estimated online. In addition, for online implementation, 
\begin{eqnarray}
\tilde{\bm{y}}_k=\tilde U^\text{T} {\bm{y}}
\end{eqnarray}
can be used where the left singular vector is assumed to be fixed using the sample data. After this process, the eigenvalues and eigenmodes are computed by solving the eigenvalue problem of $A$.

Finally, in the third step (step 3), the original dimension of the eigenmode is obtained by multiplying matrix $U$ after obtaining the right eigenvector of the reduced system by EKFDMD. 
\begin{eqnarray}
{\bm{x}}_k=\tilde U \tilde{\bm{x}}_k, \label{eq:recov}
\end{eqnarray}

Again, note that we can use the same formulation in Eqs.~\ref{eq:trun} through \ref{eq:recov} for an online situation in which the left singular vector (spatial mode) $\tilde U$ is known in advance. This is similar to KFDMD\cite{Nonomura2018} proposed previously. In this case, a fully online algorithm can be obtained. However, if the POD mode is not known in advance and must be estimated, then an online POD method or other methods are required. If the spatial POD modes change with time as in the case of online POD, then the projected coefficients are not consistent in time. Furthermore, the POD modes are sometimes activated or deactivated in the online POD algorithm. Thus, it appears to be difficult to straightforwardly extend the EKFDMD to a method combined with the online POD, and this is left for a future study.\color{black}

In the present paper, Eq. \ref{eq:trun} is adopted for the truncated POD. This procedure is used for many-DoF problems (n$>$30) and is not used unless otherwise mentioned.  In the case of noisy dataset, it should be noted that an accurate estimate of the mode coefficient does not necessarily mean an accurate representation of the full state because the spatial POD mode contains noise as shown later. However, despite the imperfect estimation of POD modes, eigenvalue and reconstructed data by EKFDMD are sufficiently accurate, which is also shown later. \color{black} 

\subsection{Implementation of the EKFDMD algorithm}
\label{sec:Imple}
Here, the EKFDMD algorithm is briefly summarized. After initialization, the prediction (a priori estimation) and update steps are alternately performed. 
\subsubsection*{Initialization}
\begin{enumerate}
	\item If the DoF is large, trPOD is applied to the data.
	\item Set $\bm{\theta}=\textrm{vec}(I)$ and $P_{0|0}=\gamma I$. Here, $\gamma$ is large. (In the present study, we set $\gamma=1,000$).
\end{enumerate}

\subsubsection*{Prediction step}
\begin{enumerate}
	\item $x_{k|k-1}$ are predicted by Eqs. \ref{eq:predicttheta} and \ref{eq:f}, while $a_{ij,k|k-1}$ are predicted to be the same as $a_{ij,k-1|k-1}$.
	\item $P_{k|k-1}$ is predicted by Eqs. \ref{eq:predictP} and \ref{eqn:Fmatrix}.
\end{enumerate}

\subsubsection*{Update step}
\begin{enumerate}
	\item Kalman gain $K$ is computed by Eqs. \ref{eq:updates} and \ref{eq:updatekg}.
	\item $\theta_{k|k}$ is updated by Eqs. \ref{eq:deltatheta} and \ref{eq:updatetheta}, and matrix $A$ is obtained using $\theta_{k|k-1}$. 
	\item $P_{k|k}$ is updated by Eq. \ref{eq:updateP}.
\end{enumerate}

\section{Numerical Experiments and Discussion}

The EKFDMD algorithm described in Section \ref{sec:algorithm} is adopted in the numerical experiments below. 

\subsection{Problem with a small number of DoFs without system noise}

First, the performance of EKFDMD is investigated for the standard problem, in comparison with the standard DMD, KFDMD, tlsDMD, and optDMD. 
The problem is approximately the same as that considered in the previous study.\cite{Hemati2017} This problem is modified slightly to involve the process noise in discretized form for the next subsection, although only the observation noise is first considered in this subsection.

The discretized eigenvalues are assumed to be positioned at $\lambda_1 = \exp\left[\left( \pm 2\pi i \Delta t \right)\right]$, $\lambda_2 = \exp\left[\left( \pm 5\pi i \Delta t \right)\right]$, and $\lambda_3 = \exp\left[\left( -0.3 \pm 11\pi i\right)\Delta t\right]$, where $\Delta t = 0.01$. The corresponding continuous eigenvalues are $\omega_1 = \left( \pm 2\pi i \right)$, $\omega_2 = \left( \pm 5\pi i \right)$, and $\omega_3 = \left( -0.3 \pm 11\pi i\right)$.
The number of DoFs of this system is $d=6$. The original data $\bm{f}$ were computed in the previous study as
\begin{eqnarray}
\frac{\text{d}\bm{f}}{\text{d}t}&=&B\bm{f} \label{eq:fcont}\\
B&=&\left(\begin{array}{cccccc}
 |\text{Re}(\omega_1)| & |\text{Im}(\omega_1)| & 0 & 0  & 0  & 0\\
-|\text{Im}(\omega_1)| & |\text{Re}(\omega_1)|  & 0 & 0  & 0  & 0\\
0  & 0& |\text{Re}(\omega_2)| & |\text{Im}(\omega_2)| & 0 & 0 \\
0  & 0&-|\text{Im}(\omega_2)| & |\text{Re}(\omega_2)|  & 0 & 0  \\
0 & 0  & 0  & 0&  |\text{Re}(\omega_3)| & |\text{Im}(\omega_3)|  \\
0 & 0  & 0  & 0& -|\text{Im}(\omega_3)| & |\text{Re}(\omega_3)|  \\
\end{array}
\right).
\end{eqnarray}
However, the above formulation cannot treat system noise. Therefore, the system is integrated for each time step size, and discretized system noise is added as follows:
\begin{eqnarray}
\bm{f}_{k+1}=e^{B\Delta t}\bm{f}_k+{\bm{v}}'_k,\label{eq:feBDt}
\end{eqnarray}
where ${\bm{v}}'$ is the system noise for the original system. Equation \ref{eq:feBDt} exactly corresponds to the solution of Eq. \ref{eq:fcont} for the condition in which $\bm{v}_k$ is absent. In this subsection, no system noise is considered with $\bm{v}_k=0$. 

\begin{figure}
	\centering
	\subfigure[$\sigma_w^2=0.0001$.       ]{\includegraphics[width=5cm]{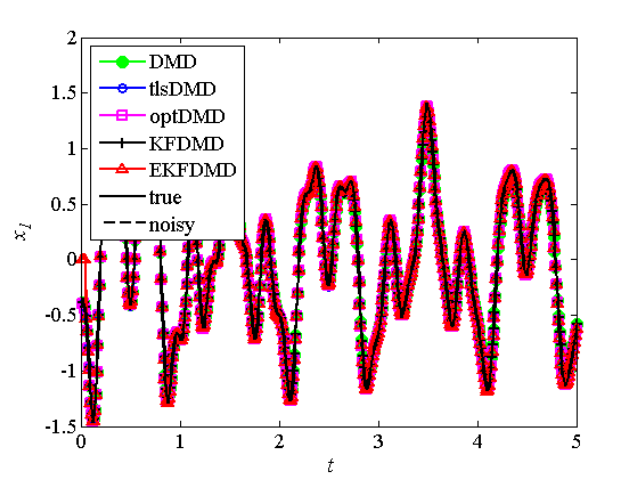}}
	\subfigure[$\sigma_w^2=0.001$.        ]{\includegraphics[width=5cm]{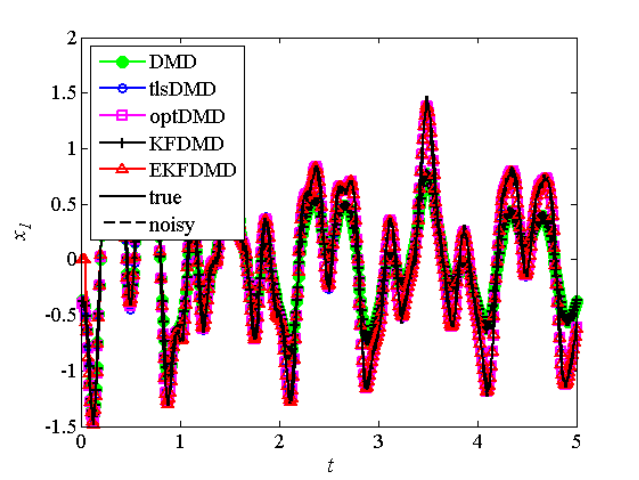}}\\
	\subfigure[$\sigma_w^2=0.01$.         ]{\includegraphics[width=5cm]{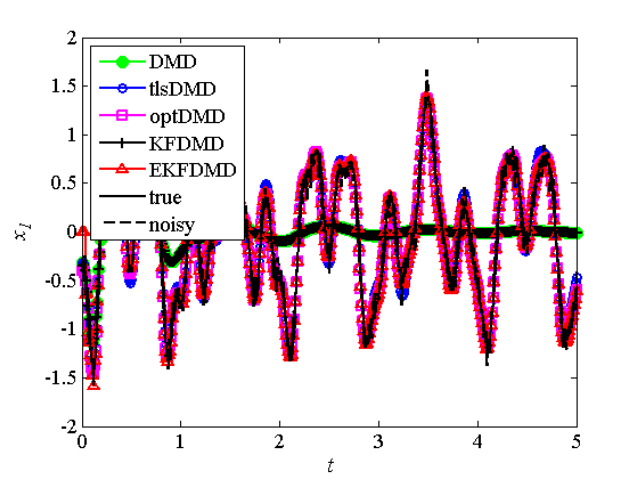}}
	\subfigure[$\sigma_w^2=0.1$.          ]{\includegraphics[width=5cm]{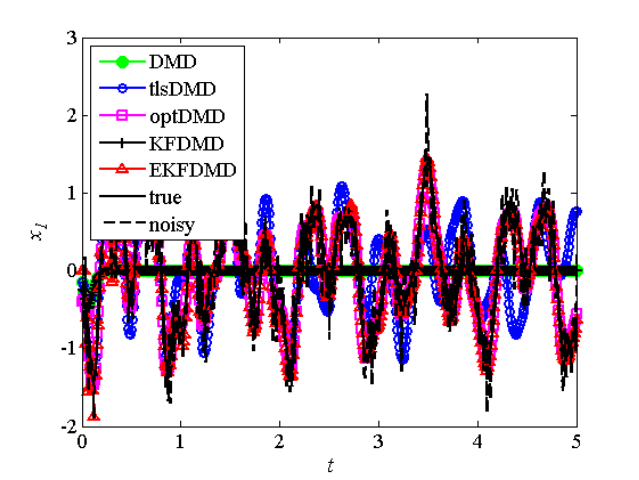}}\\		
	\subfigure[$\sigma_w^2=0.01$,  tlsDMD.]{\includegraphics[width=5cm]{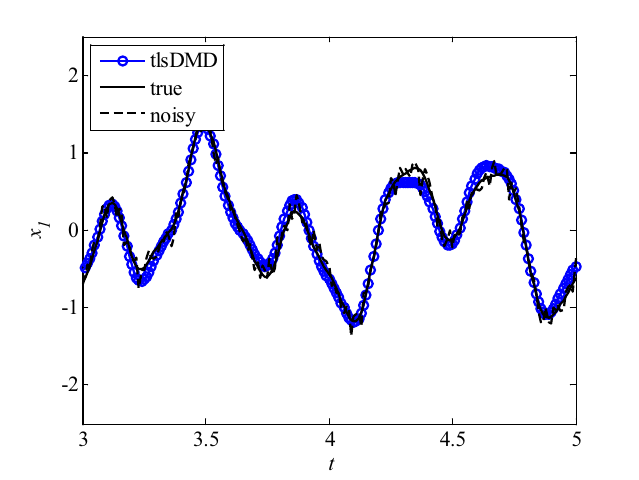}}
	\subfigure[$\sigma_w^2=0.01$,  optDMD.]{\includegraphics[width=5cm]{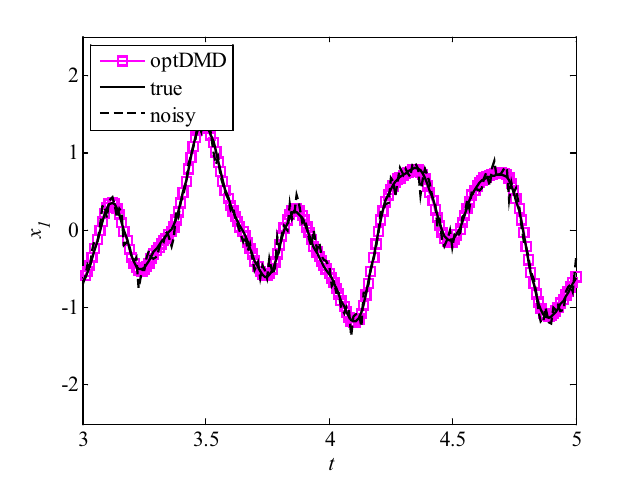}}
	\subfigure[$\sigma_w^2=0.01$,  EKFDMD.]{\includegraphics[width=5cm]{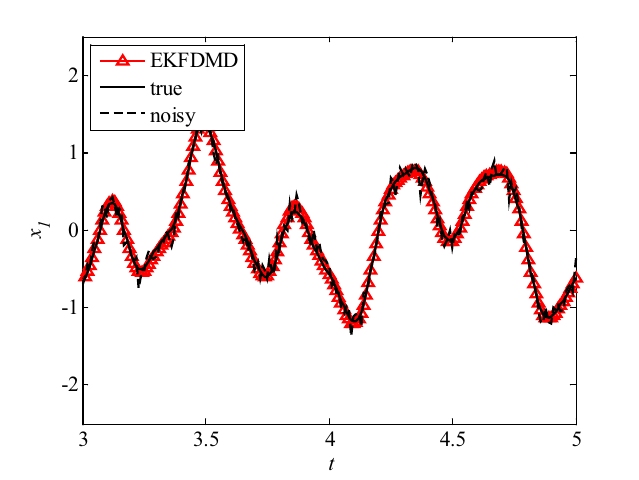}}\\
	\subfigure[$\sigma_w^2=0.1$,   tlsDMD.]{\includegraphics[width=5cm]{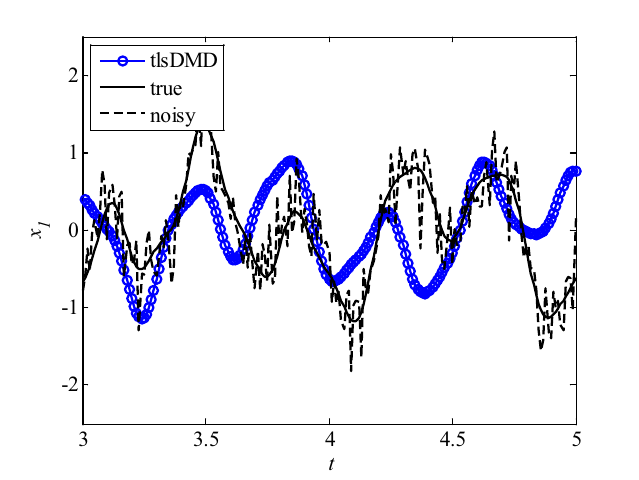}}
	\subfigure[$\sigma_w^2=0.1$,   optDMD.]{\includegraphics[width=5cm]{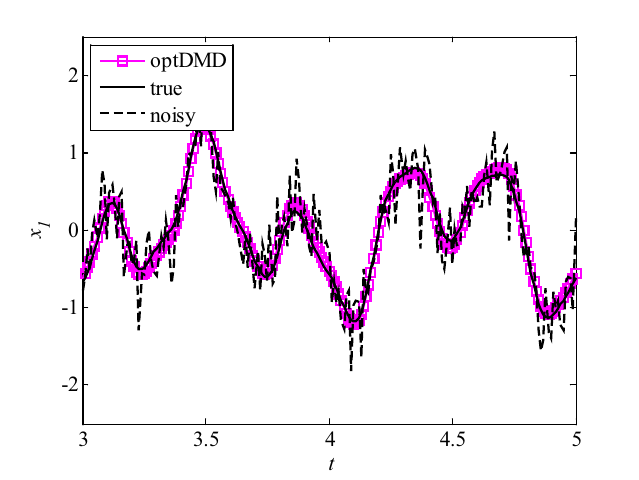}}
	\subfigure[$\sigma_w^2=0.1$,   EKFDMD.]{\includegraphics[width=5cm]{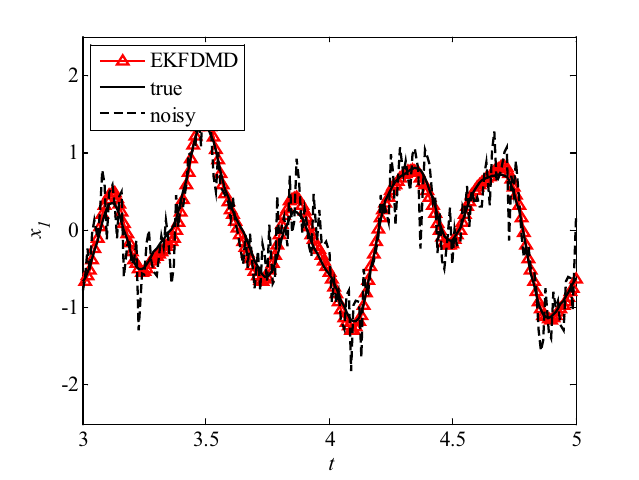}}
	\vspace{-0.2cm}
	\caption{Time histories of the first node of the reconstructed data for a problem with a small number of DoFs without system noise.}
	\label{fig:history_small_onoise}
\end{figure}

The number of DoFs of this system is $d=6$, which is expanded to snapshot data of $r=16$ DoFs by applying the $Q_\text{QR}$ matrix of QR decomposition of a random matrix. Note that this problem was originally extended to $r=400$ DoF, but the number of DoFs is limited in the present study because of the computational costs of EKFDMD, as mentioned above. In this process, a random matrix $T$ of $r \times d$ dimensions in which each of the components is a random number of $\mathcal{N}(0,1)$ is transformed into $T=Q_\text{QR}R_\text{QR}$ by QR decomposition, and the original data $\bm{f}_k$ of dimension $d$ are extended to $\bm{x}_k$ of dimension $r$ by multiplication by matrix $Q_\text{QR}$, as follows: 
\begin{eqnarray}
\bm{x}_{k}=Q_\text{QR}\bm{f}_k.
\end{eqnarray}
Then, $\bm{y}$ data matrices are created by adding white observation noise to the original $\bm{x}$ data matrix, where the noise $\bm{w}_k$ is expressed as $\mathcal{N}\left(0,\sigma^2\right)$. 
\begin{eqnarray}
\bm{y}_{k}=\bm{x}_k+\bm{w}_k.
\end{eqnarray}
Here, the variance ($\sigma_w^2$) is varied as $0.0001, 0.001, 0.01$, and $0.1$ (the noise strengths of which are shown by the solid and dotted lines in Fig.~\ref{fig:history_small_onoise}). A total of 500 snapshots are given, and the eigenvalues of matrix $A$ in the final stage are analyzed.

For the initial adjustable parameters of the Kalman filter, the diagonal parts of the variance matrix are set to be $10^3$. The diagonal elements of $Q$ and $R$ are set to be $0$ and $\sigma_w^2$, respectively, and the nondiagonal elements of $Q$ and $R$ are set to be 0 in this subsection. The assumption of $Q=0$ corresponds to providing the information that the system noise is absent and the system is temporally constant.

\begin{figure}
\centering
\subfigure[$\sigma_w^2=0.0001$.]{\includegraphics[width=5cm]{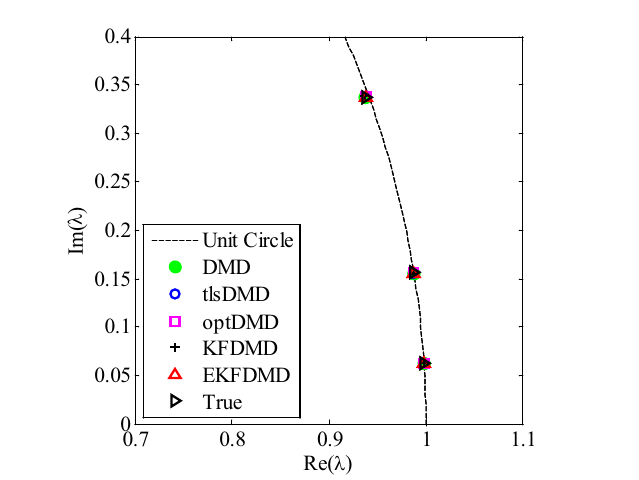}}
\subfigure[$\sigma_w^2=0.001 $.]{\includegraphics[width=5cm]{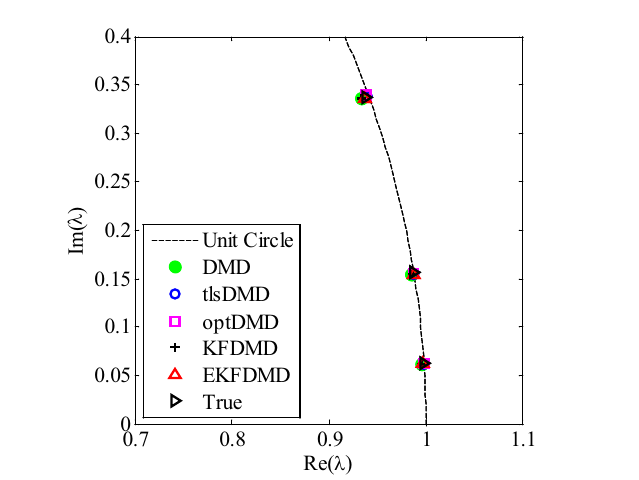}}\\
\subfigure[$\sigma_w^2=0.01  $.]{\includegraphics[width=5cm]{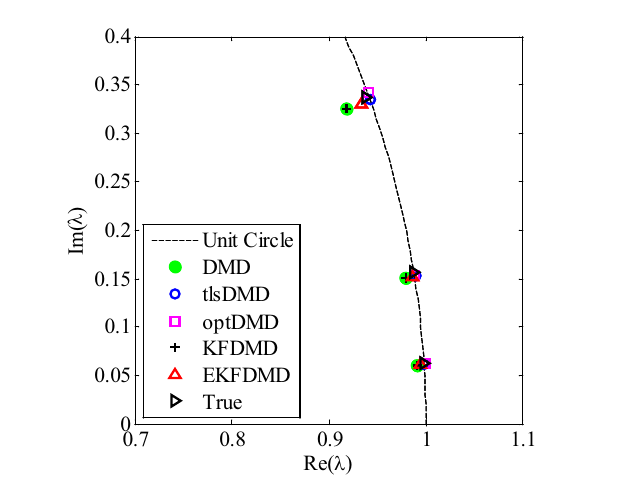}}
\subfigure[$\sigma_w^2=0.1   $.]{\includegraphics[width=5cm]{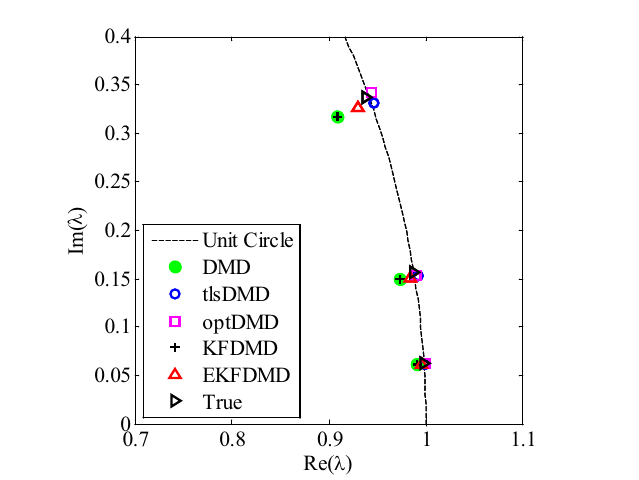}}
\vspace{-0.2cm}
\caption{Eigenvalues for a problem with a small number of DoFs without system noise. The algorithms are almost identical in (a) and (b).}
\label{fig:eigen_small_onoise}
\end{figure}

\begin{figure}
	\centering
    \subfigure[$\sigma_w^2=0.0001$.]{\includegraphics[width=5cm]{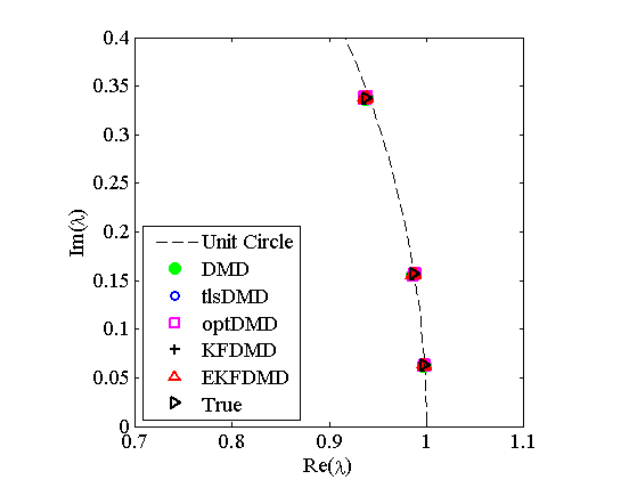}}
	\subfigure[$\sigma_w^2=0.001$.]{\includegraphics[width=5cm]{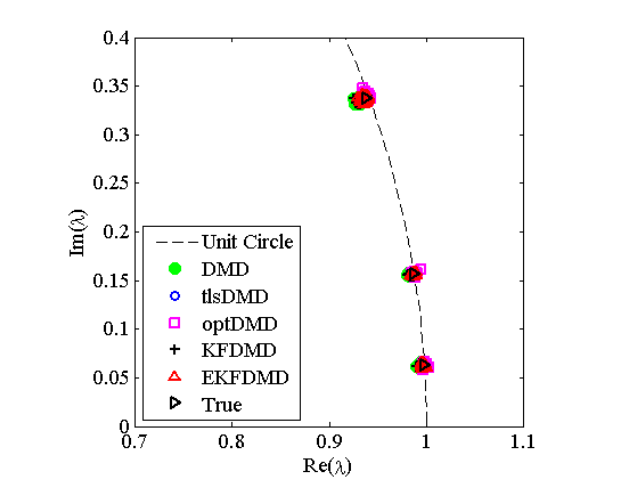}}
	\subfigure[$\sigma_w^2=0.01$.]{\includegraphics[width=5cm]{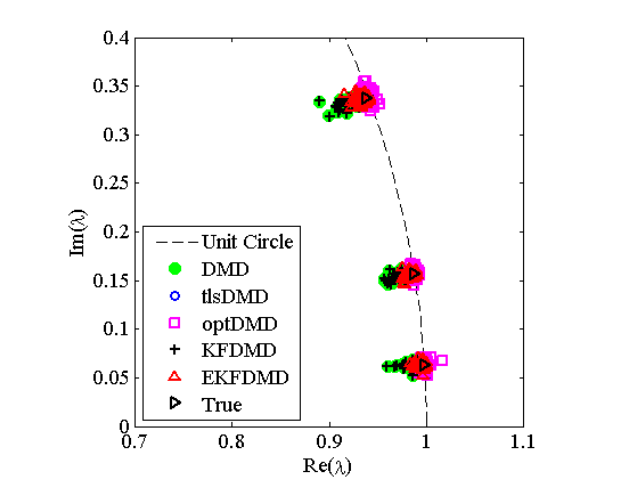}}\\
	\subfigure[$\sigma_w^2=0.1$ without EKFDMD.]{\includegraphics[width=5cm]{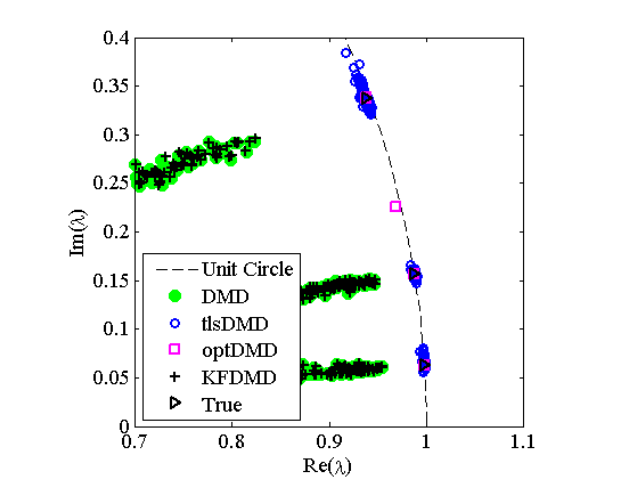}}
	\subfigure[$\sigma_w^2=0.1$ without optDMD.]{\includegraphics[width=5cm]{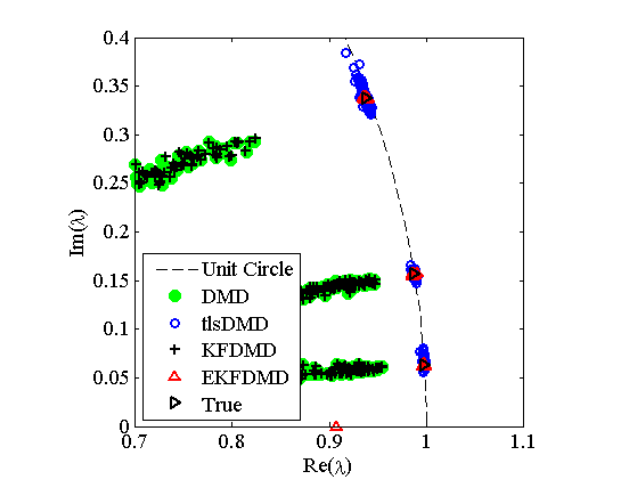}}
	\vspace{-0.2cm}
	\caption{Eigenvalues for multiple runs of a problem with a small number of DoFs without system noise, where the seed for the random number is different for multiple runs.}
	\label{fig:eigen_small_onoise_multi}
\end{figure}

The results for the noisy data while changing the noise level are discussed. Figures \ref{fig:eigen_small_onoise} and \ref{fig:eigen_small_onoise_multi} show the eigenvalues estimated in the representative case and in all of the 100 cases we examined by changing the random number seed, respectively\color{black}. The results of the estimated eigenvalues in Figs.~\ref{fig:eigen_small_onoise} and \ref{fig:eigen_small_onoise_multi} show that DMD and KFDMD do not work well for accurate estimation of the eigenvalues of the system for the case in which the noise level is high. On the other hand, tlsDMD works better than DMD and KFDMD. Furthermore, optDMD and EKFDMD appear to work the best for estimation of the eigenvalues. This might be because optDMD and EKFDMD denoises the data, and a more accurate eigenvalue of the system can be obtained by the denoised data. The system identification performance of EKFDMD appears to be better than that of tlsDMD. 

\begin{figure}
	\centering
	\subfigure[$\lambda_1$.]{\includegraphics[width=5cm]{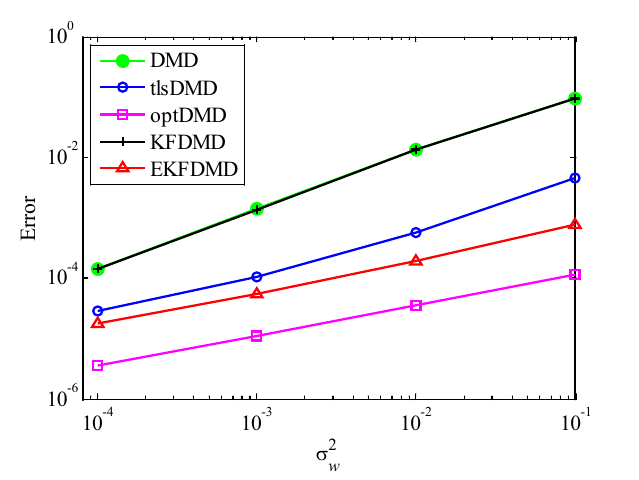}}
	\subfigure[$\lambda_2$.]{\includegraphics[width=5cm]{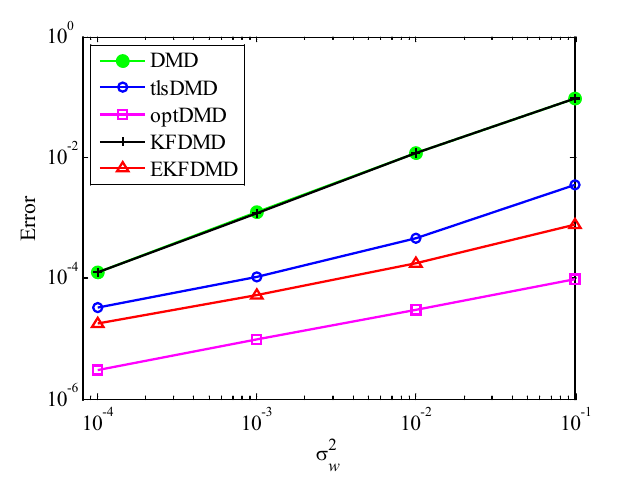}}
	\subfigure[$\lambda_3$.]{\includegraphics[width=5cm]{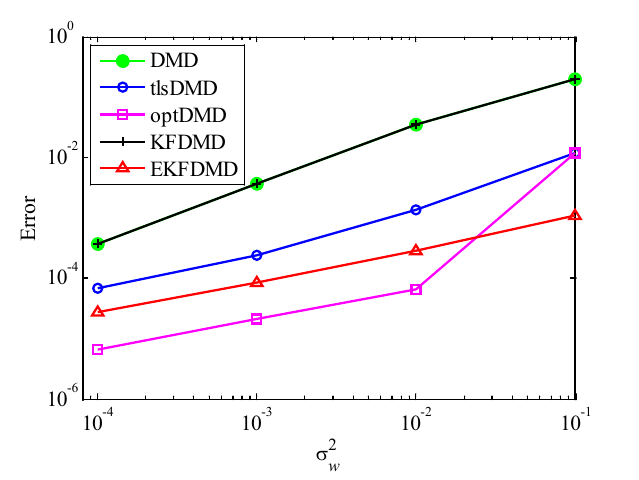}}\\
	\caption{Errors in the eigenvalues for multiple runs of a problem with a small number of DoFs without system noise.}
	\label{fig:error_eigen_small_onoise}
\end{figure}

 The above characteristics are discussed with the quantitative data. Figure \ref{fig:error_eigen_small_onoise} shows the error of eigenvalues\color{black}. The errors in the eigenvalues are defined by the norm of the closest computed eigenvalue to the true eigenvalue specified. Here, outliers were not removed in this process. The error in the eigenvalues decreases with decreasing noise strength for all methods. This plot quantitatively shows that the error basically decreases with the order of DMD as well as KFDMD, tlsDMD, EKFDMD, and optDMD. The system noise is not considered in the present problem setting, and therefore optDMD can give the best-fit curve for the all of the data points, owing to its offline procedures. On the other hand, EKFDMD incrementally updates the information and cannot use all of the data at once. Therefore, it is reasonable that optDMD works slightly better than EKFDMD.

Data reconstruction is then considered. In addition, as noted previously, EKFDMD is expected to be able to denoise the data. Figure \ref{fig:history_small_onoise}, which illustrates the time-series of the true data, the observation (noisy) data and the reconstructed data of DMD, tlsDMD, KFDMD, optDMD, and EKFDMD. This plot reveals that DMD and KFDMD cannot predict the oscillation because they estimate the dumping oscillation due to the noise included in the observation data. Moreover, tlsDMD can predict the oscillation for the weaker noise level. Although tlsDMD can predict neutral oscillation for a stronger noise level, as shown in Fig.~\ref{fig:history_small_onoise}, the phase of oscillation of reconstructed data is very different from the true value. On the other hand, optDMD and EKFDMD can successfully denoise the data, even though the noise level is very high. 

\begin{figure}
	\centering
	{\includegraphics[width=5cm]{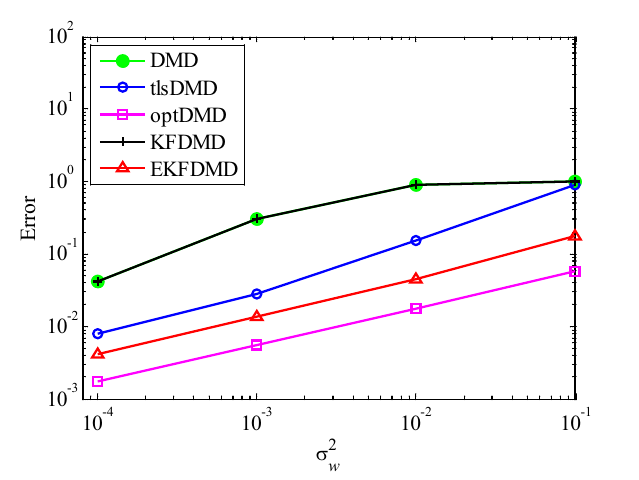}}
    \caption{Errors in the reconstructed data for multiple runs of a problem with a small number of DoFs without system noise.}
	\label{fig:error_history_small_onoise}
\end{figure}

The error level of the reconstructed data is quantitatively discussed in term of Fig.~\ref{fig:error_history_small_onoise}, which shows the following normalized error:
\begin{eqnarray}
E_{\text{reconst}}=\frac{\|X_{\text{reconst},101:m}-X_{101:m}\|^2_F}{\|X_{101:m}\|^2_F}.
\end{eqnarray}
Figure \ref{fig:error_history_small_onoise} shows that the error decreases with the order of DMD, KFDMD, tlsDMD, EKFDMD, and optDMD, similar to those in the eigenvalues. 
This trend also shows that EKFDMD works reasonably for simultaneous system identification and denoising of the data by running the algorithm online. The better performance of optDMD, as compared to EKFDMD, originates from their online or offline characteristics.

Although we are interested in the performance for the case in which system noise is present, we hereinafter discuss the effects of parameters on this problem without system noise, before discussing the problem with system noise in Subsection~\ref{sec:testSPO}.
\color{black}


\subsubsection{Effect of the number of snapshots $m$}
\begin{figure}
	\centering
	\subfigure[$\lambda_1$.]{\includegraphics[width=5cm]{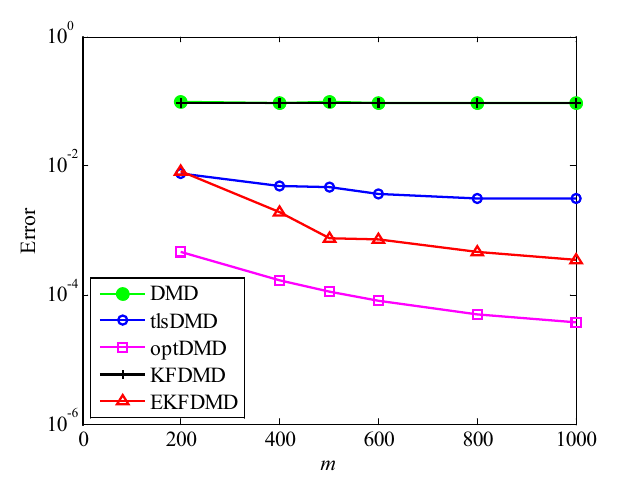}}
	\subfigure[$\lambda_2$.]{\includegraphics[width=5cm]{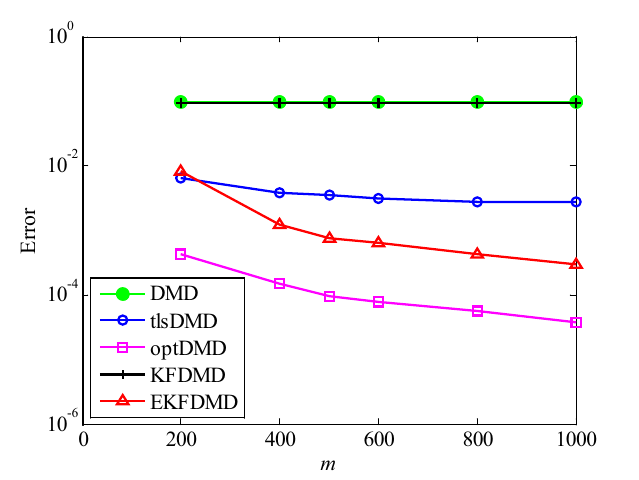}}
	\subfigure[$\lambda_3$.]{\includegraphics[width=5cm]{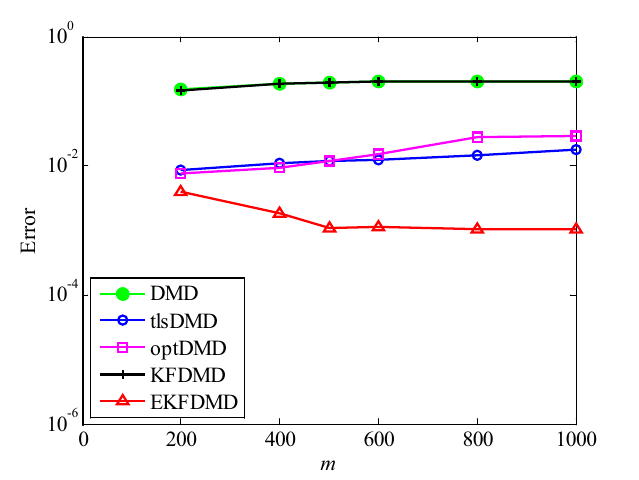}}		
	\vspace{-0.2cm}
	\caption{Effect of $m$ on errors in the eigenvalues for multiple runs of a problem with a small number of DoFs without system noise.}
	\label{fig:error_eigen_small_onoise_m}
\end{figure}

Here, the parameter effects for the problem without system noise are considered. First, the effect of the number of snapshots $m$ is investigated. Similar to the previous discussion, the errors in the eigenvalues and reconstructed data for DMD, tlsDMD, KFDMD, optDMD, and EKFDMD are calculated for various values of $m$ for data of $\sigma_w^2=0.1$. These errors are evaluated by 100 runs and are averaged for each algorithm. 
The error in eigenvalues in Fig.~\ref{fig:error_eigen_small_onoise_m} shows that the errors of tlsDMD, EKFDMD, and optDMD basically decrease (except for some bumps), while those of DMD and KFDMD do not. Interestingly, the error of EKFDMD decreases more rapidly and is larger than that of tlsDMD for $m\le200$ but smaller for $m\ge300$. This is because EKFDMD is an online algorithm and its accuracy in the early stage is not sufficiently high, but increases rapidly as more successive data are obtained. Note that both tlsDMD and optDMD algorithms are offline algorithms.

\begin{figure}
	\centering
	{\includegraphics[width=5cm]{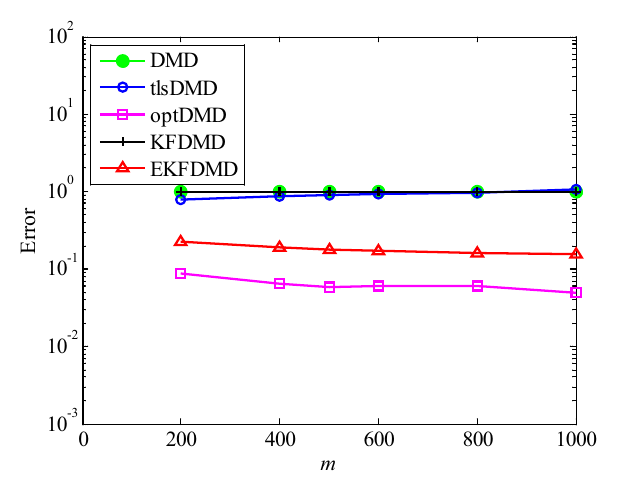}}
	\vspace{-0.2cm}
    \caption{Effect of $m$ on errors in the reconstructed data for multiple runs of a problem with a small number of DoFs without system noise.}
	\label{fig:error_history_small_onoise_m}
\end{figure}

Then, the errors in reconstructed data shown in Fig.~\ref{fig:error_history_small_onoise_m} are discussed. The errors of DMD, KFDMD, and tlsDMD do not change. The errors of DMD and KFDMD do not decrease because they cannot better predict the eigenvalues for the case in which $m$ increases, and the errors of tlsDMD do not decrease, despite the decrease in the error in the eigenvalues, because the reconstructed data with tlsDMD have a different phase due to the very strong observation noise in the initial snapshot, as discussed previously. On the other hand, the errors of EKFDMD and optDMD decrease because both algorithms find the best-fit data for reconstruction and the accuracy of this data increases by using the information of an increased number of snapshots. 

\subsubsection{Effect of mismatched error level for $R$}
\begin{figure}
	\centering
	\subfigure[$\lambda_1$.]{\includegraphics[width=5cm]{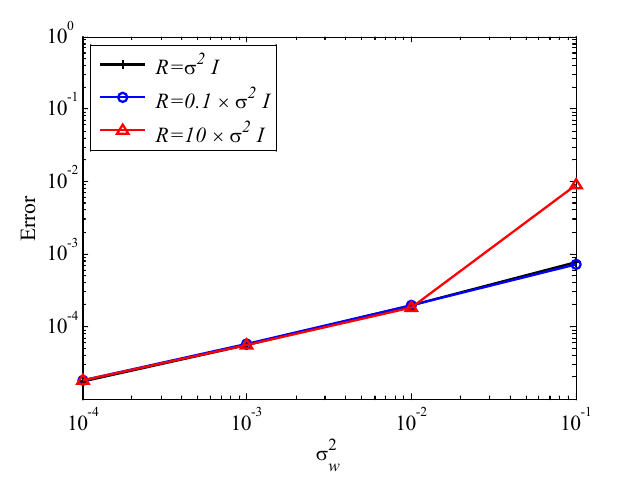}}
	\subfigure[$\lambda_2$.]{\includegraphics[width=5cm]{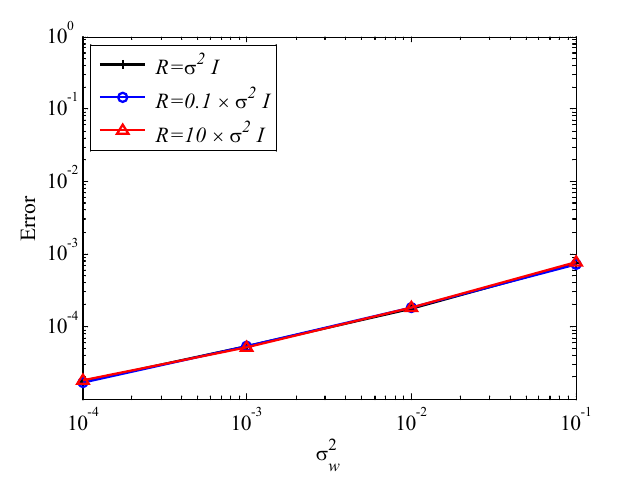}}
	\subfigure[$\lambda_3$.]{\includegraphics[width=5cm]{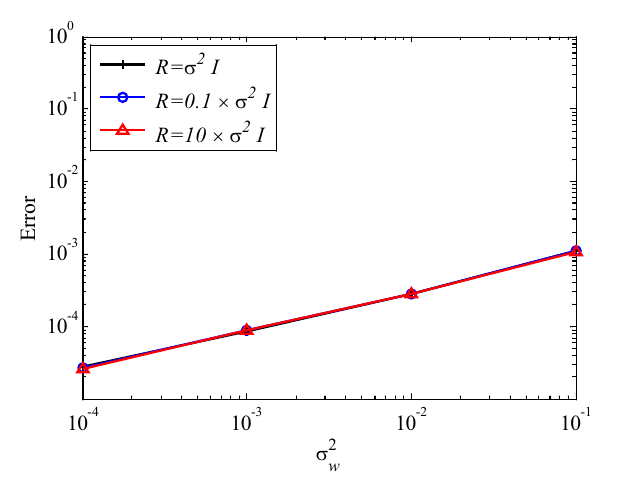}}	
	\vspace{-0.2cm}
	\caption{Effect of $R$ on errors in the eigenvalues for multiple runs of a problem with a small number of DoFs without system noise.}
	\label{fig:error_eigen_small_noise_R}
\end{figure}

\begin{figure}
	\centering
	{\includegraphics[width=5cm]{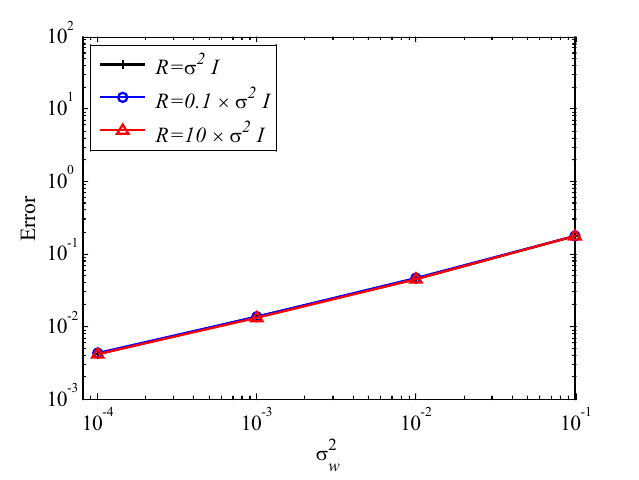}}
	\vspace{-0.2cm}
    \caption{Effect of $R$ on errors in reconstructed data for multiple runs of a problem with a small number of DoFs without system noise.}
	\label{fig:error_history_small_noise_R}
\end{figure}

Next, the effect of mismatched $R$ settings is discussed, while the system error is absent and $Q$ is set to be ${0}$. In the present study, we investigate the mismatched cases of $R=10\sigma^2_wI$ and $R=0.1\sigma^2_wI$, as well as the matched case of $R=\sigma^2_wI$, the results of which are presented in the previous sections. The number of snapshots $m$ is set to be 500. The errors are evaluated by 100 runs and are averaged for each case, similar to previous cases. The errors of EKFDMD in eigenvalues and reconstructed data for the case in which $R$ is mismatched are shown in Figs.~\ref{fig:error_eigen_small_noise_R} and \ref{fig:error_history_small_noise_R}, respectively. These figures show that the mismatched $R$ does not affect the results, except for the strong-observation-noise case ($\sigma_w^2=0.1$), because the balance of $R$ and $Q$ changes the behavior of Kalman filter, whereas a change in $R$ under the condition of $Q=0$ does not affect the behavior of Kalman filter. 


\clearpage

\subsection{Problem with a small number of DoFs with system noise}
\label{sec:testSPO}
Next, we consider a problem with system noise. In this problem, ${\bm{v}}'$ is assumed to be $\mathcal{N}(0,n\sigma^2_v/6)$, resulting in ${\bm{v}}$ being  $\mathcal{N}(0,\sigma^2_v)$, and we vary $\sigma^2_v$= $\sigma^2_w$ as 0.1, 0.01, 0.001, and 0.0001. A hyperparameter $Q$ is set to be
\begin{eqnarray}
Q=\left[\begin{array}{cc}
Q_{1,1} & Q_{1,2}\\
Q_{2,1} & Q_{2,2}\\
\end{array} 
\right]
=
\left[\begin{array}{cc}
\sigma^2_v I_{n\times n} & 0\\
0 & 0\\
\end{array} 
\right]
\end{eqnarray}
and $R$ is set to be $\sigma^2_w I$.
The number of snapshots $m$ is set to be 500, and a total of 100 runs are conducted for each case. 

\begin{figure}
	\centering
	\subfigure[$\sigma_w^2=0.0001$.]{\includegraphics[width=5cm]{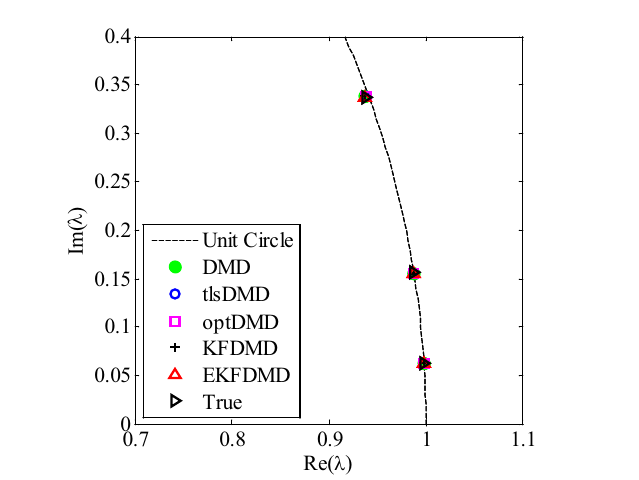}}
	\subfigure[$\sigma_w^2=0.001 $.]{\includegraphics[width=5cm]{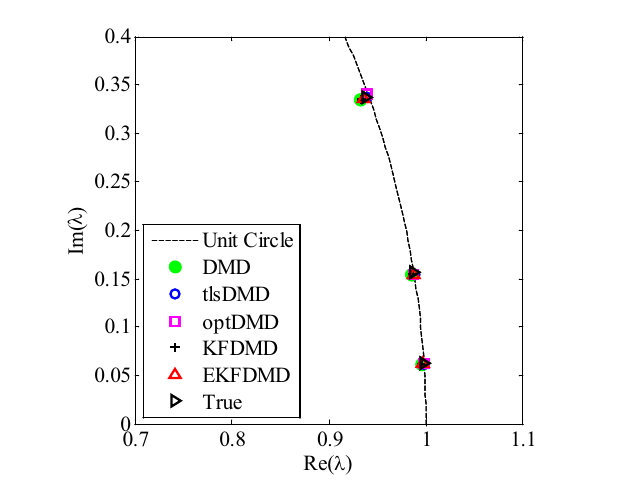}}\\
	\subfigure[$\sigma_w^2=0.01  $.]{\includegraphics[width=5cm]{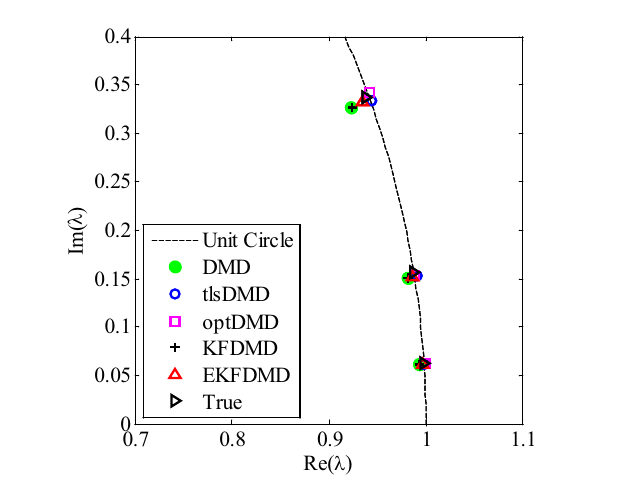}}
	\subfigure[$\sigma_w^2=0.1   $.]{\includegraphics[width=5cm]{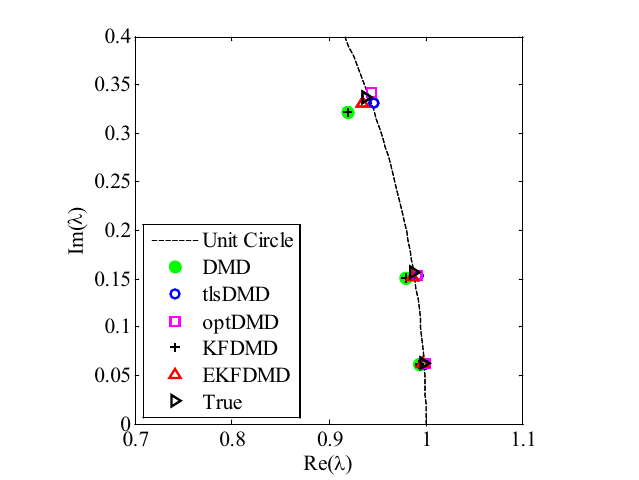}}
	\vspace{-0.2cm}
	\caption{Eigenvalues for a problem with a small number of DoFs with system noise. The algorithms are almost identical in (a) and (b), and tlsDMD, optDMD, and EKFDMD are almost identical in (c) and (d).}
	\label{fig:eigen_small_sonoise}
\end{figure}

\begin{figure}
	\centering
	\subfigure[$\sigma_w^2=0.0001$.]{\includegraphics[width=5cm]{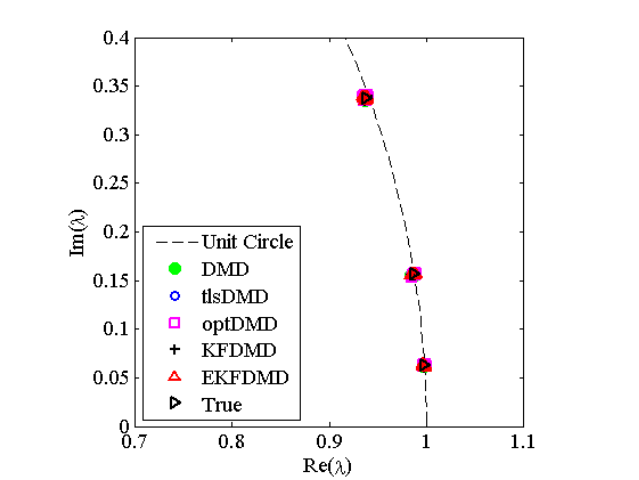}}
	\subfigure[$\sigma_w^2=0.001$.]{\includegraphics[width=5cm]{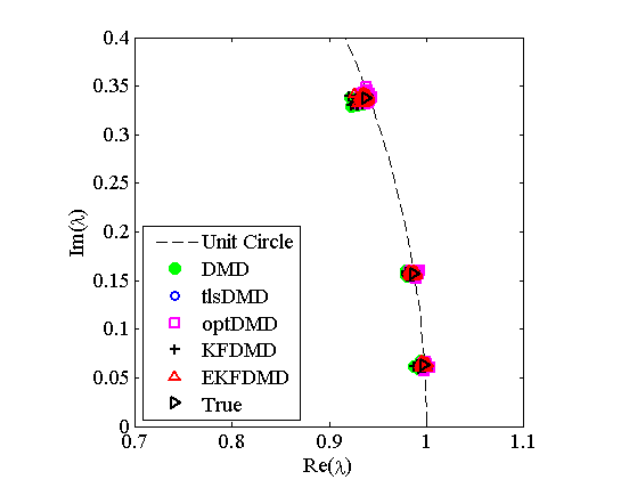}}
	\subfigure[$\sigma_w^2=0.01$.]{\includegraphics[width=5cm]{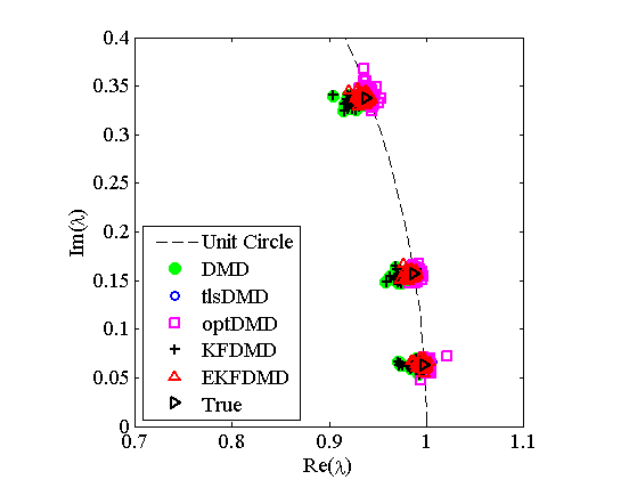}}\\
	\subfigure[$\sigma_w^2=0.1$ without optDMD and EKFDMD.]{\includegraphics[width=5cm]{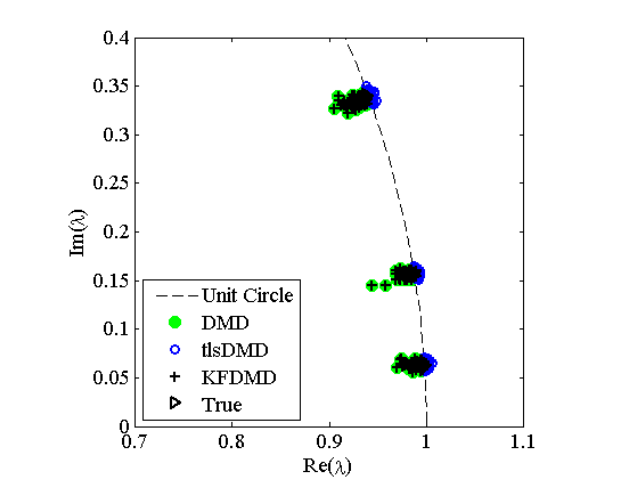}}
	\subfigure[$\sigma_w^2=0.1$ without EKFDMD.]{\includegraphics[width=5cm]{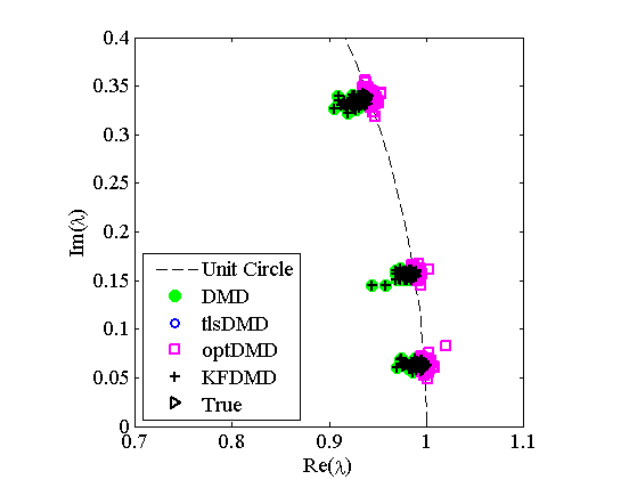}}
	\subfigure[$\sigma_w^2=0.1$ without optDMD.]{\includegraphics[width=5cm]{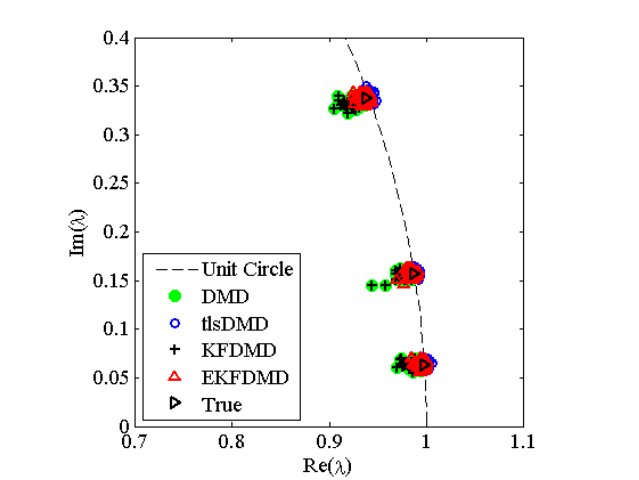}}
	\vspace{-0.2cm}
	\caption{Eigenvalues for multiple runs of a problem with a small number of DoFs with system noise, where the seed for the random number is different for multiple runs.}
	\label{fig:eigen_small_sonoise_multi}
\end{figure}

\begin{figure}
	\centering
	\subfigure[$\lambda_1$.]{\includegraphics[width=5cm]{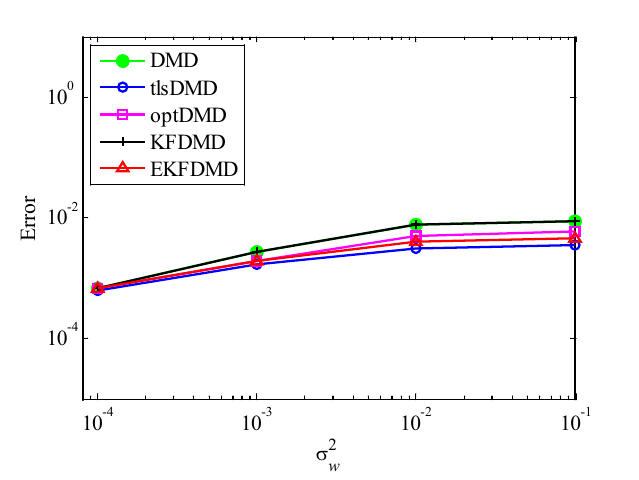}}
	\subfigure[$\lambda_2$.]{\includegraphics[width=5cm]{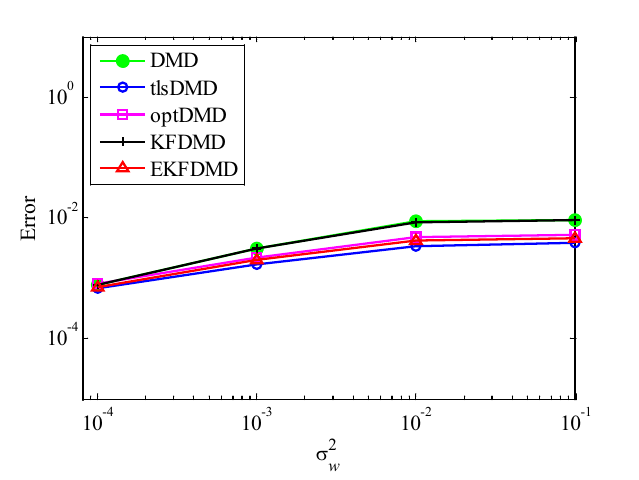}}
	\subfigure[$\lambda_3$.]{\includegraphics[width=5cm]{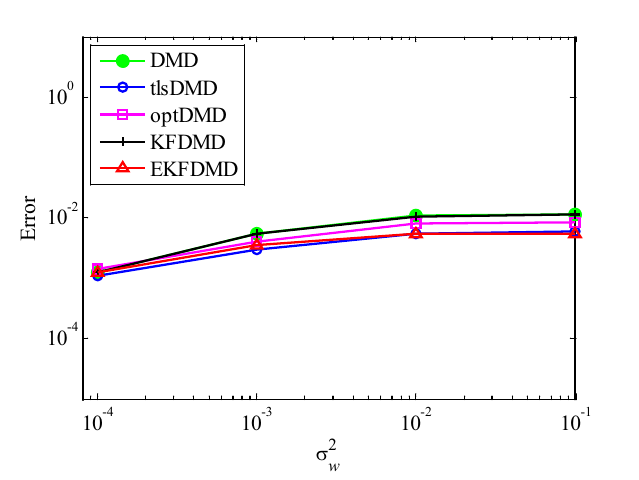}}\\
	\caption{Errors in the eigenvalues for multiple runs of a problem with a small number of DoFs without system noise.}
	\label{fig:error_eigen_small_sonoise}
\end{figure}

Figures \ref{fig:eigen_small_sonoise} and \ref{fig:eigen_small_sonoise_multi} show the eigenvalues estimated in the representative case and in all 100 cases we examined by changing the seed of the random numbers, respectively. Figures \ref{fig:eigen_small_sonoise} and \ref{fig:eigen_small_sonoise_multi} show that DMD and KFDMD do not work well for the accurate estimation of the eigenvalues of the system for the case in which the noise level is high, although its accuracy is somehow improved compared with the case without the system noise. On the other hand, tlsDMD, optDMD and EKFDMD appear to work better than DMD or KFDMD. This might be because denoising algorithms for estimation of eigenvalues of tlsDMD, optDM, and EKFDMD works well for these data, and a more accurate eigenvalue of the system can be obtained. The system identification performance of EKFDMD appears to be as good as that of tlsDMD and optDMD in these plot. Finally, the errors of eigenvalue estimation are shown in Fig.~\ref{fig:error_eigen_small_sonoise}. Figure \ref{fig:error_eigen_small_sonoise} shows that tlsDMD, optDMD, and EKFDMD work better than DMD and KFDMD. Among tlsDMD, optDMD, and EKFDMD, tlsDMD works slightly better for $\lambda_1$ and $\lambda_2$, whereas the performance of EKFDMD is similar to that of tlsDMD for $\lambda_3$. This result illustrates that the system identification performances of tlsDMD, optDMD, and EKFDMD are approximately the same for the case in which system noise is present.

\begin{figure}
	\centering
	\subfigure[$\sigma_w^2=0.0001$.       ]{\includegraphics[width=5cm]{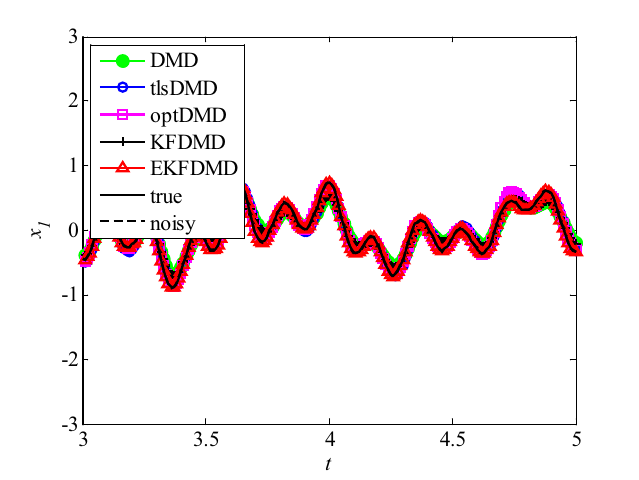}}
	\subfigure[$\sigma_w^2=0.001$.        ]{\includegraphics[width=5cm]{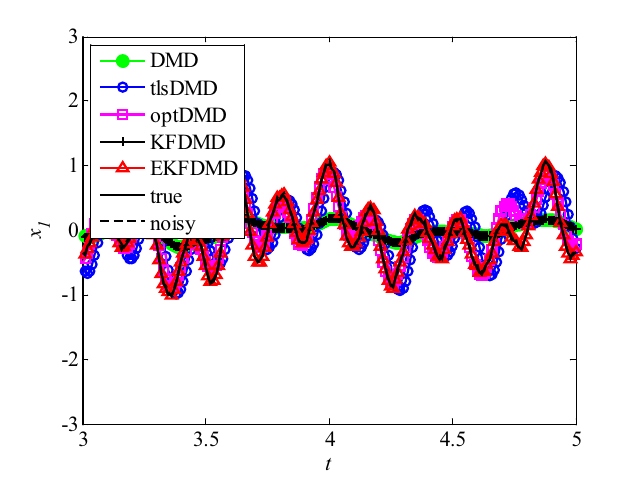}}\\
	\subfigure[$\sigma_w^2=0.01$.         ]{\includegraphics[width=5cm]{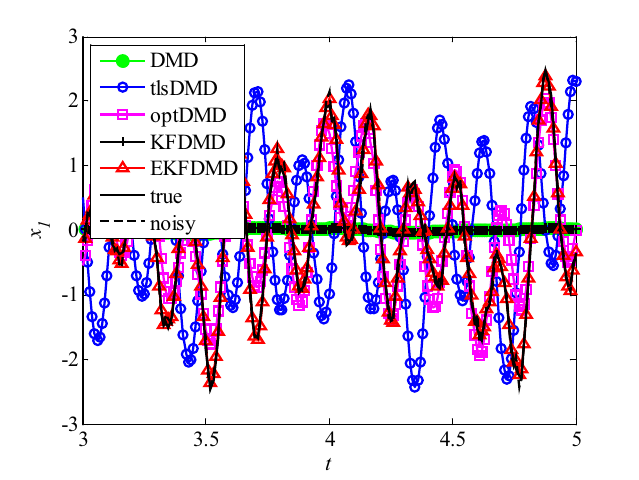}}
	\subfigure[$\sigma_w^2=0.1$.          ]{\includegraphics[width=5cm]{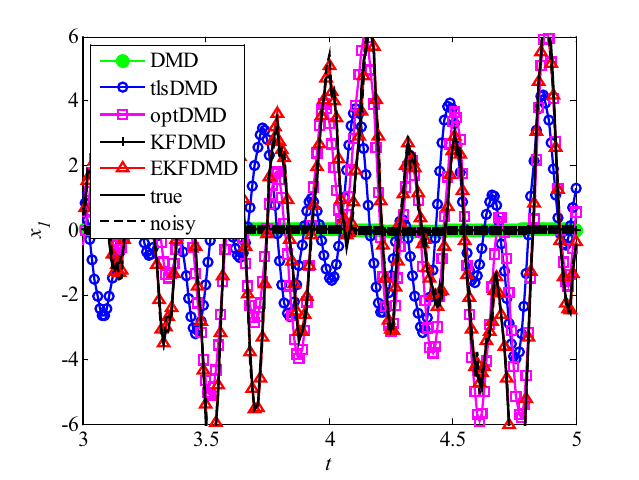}}\\		
	\subfigure[$\sigma_w^2=0.01$,  tlsDMD.]{\includegraphics[width=5cm]{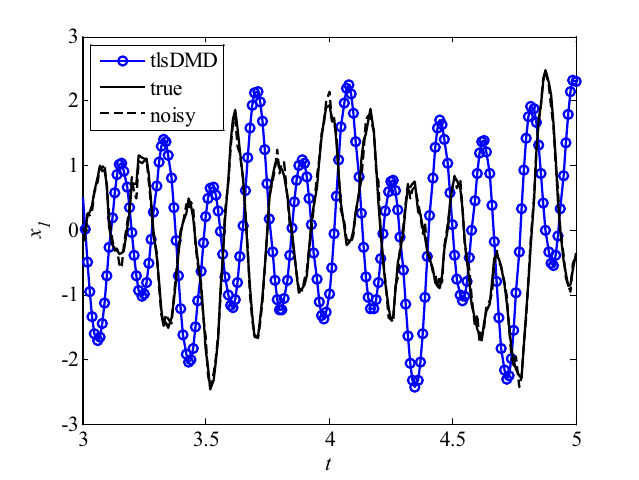}}
	\subfigure[$\sigma_w^2=0.01$,  optDMD.]{\includegraphics[width=5cm]{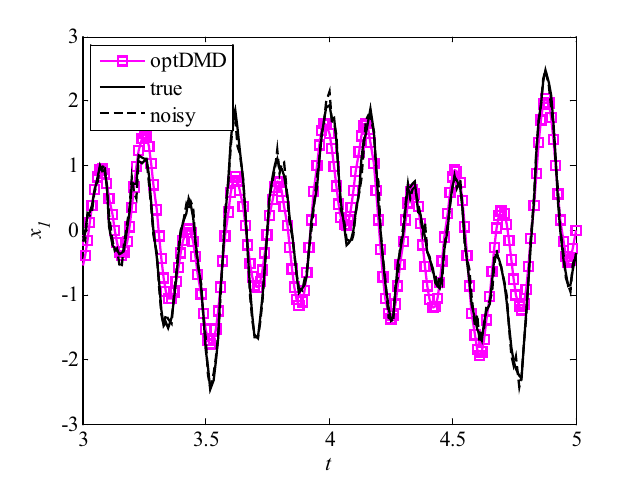}}
	\subfigure[$\sigma_w^2=0.01$,  EKFDMD.]{\includegraphics[width=5cm]{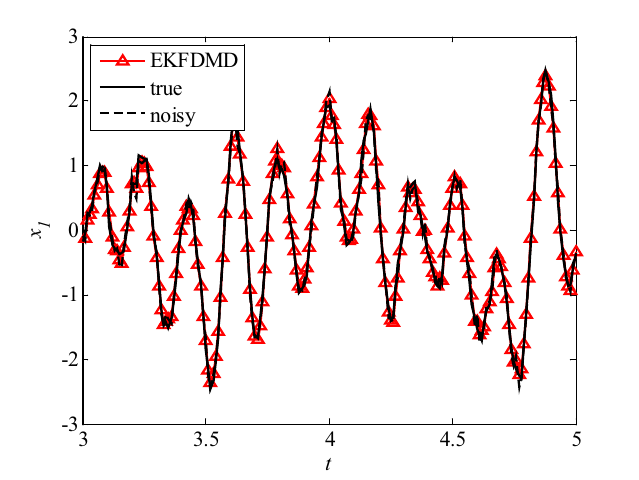}}\\
	\subfigure[$\sigma_w^2=0.1$,   tlsDMD.]{\includegraphics[width=5cm]{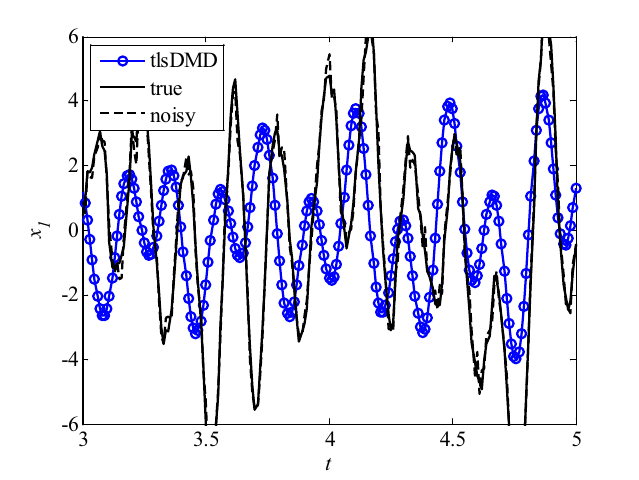}}
	\subfigure[$\sigma_w^2=0.1$,   optDMD.]{\includegraphics[width=5cm]{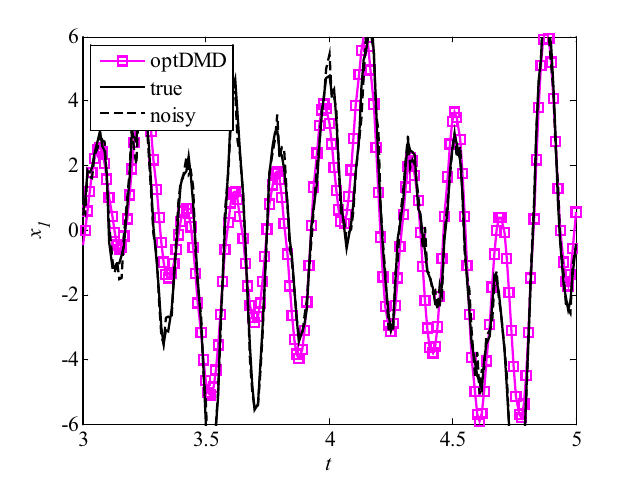}}
	\subfigure[$\sigma_w^2=0.1$,   EKFDMD.]{\includegraphics[width=5cm]{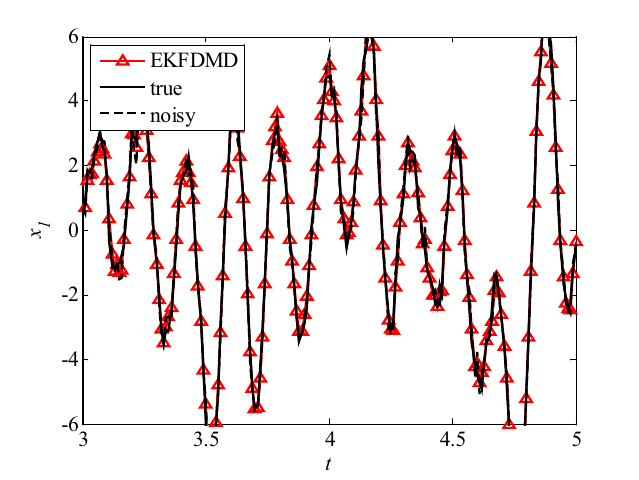}}
	\vspace{-0.2cm}
	\caption{Reconstructed data of the first node for a problem with a small number of DoFs with system noise.}
	\label{fig:history_small_sonoise}
\end{figure}

\begin{figure}
	\centering
	{\includegraphics[width=5cm]{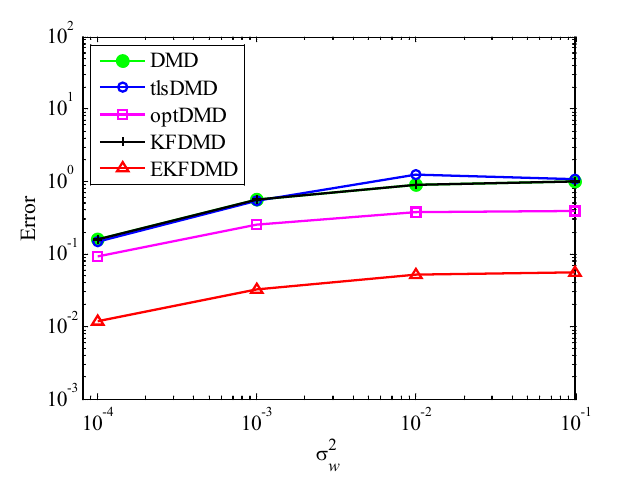}}
	\caption{Errors in the reconstructed data for multiple runs of a problem with a small number of DoFs with system noise.}
	\label{fig:error_history_small_sonoise}
\end{figure}

Then, reconstruction using these algorithms, as shown in Fig.~\ref{fig:history_small_sonoise}, is discussed. Similar to the cases without system noise, data reconstructed by DMD and KFDMD are dumped in the early stage. This is again because the these algorithms predict dumping modes. The data reconstructed by tlsDMD have good amplitude of oscillations, but their phases do not match well with those of the original data. Although the data reconstructed by optDMD have good amplitude and phase, the data around peaks are sometimes not reconstructed. These errors around peaks in the reconstruction data obtained using optDMD are caused by system noise in the data because optDMD cannot handle system noise. Unlike the algorithm described above, the data reconstructed by EKFDMD shows excellent agreement with the original data. This is because EKFDMD can handle data with system noise. This characteristic can be used for simultaneous online system identification and denoising of data containing system noise. The error in the reconstructed data shown in Fig.~\ref{fig:error_history_small_sonoise} clearly shows this characteristic.


\subsubsection{Effects of the balance of system and observation noises}
In this subsubsection, the effects of the balance of system and observation noises in the observation data are discussed. System noise variance $\sigma_v^2$ is set to be 10$\sigma_w^2$ and 0.1$\sigma_w^2$. Here, $Q$ and $R$ are correctly given in this problem. In both cases, test cases with $\sigma^2_w$ of 0.1, 0.01, 0.001, and 0.0001 are conducted, and the results of 100 runs with different seeds for random numbers are averaged for error characteristics.

\begin{figure}
	\centering
	\subfigure[$\lambda_1$.]{\includegraphics[width=5cm]{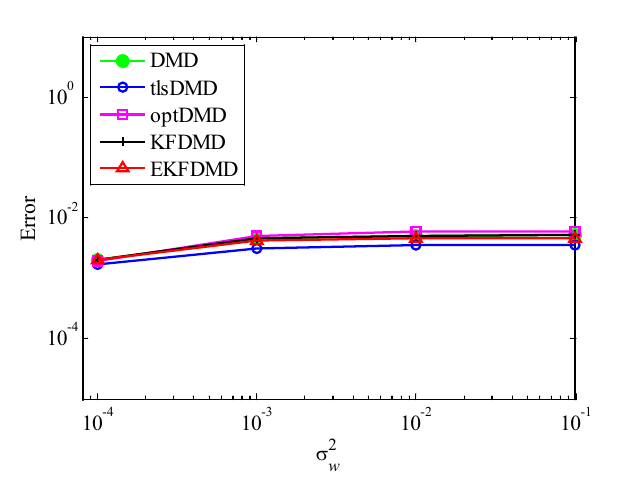}}
	\subfigure[$\lambda_2$.]{\includegraphics[width=5cm]{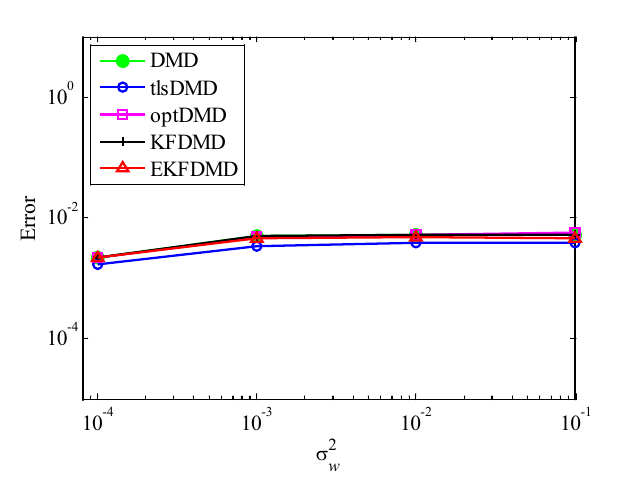}}
	\subfigure[$\lambda_3$.]{\includegraphics[width=5cm]{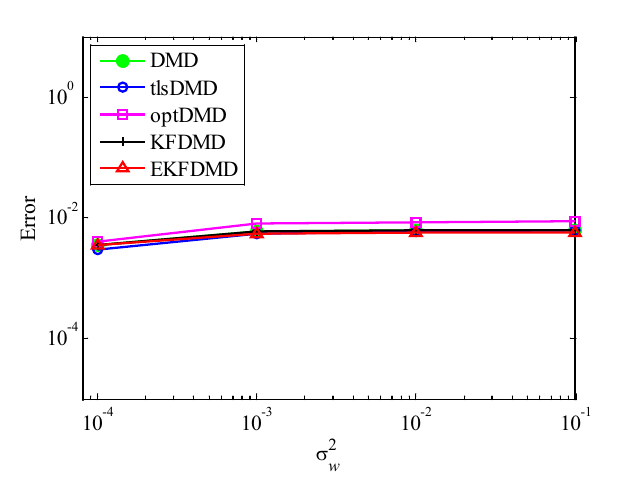}}\\
	\caption{Errors in the eigenvalues for multiple runs of a problem with a small number of DoFs without system noise for the case in which $\sigma^2_v=10\sigma^2_w$.}
	\label{fig:error_eigen_small_sonoise10}
\end{figure}

\begin{figure}
	\centering
	\includegraphics[width=5cm]{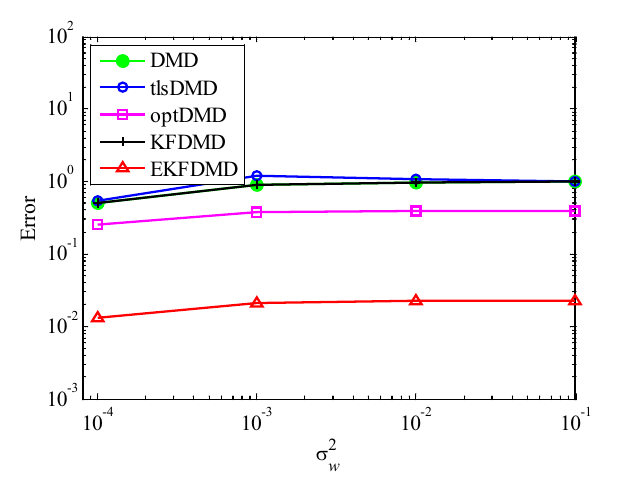}
	\caption{Errors in the reconstructed data for multiple runs of a problem with a small number of DoFs with system noise for the case in which $\sigma^2_v=10\sigma^2_w$.}
	\label{fig:error_history_small_sonoise10}
\end{figure}

First, the case with strong system noise $\sigma_v^2=10\sigma_w^2$ is discussed. The errors in the estimated eigenvalues shown in Fig.~\ref{fig:error_eigen_small_sonoise10} indicate that the errors of all of the algorithms are almost the same and the error does not decrease with decreasing noise level. This figure shows that advanced DMD methods do not significantly improve the estimation of eigenvalues for data with strong system noise. The error in the reconstructed data is shown in Fig.~\ref{fig:error_history_small_sonoise10}. This figure shows that the error of EKFDMD is much less than the errors of the other algorithms. This indicates that EKFDMD can be used for noise reduction for the case in which the system noise is stronger than the observation noise. 

\begin{figure}
	\centering
	\subfigure[$\lambda_1$.]{\includegraphics[width=5cm]{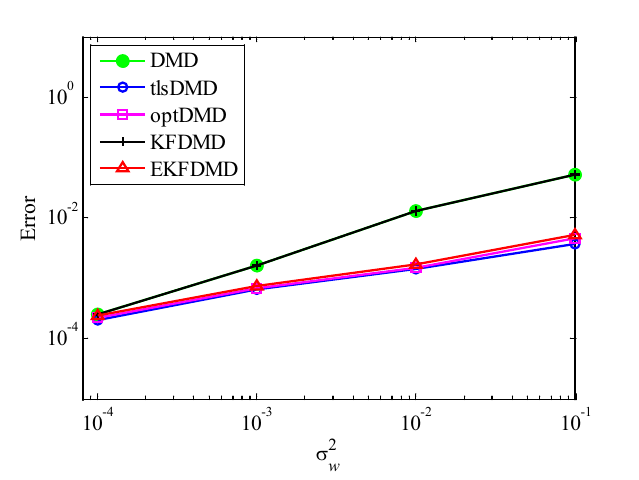}}
	\subfigure[$\lambda_2$.]{\includegraphics[width=5cm]{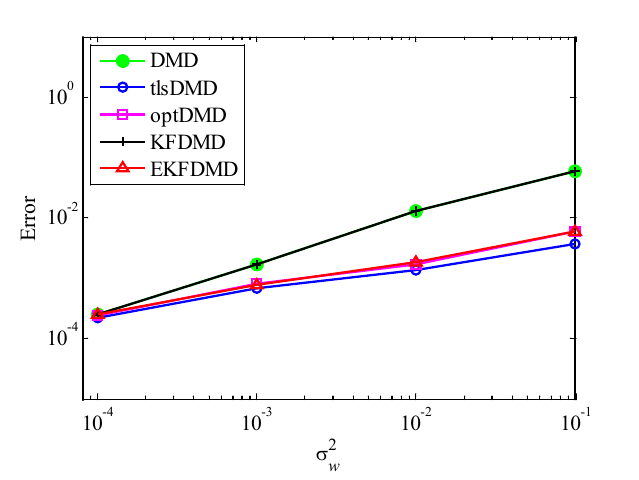}}
	\subfigure[$\lambda_3$.]{\includegraphics[width=5cm]{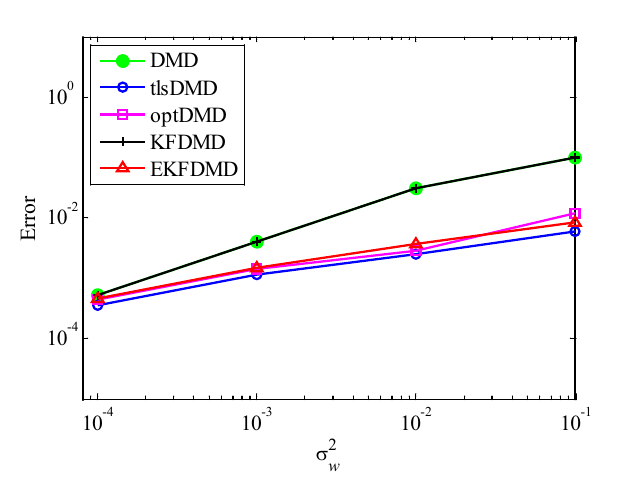}}\\
	\caption{Errors in the eigenvalues for multiple runs of a problem with a small number of DoFs with system noise for the case in which $\sigma^2_v=0.1\sigma^2_w$.}
	\label{fig:error_eigen_small_sonoise01}
\end{figure}

\begin{figure}
	\centering
	{\includegraphics[width=5cm]{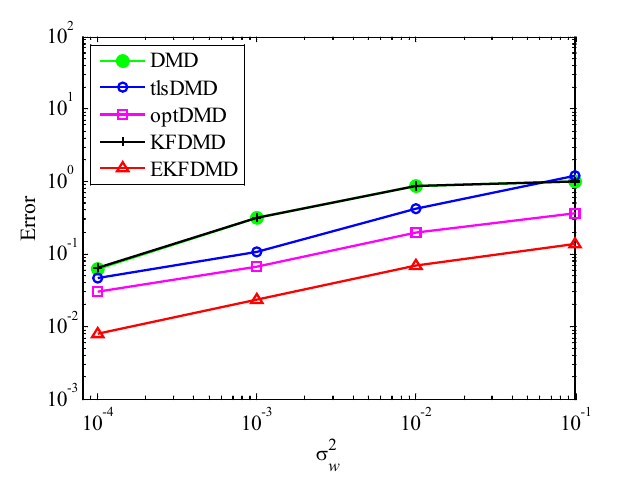}}
	\caption{Errors in the reconstructed data for multiple runs of a problem with a small number of DoFs with system noise for the case in which $\sigma^2_v=0.1\sigma^2_w$.}
	\label{fig:error_history_small_sonoise01}

\end{figure}

Then, the case with the weaker system noise $\sigma_v^2=0.1\sigma_w^2$ is discussed. Again, $Q$ and $R$ are correctly given in this problem. The error plots in Fig.~\ref{fig:error_eigen_small_sonoise01} show that the errors of tlsDMD, optDMD, and EKFDMD are approximately the same and are lower than those of DMD and KFDMD. This figure illustrates that advanced DMD methods improve the estimation ability of eigenvalues. The error in the reconstructed data is shown in Fig.~\ref{fig:error_history_small_sonoise01}. This plot indicates that the errors decrease in the order of DMD and KFDMD (same as that of DMD), tlsDMD, optDMD, and EKFDMD. The figure also shows that EKFDMD performs better than optDMD, even if weaker system noise is present. This fact indicates that EKFDMD can be used for noise reduction in the range we investigated for the case in which system noise is present, regardless of its strength.

\subsubsection{Effects of mismatched error level for $Q$ and $R$}
\begin{figure}
	\centering
	\subfigure[$\lambda_1$.]{\includegraphics[width=5cm]{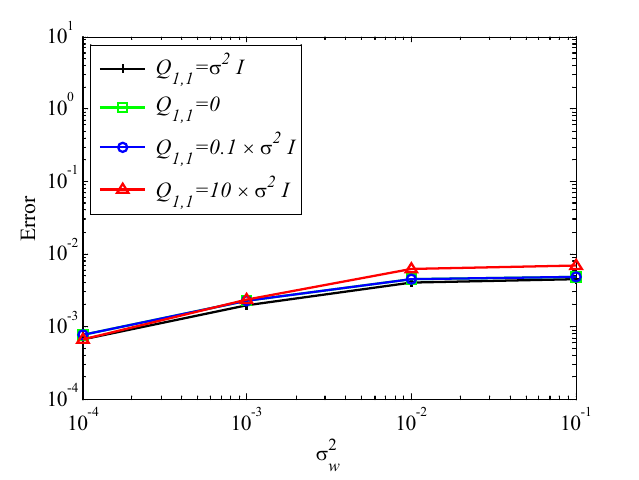}}
	\subfigure[$\lambda_2$.]{\includegraphics[width=5cm]{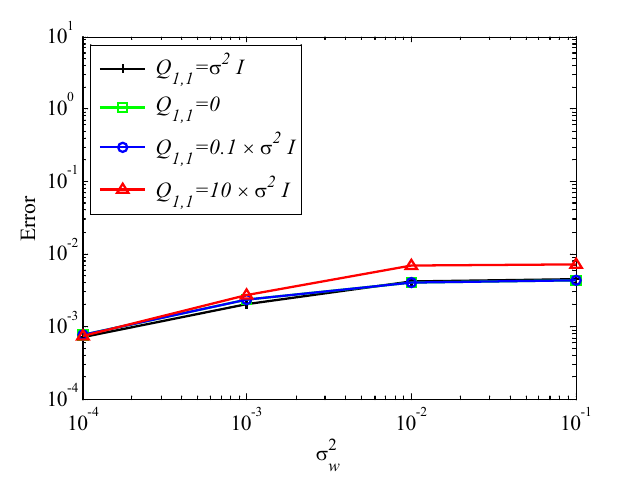}}
	\subfigure[$\lambda_3$.]{\includegraphics[width=5cm]{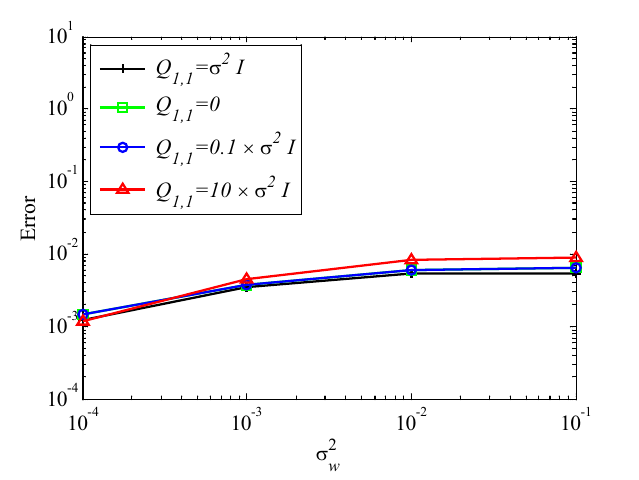}}\\
	\caption{Effect of mismatched $Q$ on the errors in the eigenvalues for multiple runs of a problem with a small number of DoFs with system noise.}
	\label{fig:error_eigen_small_sonoise_misQ}
\end{figure}

\begin{figure}
	\centering
	{\includegraphics[width=5cm]{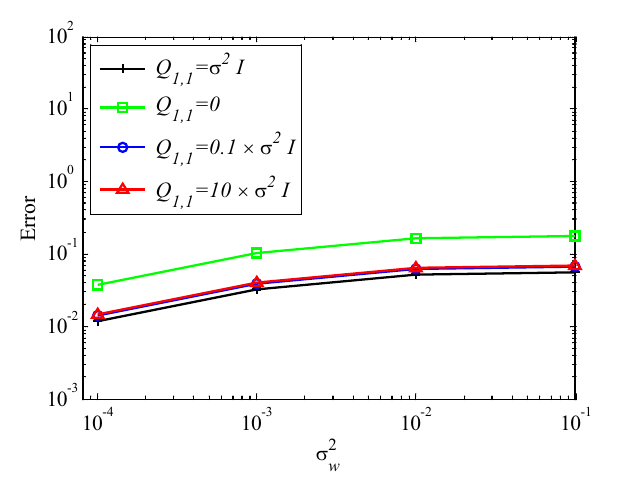}}
	\caption{Effect of mismatched $Q$ on the errors in the reconstructed data for multiple runs of a problem with a small number of DoFs with system noise.}
	\label{fig:error_history_small_sonoise_misQ}
\end{figure}

In this subsubsection, the effects of mismatched selection of $Q$ and $R$ are discussed. The system noise variance $\sigma_v^2$ is set to be the same as $\sigma_w^2$. First, the effect of mismatched $Q$ is discussed. Figure \ref{fig:error_eigen_small_sonoise_misQ} shows that mismatched $Q$ does not significantly affect the error in the estimated eigenvalues, although the result with the appropriate setting (matched $Q$ of $Q_{1,1}=\sigma_w^2I$) exhibits the best performance. Figure \ref{fig:error_history_small_sonoise_misQ} shows the errors in reconstructed data with the mismatched $Q$. In this case, if $Q$ is assumed to be zero, which corresponds to the assumption of no system noise, then the error becomes noticeably larger. On the other hand, if $Q$ is set to be 10 times or 0.1 times larger than the appropriate value, then the results are not significantly degraded. This indicates that the setting of $Q$ does not significantly affect the results if the system noise is considered and $Q$ is appropriately set to be within the order of $\sigma^2_v$.

\begin{figure}
	\centering
	\subfigure[$\lambda_1$.]{\includegraphics[width=5cm]{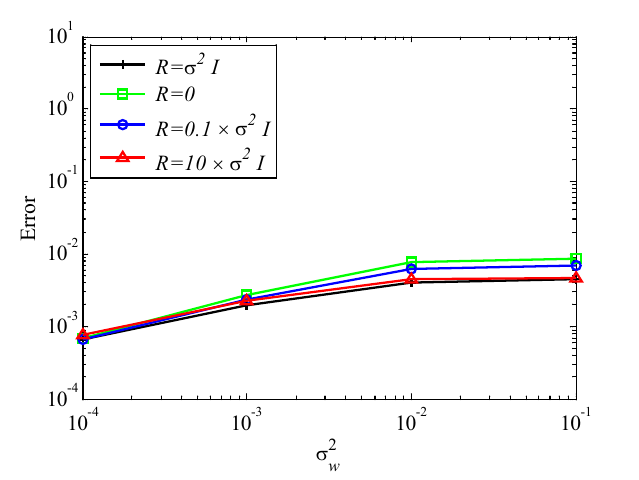}}
	\subfigure[$\lambda_2$.]{\includegraphics[width=5cm]{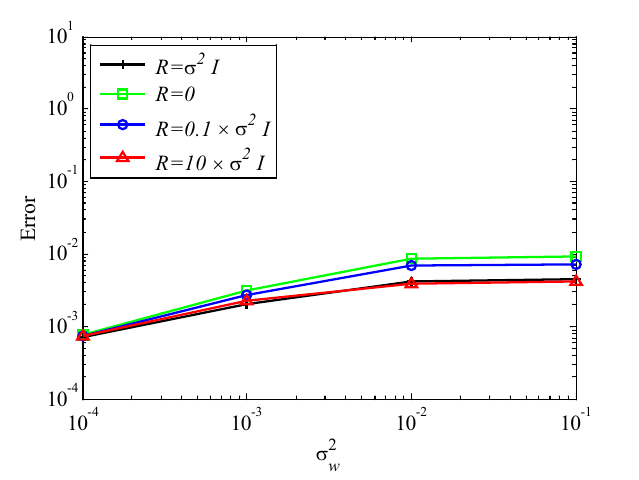}}
	\subfigure[$\lambda_3$.]{\includegraphics[width=5cm]{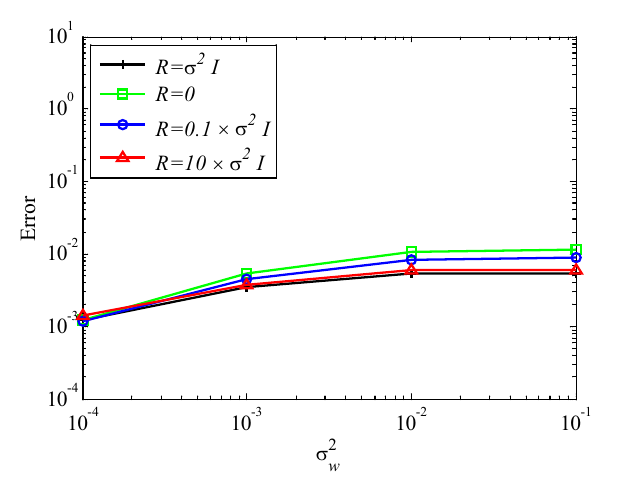}}\\
	\caption{Effect of mismatched $R$ on the errors in the eigenvalues for multiple runs of a problem with a small number of DoFs with system noise.}
\label{fig:error_eigen_small_sonoise_misR}
\end{figure}

\begin{figure}
	\centering
	{\includegraphics[width=5cm]{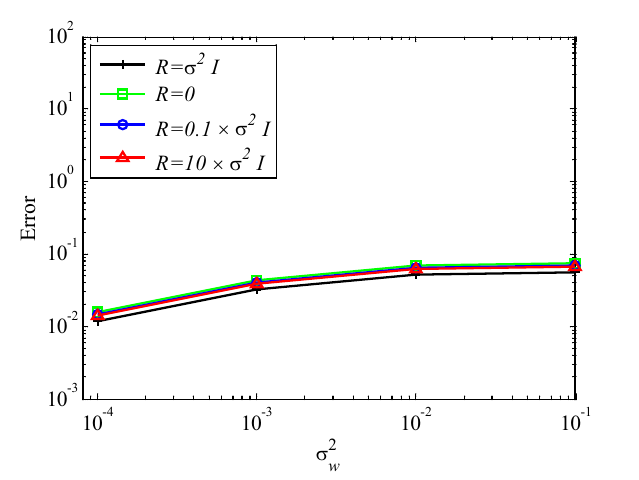}}
	\caption{Effect of mismatched $R$ on the errors in the reconstructed data for multiple runs of a problem with a small number of DoFs with system noise.}
\label{fig:error_history_small_sonoise_misR}
\end{figure}

Then, the effect of mismatched $R$ is discussed. The error in estimated eigenvalues shown in Fig.~\ref{fig:error_eigen_small_sonoise_misR} illustrates that the mismatched $R$ does not significantly change the error, although errors for smaller $R$ or $R$=0 become slightly larger. Figure \ref{fig:error_history_small_sonoise_misR} shows the errors in reconstructed data with mismatched $R$. In this case, mismatched $R$ does not significantly affects the results. This result shows that the setting of $R$ does not significantly affect the results, similar to the mismatched $Q$ cases.

\begin{figure}
	\centering
	\subfigure[$\lambda_1$.]{\includegraphics[width=5cm]{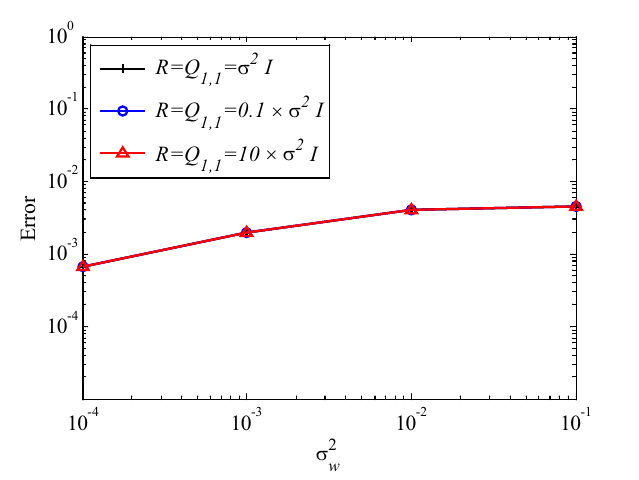}}
	\subfigure[$\lambda_2$.]{\includegraphics[width=5cm]{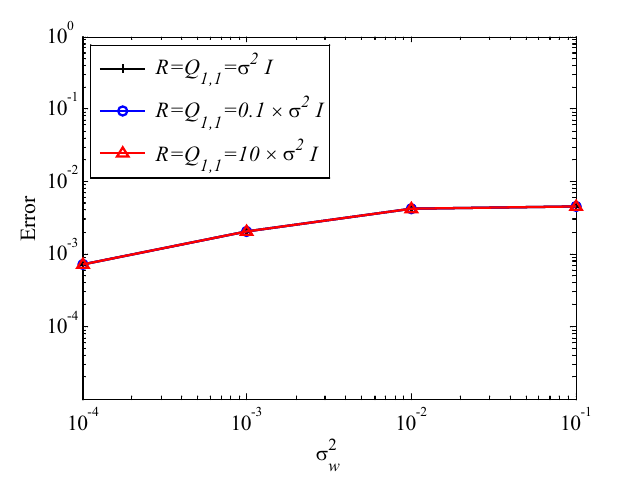}}
	\subfigure[$\lambda_3$.]{\includegraphics[width=5cm]{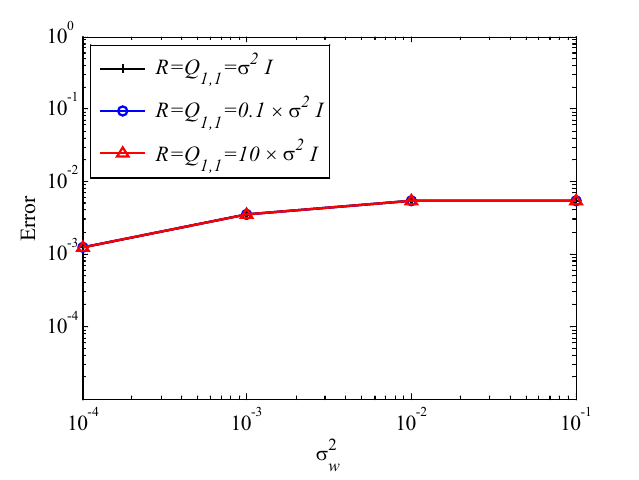}}\\
	\caption{Effects of mismatched $Q$ and $R$ on the errors in the reconstructed data for multiple runs of a problem with a small number of DoFs with system noise.}
\label{fig:error_eigen_small_sonoise_misQR}
\end{figure}

\begin{figure}
	\centering
	{\includegraphics[width=5cm]{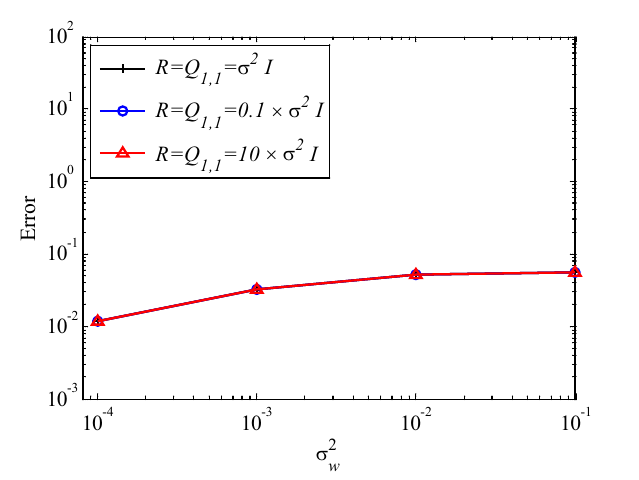}}
	\caption{Effects of mismatched $Q$ and $R$ on the errors in the reconstructed data for multiple runs of a problem with a small number of DoFs with system noise.}
\label{fig:error_history_small_sonoise_misQR}
\end{figure}

Finally, the effects of mismatched $Q$ and $R$, but with the condition $Q_{1,1}=R$, are investigated. The errors in the estimated eigenvalues and reconstructed data for the cases in which $Q_{1,1}=R=10\sigma_w^2I=10\sigma_v^2I$ and $Q_{1,1}=R=0.1\sigma_w^2I=0.1\sigma_v^2I$ are shown in Figs.~\ref{fig:error_eigen_small_sonoise_misQR} and \ref{fig:error_history_small_sonoise_misQR}, respectively. These figures show that the results do not change for the case in which the ratio of $Q_{1,1}$ to $R$ does not change. As noted earlier, the ratio of $Q$ and $R$ should be carefully chosen in order to achieve accurate estimation. 
\color{black}

\clearpage

\subsection{Problem with a moderate number of DoFs without system noise }

Next, a similar problem, but with the number of DoFs extended to 200 by the same procedure, is adopted with the same noise levels. In this case, the computational cost is very high, and we conducted trPOD as a preconditioner. In this problem, first, the number of DoFs is reduced from 200 to 10 by trPOD, and the reduced data are processed by EKFDMD. On the other hand, for the purpose of comparison, DMD and tlsDMD are applied directly to the data for 200 DoFs in order to reduce the number of DoFs to 10 because these algorithms can treat a data matrix of this size within a reasonable computational time by inherently involving truncated SVD (same as trPOD). In this problem, 500 samples were given. Similar to the previous example, the diagonal elements of the covariance matrix were set to be $10^3$ in the initial condition. The diagonal elements of $Q$ and $R$ are set to be 0 and $\sigma_w^2$, respectively, and their nondiagonal elements are set to be 0.

\begin{figure}[!htb]
	\centering
	\includegraphics[width=6cm]{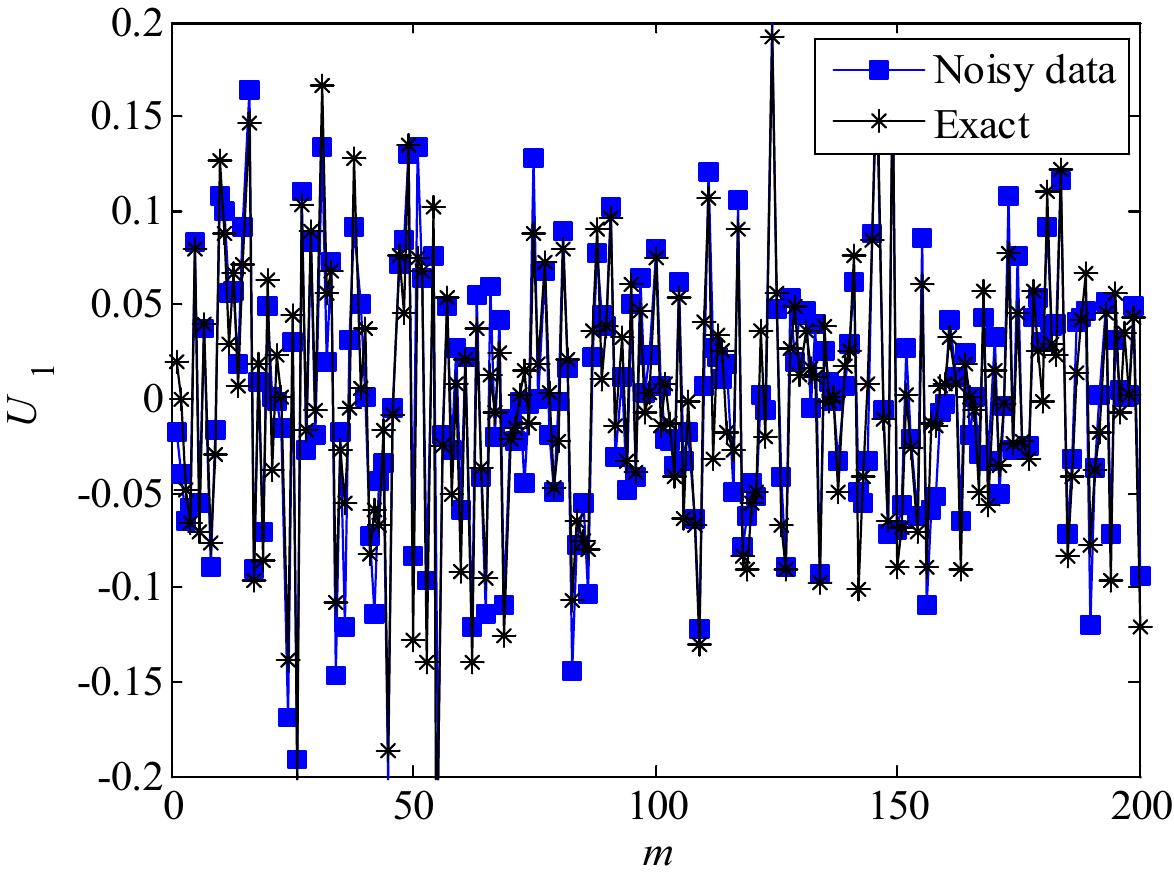}
	\vspace{-0.2cm}
	\caption{First POD mode of original and noisy data for a problem with a moderate number of DoFs. Here, the first POD modes of the most noisy case ($\sigma^2_w=0.1$) are shown.}
	\label{fig:pod_large}
\end{figure}

The results of trPOD are shown in Fig.~\ref{fig:pod_large}, where the first POD spatial mode obtained by data without noise and that obtained by data with noise are plotted together. Note that the mode of the node distribution in snapshots is referred to as the POD spatial mode, which is analogous to fluid analysis.  This plot indicates that the noise level is very high and that the estimation of the POD spatial mode is not accurate. However, the contaminated POD modes obtained by data with noise are used for EKFDMD. 
\color{black}

\begin{figure}
	\centering
	\subfigure[$\sigma_w^2=0.0001$.]{\includegraphics[width=5cm]{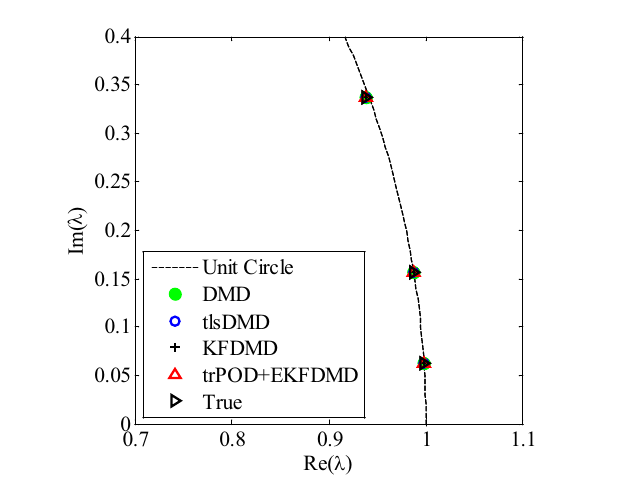}}
	\subfigure[$\sigma_w^2=0.001 $.]{\includegraphics[width=5cm]{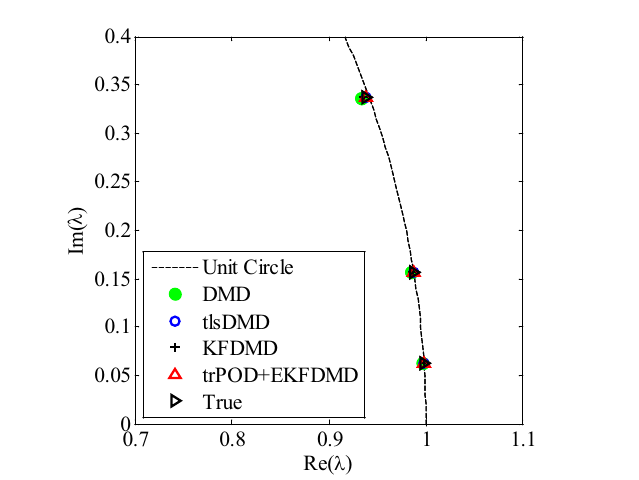}}\\
	\subfigure[$\sigma_w^2=0.01  $.]{\includegraphics[width=5cm]{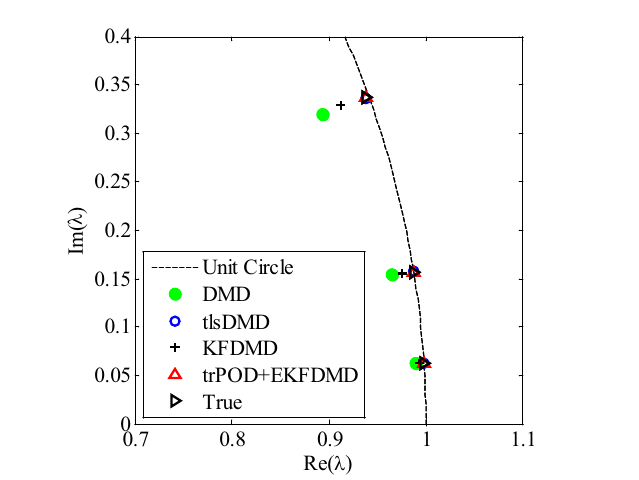}}
	\subfigure[$\sigma_w^2=0.1   $.]{\includegraphics[width=5cm]{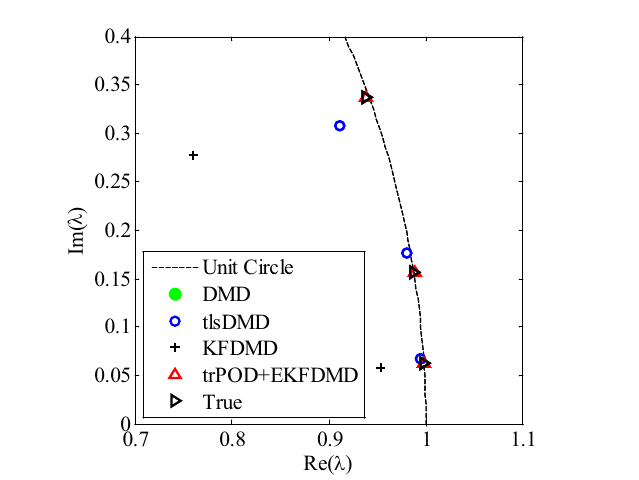}}
	\vspace{-0.2cm}
	\caption{Eigenvalues for a problem with a small number of DoFs without system noise. Here, rank $r$ is set to be 10. The algorithms are almost identical in (a) and (b), and optDMD, and trPOD+EKFDMD are almost identical in (c) and (d).}
	\label{fig:eigen_medium_onoise}
\end{figure}

\begin{figure}
	\centering
	\subfigure[$\sigma_w^2=0.0001$.]{\includegraphics[width=5cm]{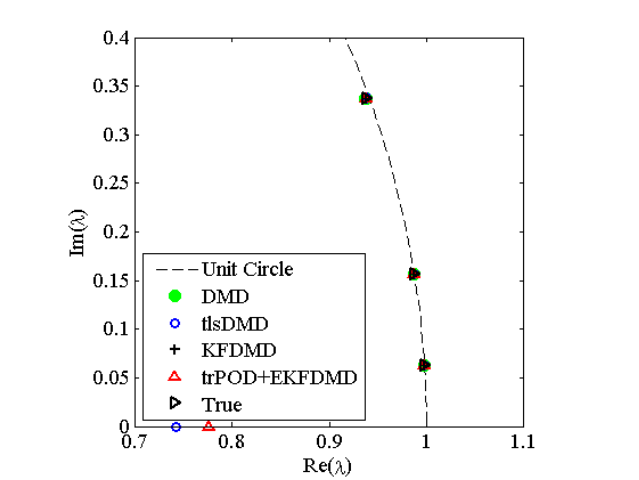}}
	\subfigure[$\sigma_w^2=0.001$.]{\includegraphics[width=5cm]{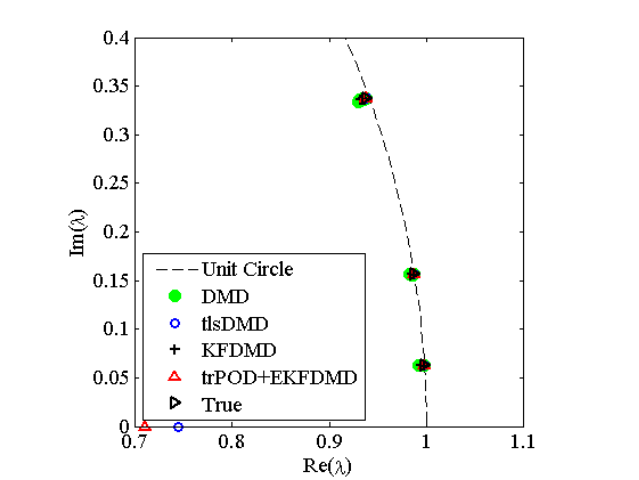}}\\
	\subfigure[$\sigma_w^2=0.01$.]{\includegraphics[width=5cm]{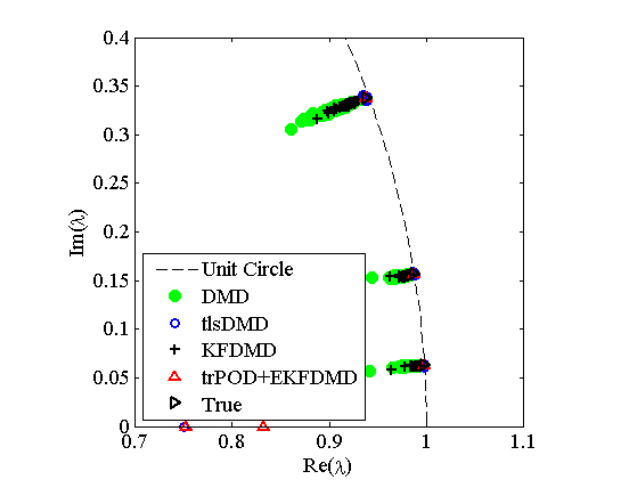}}
	\subfigure[$\sigma_w^2=0.1$. ]{\includegraphics[width=5cm]{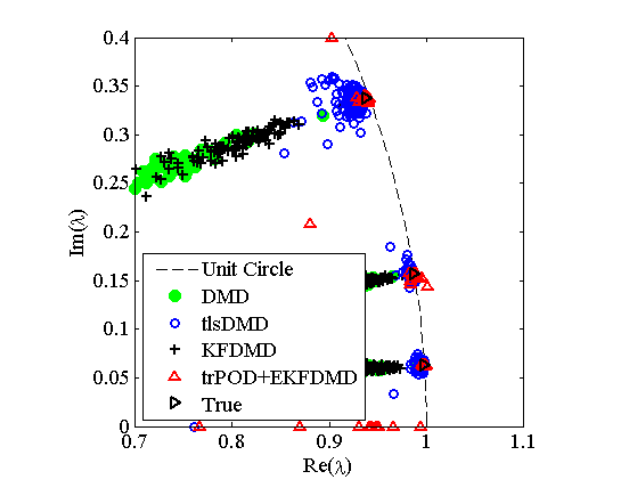}}
	\vspace{-0.2cm}
	\caption{Eigenvalues for multiple runs of a problem with a moderate number of DoFs without system noise, where the seed for the random number is different for multiple runs. Here, rank $r$ is set to be 10. }
	\label{fig:eigen_medium_onoise_multi}
\end{figure}

\begin{figure}
	\centering
	\subfigure[$\lambda_1$.]{\includegraphics[width=5cm]{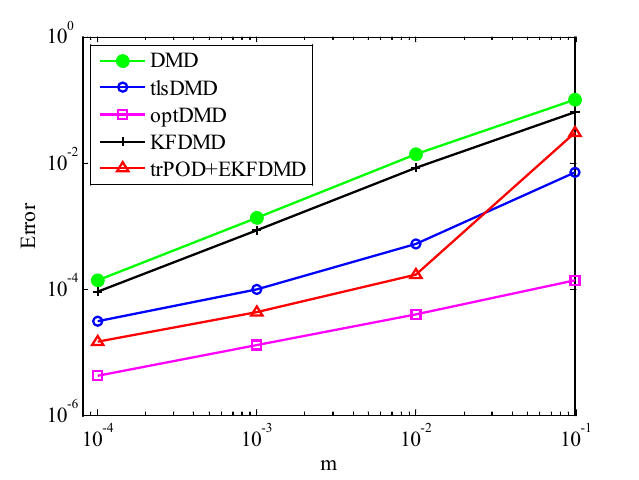}}
	\subfigure[$\lambda_2$.]{\includegraphics[width=5cm]{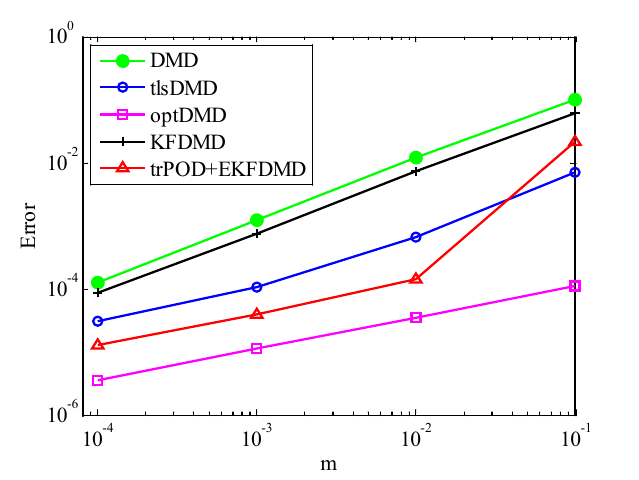}}
	\subfigure[$\lambda_3$.]{\includegraphics[width=5cm]{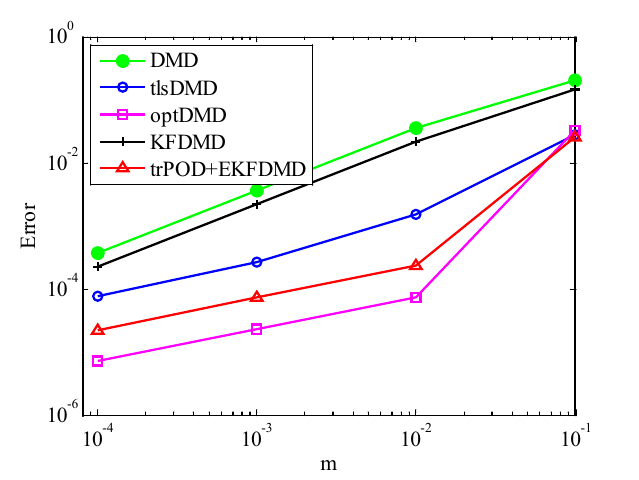}}\\
	\caption{Errors in the eigenvalues for multiple runs of a problem with a moderate number of DoFs without system noise for the case in which $\sigma^2_v=0.1\sigma^2_w$.}
	\label{fig:error_eigen_medium_onoise}
\end{figure}

\begin{figure}
	\centering
	\subfigure[$\sigma_w^2=0.0001$.]{\includegraphics[width=5cm]{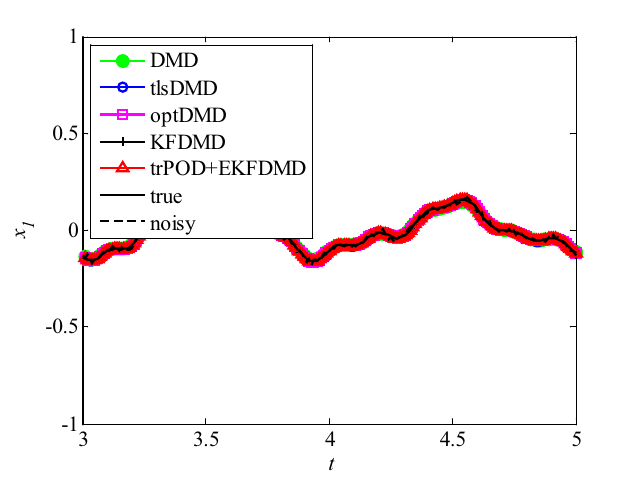}}
	\subfigure[$\sigma_w^2=0.001$. ]{\includegraphics[width=5cm]{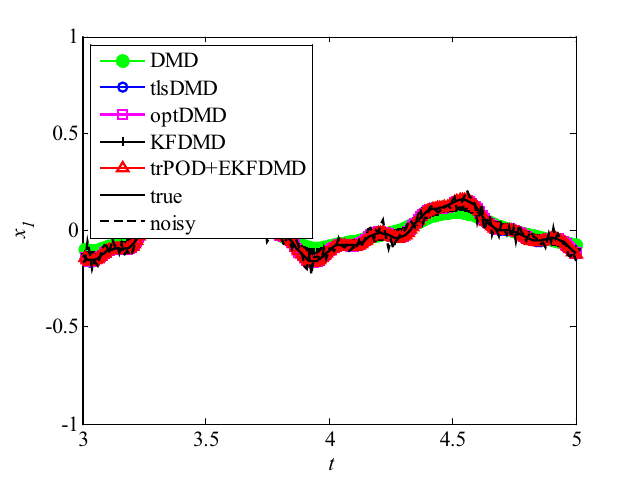}}\\
	\subfigure[$\sigma_w^2=0.01$.  ]{\includegraphics[width=5cm]{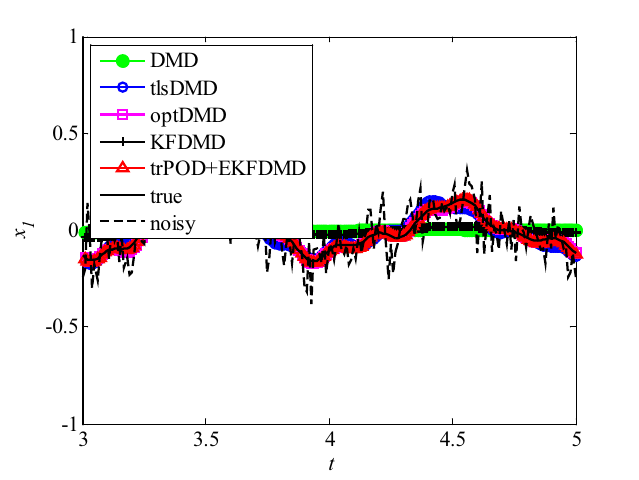}}
	\subfigure[$\sigma_w^2=0.1$.   ]{\includegraphics[width=5cm]{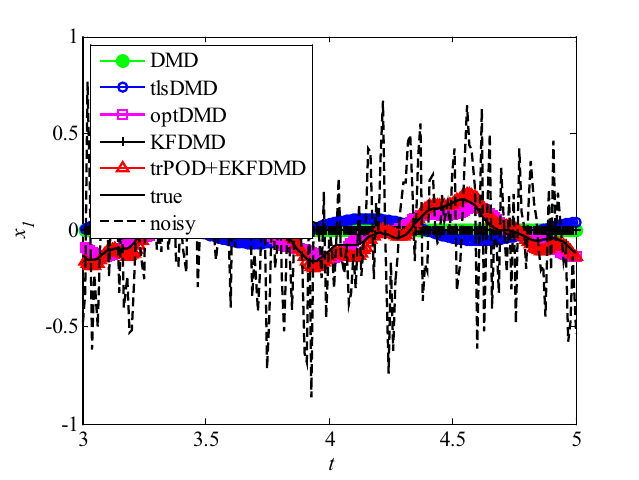}}\\		
	\subfigure[$\sigma_w^2=0.01$, tlsDMD.]{\includegraphics[width=5cm]{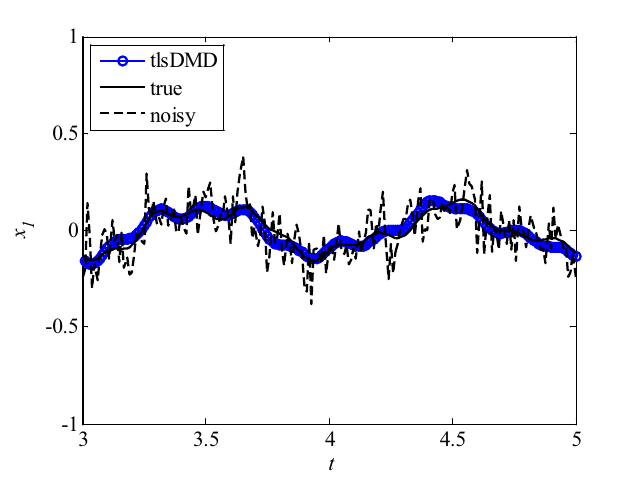}}
	\subfigure[$\sigma_w^2=0.01$, optDMD.]{\includegraphics[width=5cm]{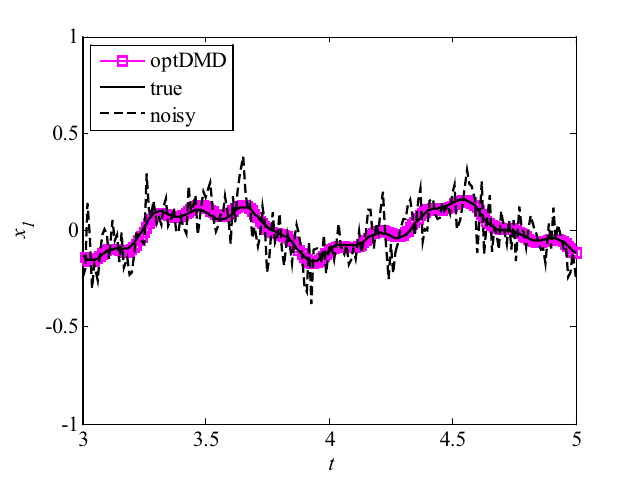}}
	\subfigure[$\sigma_w^2=0.01$, trPOD+EKFDMD.]{\includegraphics[width=5cm]{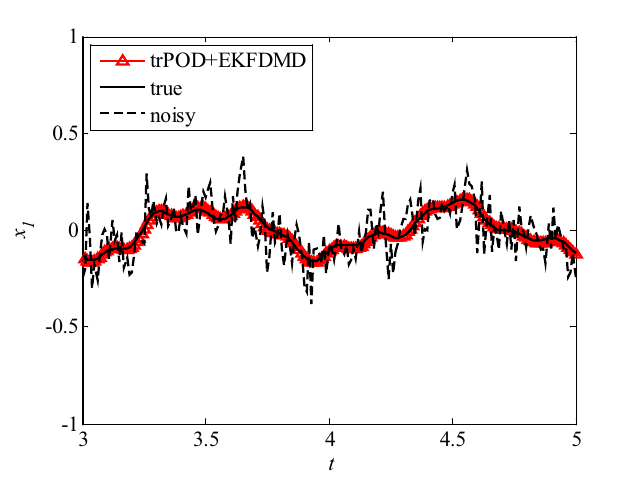}}\\
	\subfigure[$\sigma_w^2=0.1$,  tlsDMD.]{\includegraphics[width=5cm]{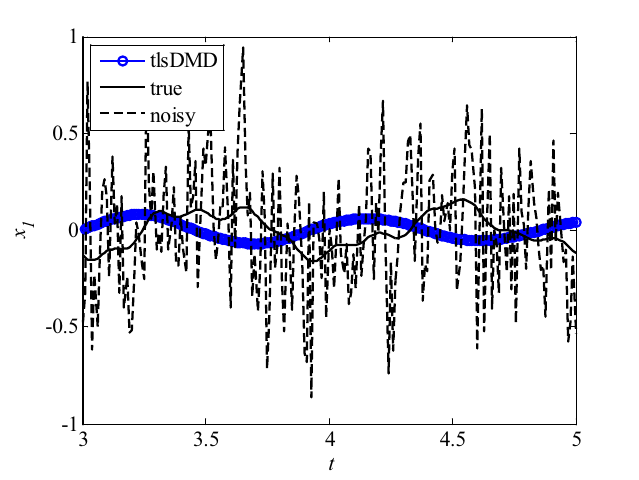}}
	\subfigure[$\sigma_w^2=0.1$,  optDMD.]{\includegraphics[width=5cm]{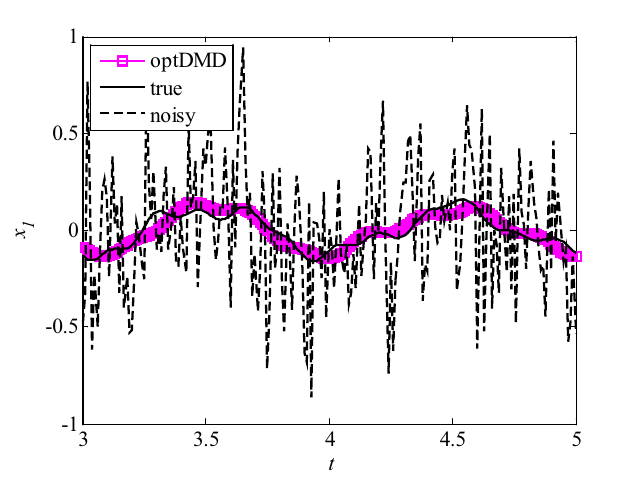}}
	\subfigure[$\sigma_w^2=0.1$,  trPOD+EKFDMD.]{\includegraphics[width=5cm]{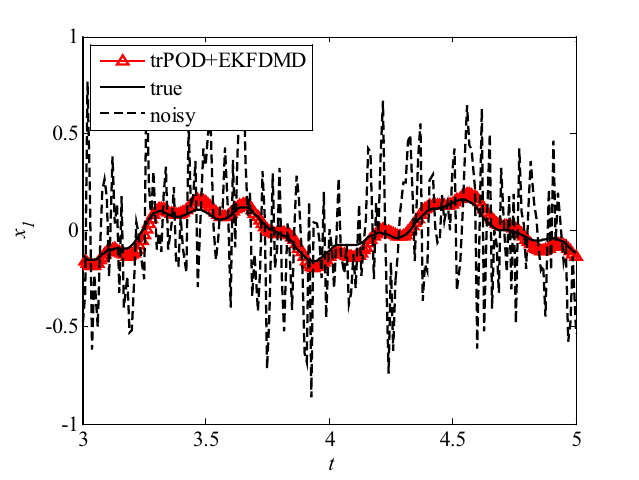}}
	\vspace{-0.2cm}
    \caption{Reconstructed data of the first node for a problem with a moderate number of DoFs without system noise.}
	\label{fig:history_medium_onoise}
\end{figure}

\begin{figure}
	\centering
	{\includegraphics[width=5cm]{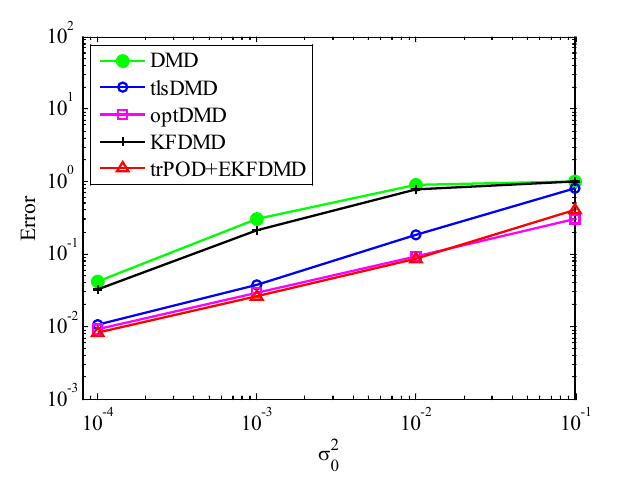}}
    \caption{Errors in the reconstructed data for multiple runs of a problem with a moderate number of DoFs without system noise. Here, rank $r$ is set to be 10.}
	\label{fig:error_history_medium_onoise}
\end{figure}

The eigenvalues and their errors for this problem are shown in Figs.~\ref{fig:eigen_medium_onoise}, \ref{fig:eigen_medium_onoise_multi}, and \ref{fig:error_eigen_medium_onoise}. Except for the condition with strong noise ($\sigma_w^2=0.1$), trPOD+EKFDMD works better than DMD, KFDMD, and tlsDMD, while optDMD works best. This characteristic does not change from the small-degree-of-freedom problem, as shown earlier. The degradation in performance of the trPOD+EKFDMD for the very noisy condition might occur because the important signal is filtered out in the POD procedure. This characteristic is relaxed by increasing the number of POD modes, as shown later herein, but the number of POD modes is in a trade-off relationship with the computational cost. The reconstructed data are then shown in Fig.~\ref{fig:history_medium_onoise}. Even if we apply POD, the reconstructed data of trPOD+EKFDMD and optDMD agree well with the original data in all the condition, whereas DMD, KFDMD, and tlsDMD fail to capture the behavior of the original data in the severe noise cases. The error in the reconstructed data is shown in Fig.~\ref{fig:error_history_medium_onoise}. As shown earlier, the error of trPOD+EKFDMD is smaller than that of tlsDMD and is larger than that of
optDMD. Thus, trPOD+EKFDMD works reasonably well in reconstructing the data even with the imperfect POD modes shown in Fig. \ref{fig:pod_large}. 
\color{black}



\subsubsection{Effect of POD truncation}
\begin{figure}
	\centering
	\subfigure[$\lambda_1$.]{\includegraphics[width=5cm]{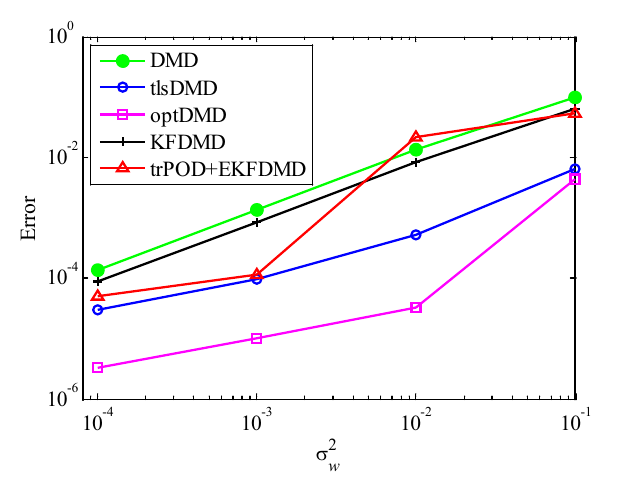}}
	\subfigure[$\lambda_2$.]{\includegraphics[width=5cm]{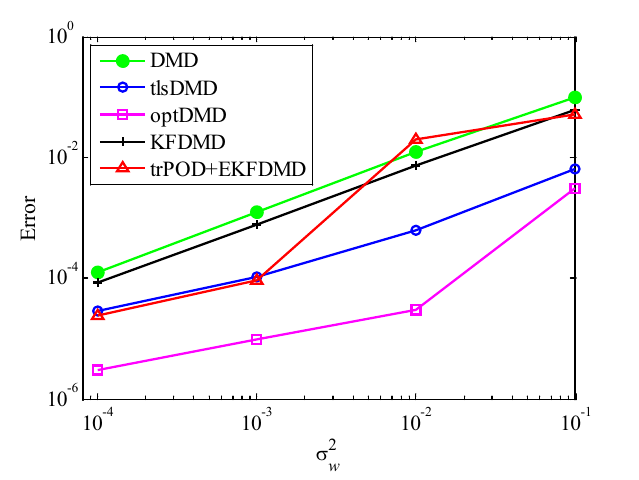}}
	\subfigure[$\lambda_3$.]{\includegraphics[width=5cm]{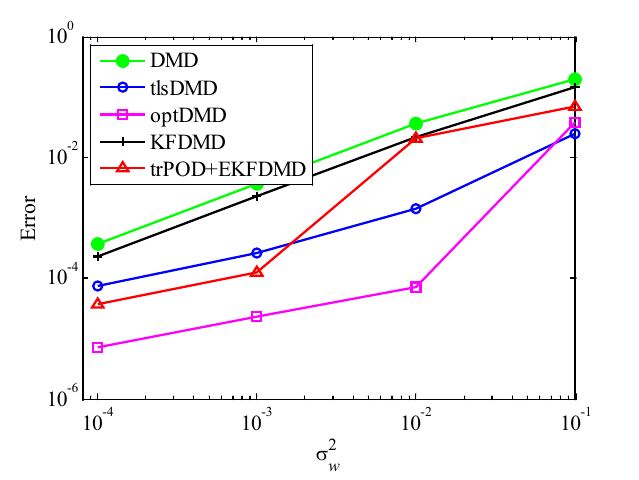}}\\
	\caption{Errors in the eigenvalues for multiple runs of a problem with a moderate number of DoFs without system noise, whereas the rank $r$ is set to be 6. Here the seed for the random numbers is different for multiple runs.}
	\label{fig:error_eigen_medium_onoise_r6}
\end{figure}

\begin{figure}
	\centering
	{\includegraphics[width=5cm]{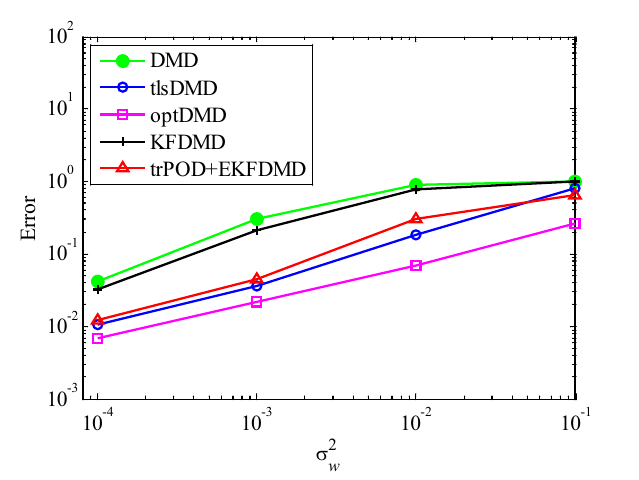}}
	\caption{Errors in the reconstructed data for multiple runs of a problem with a moderate number of DoFs without system noise, whereas the rank $r$ is set to be 6. Here, the seed for the random numbers is different for multiple runs.}
	\label{fig:error_history_medium_onoise_r6}
\end{figure}

\begin{figure}
	\centering
	\subfigure[$\lambda_1$.]{\includegraphics[width=5cm]{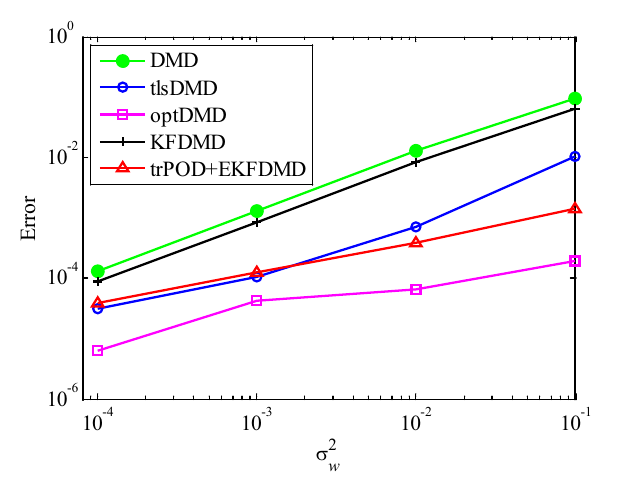}}
	\subfigure[$\lambda_2$.]{\includegraphics[width=5cm]{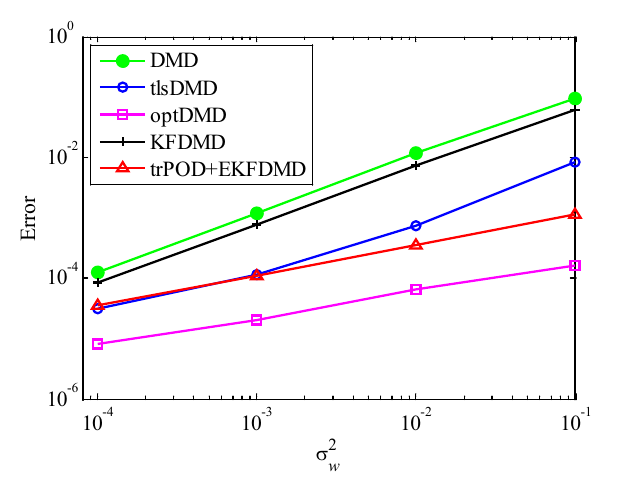}}
	\subfigure[$\lambda_3$.]{\includegraphics[width=5cm]{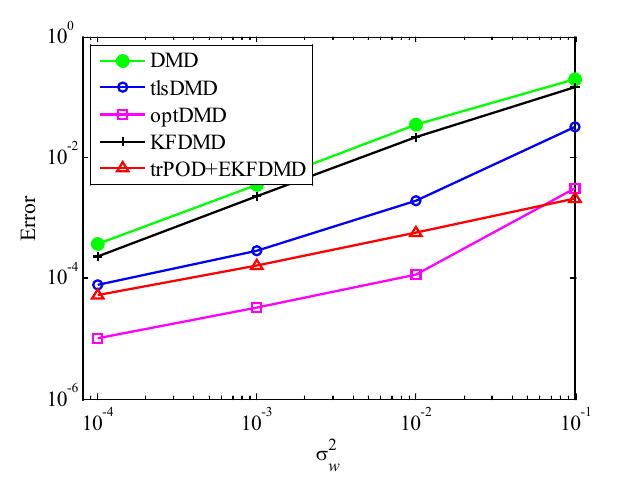}}\\
	\caption{Errors in the eigenvalues for multiple runs of a problem with a moderate number of DoFs without system noise, whereas the rank $r$ is set to be 20. Here, the seed for the random number is different for multiple runs.}
    \label{fig:error_eigen_medium_onoise_r20}
\end{figure}

\begin{figure}
	\centering
	{\includegraphics[width=5cm]{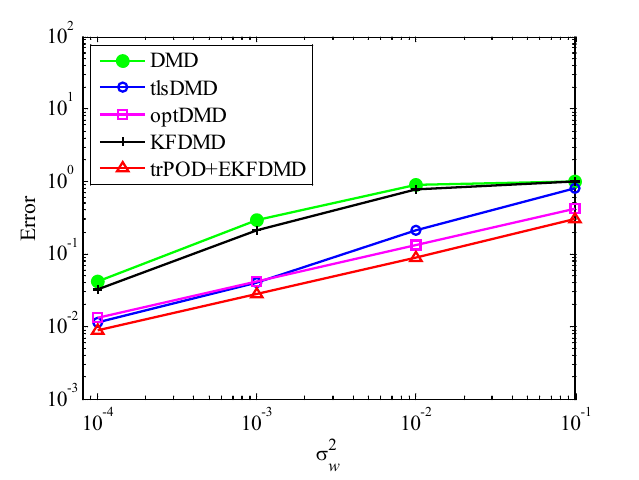}}
	\caption{Errors in the reconstructed data for multiple runs of a problem with a moderate number of DoFs without system noise, whereas the rank $r$ is set to be 20. Here, the seed for the random number is different for multiple runs.}
    \label{fig:error_history_medium_onoise_r20}
\end{figure}

For POD truncation, the rank number should be manually specified. Therefore, the effect of the rank number chosen by the user is investigated. Here, $r=6$ and $r=20$ are investigated, where the previous standard cases were computed with $r=10$, as noted earlier. The errors in the estimated eigenvalues and reconstructed data of the $r=6$ and $r=20$ conditions are shown in Figs.~\ref{fig:error_eigen_medium_onoise_r6} and \ref{fig:error_history_medium_onoise_r6} and Figs \ref{fig:error_eigen_medium_onoise_r20} and \ref{fig:error_history_medium_onoise_r20}, respectively. For the case in which system noise is absent, the errors of the estimation of eigenvalues by trPOD+EKFDMD does not work well with $r=6$ for $\sigma^2_w \ge 0.01$, and the resulting error in reconstructed data is slightly worse than that for tlsDMD for all cases with different noise levels. This might be because trPOD filters out the important signal and trPOD+EKFDMD cannot recover the original signal for strong-noise cases. On the other hand, the errors in the estimated eigenvalues of trPOD+EKFDMD with the $r=20$ setting are lower than those of tlsDMD or are approximately the same as (and sometimes slightly higher than) that of tlsDMD and the error in the reconstructed data of trPOD+EKFDMD with $r=20$ is smaller than that of tlsDMD. Therefore, using tnPOD+EKFDMD with better performance requires a larger rank. This is clear trade-off between the estimation accuracy and the computational cost.
\color{black}

\subsection{Problem with a moderate number of DoFs with system noise}
Next, we consider a similar problem in which system noise is adopted. The system noise variance $\sigma_v^2$ is set to be $\sigma_w^2$, similar to the small-DoF problem shown earlier. With regard to the EKFDMD procedure, trPOD is used as a preconditioner similar to the previous subsection. Again, in this problem, the number of DoFs is reduced from 200 to 10 by trPOD, and the reduced data are processed by EKFDMD. On the other hand, DMD, tlsDMD, and optDMD are applied directly to the data for 200 DoFs in order to reduce the number of DoFs to 10. Moreover, in this problem, 500 samples were given. The diagonal elements of the covariance matrix are set to be $10^3$ in the initial condition. The diagonal elements of $R$ and $Q_{1,1}$ are set to be $\sigma_w^2$ and $\sigma_v^2$, respectively, and the nondiagonal elements of $R$ and $Q_{1,1}$ are set to be 0. 

\begin{figure}
	\centering
 	\subfigure[$\sigma_w^2=0.0001$.]{\includegraphics[width=5cm]{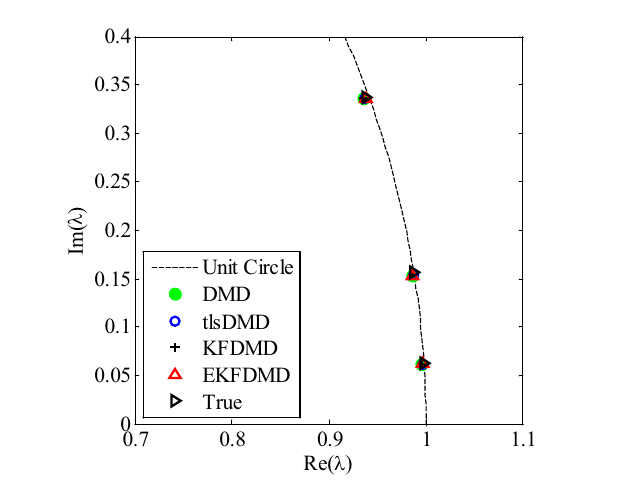}}
	\subfigure[$\sigma_w^2=0.001 $.]{\includegraphics[width=5cm]{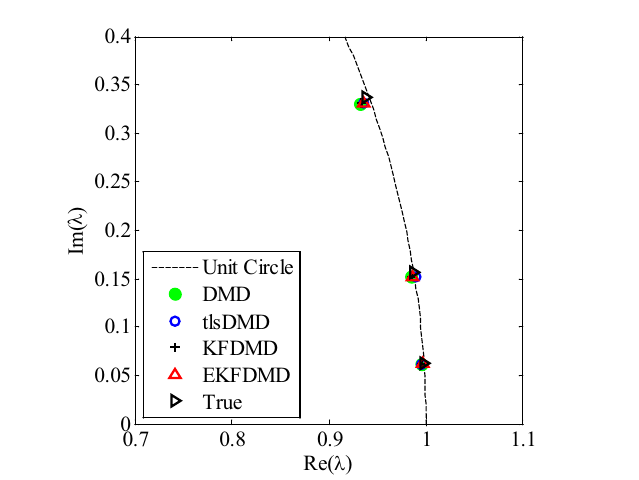}}\\
	\subfigure[$\sigma_w^2=0.01  $.]{\includegraphics[width=5cm]{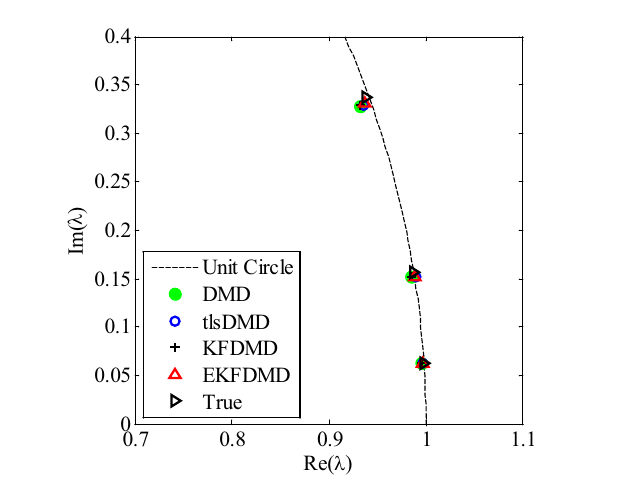}}
	\subfigure[$\sigma_w^2=0.1   $.]{\includegraphics[width=5cm]{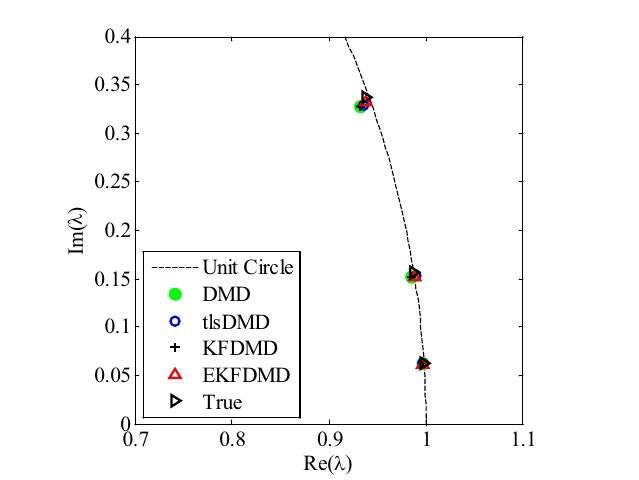}}
	\vspace{-0.2cm}
	\caption{Eigenvalues for a problem with a moderate number of DoFs with system noise. The results of all algorithms are almost identical in this plot. Here, rank $r$ is set to be 10.}
	\label{fig:eigen_medium_sonoise}
\end{figure}

\begin{figure}
	\centering
	\subfigure[$\sigma_w^2=0.0001$.]{\includegraphics[width=5cm]{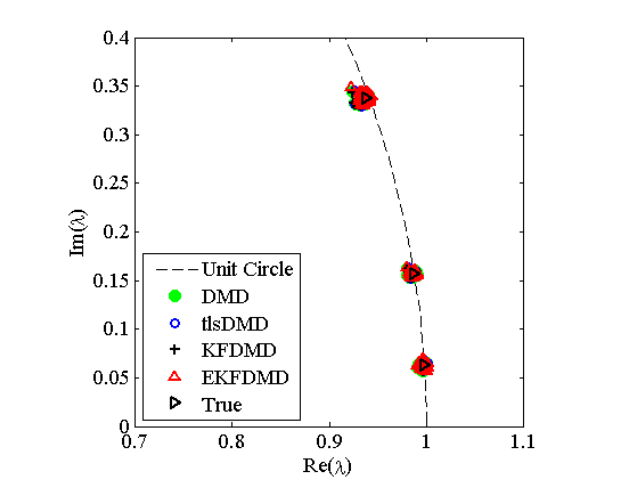}}
	\subfigure[$\sigma_w^2=0.001$.]{\includegraphics[width=5cm]{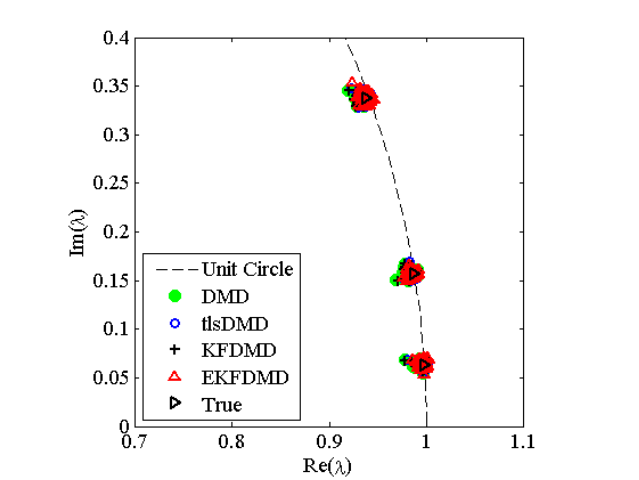}}\\
	\subfigure[$\sigma_w^2=0.01$.]{\includegraphics[width=5cm]{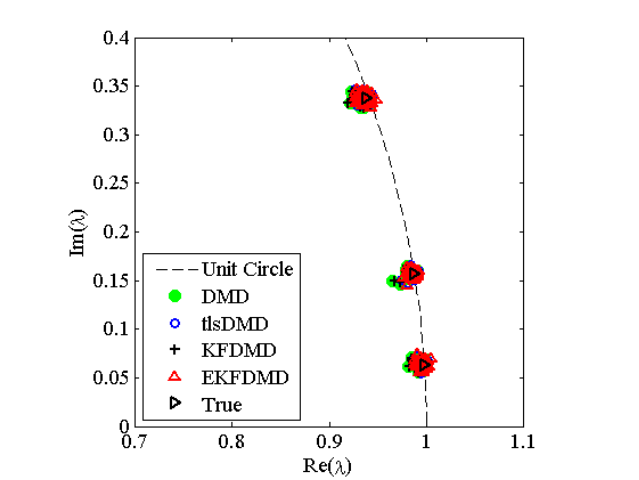}}
	\subfigure[$\sigma_w^2=0.1$ ]{\includegraphics[width=5cm]{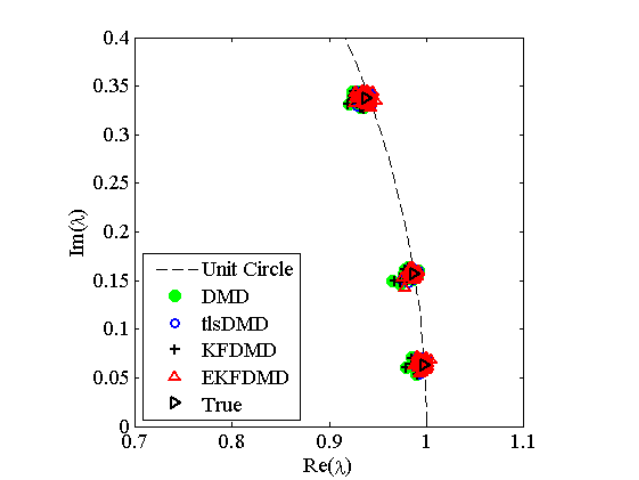}}
	\vspace{-0.2cm}
	\caption{Eigenvalues for multiple runs of a problem with a moderate number of DoFs with system noise. The results of all algorithms are almost identical in this plot. Here, rank $r$ is set to be 10.}
    \label{fig:eigen_medium_sonoise_multi}
\end{figure}

\begin{figure}
	\centering
	\subfigure[$\lambda_1$.]{\includegraphics[width=5cm]{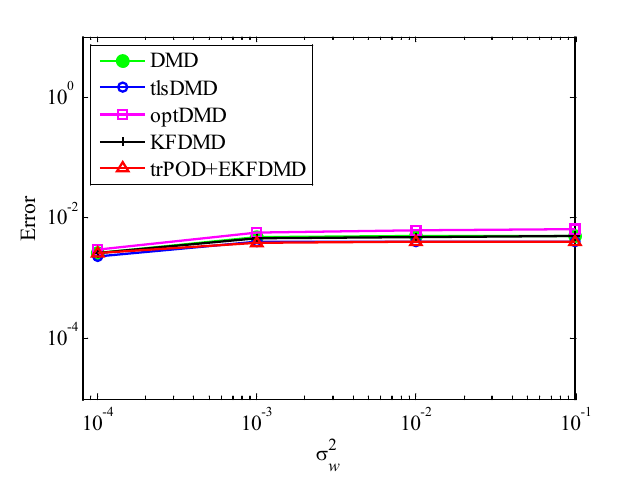}}
	\subfigure[$\lambda_2$.]{\includegraphics[width=5cm]{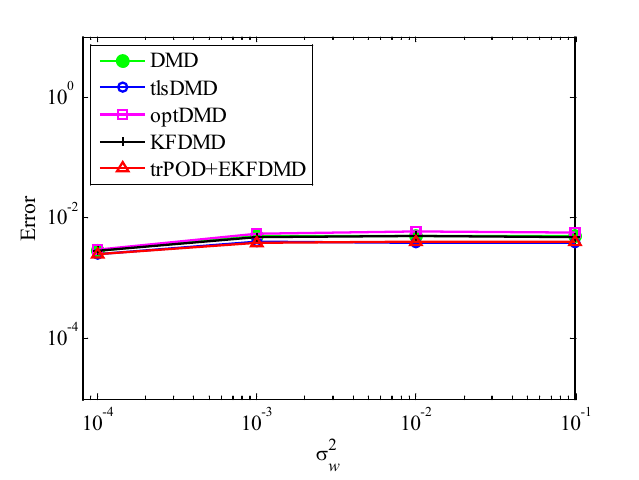}}
	\subfigure[$\lambda_3$.]{\includegraphics[width=5cm]{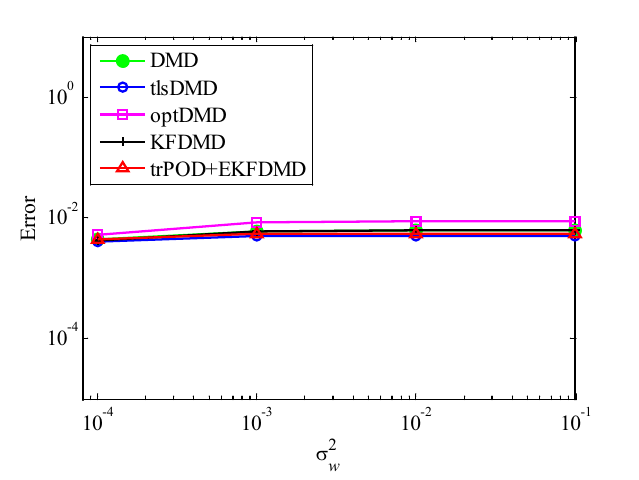}}\\
	\caption{Errors in the in the eigenvalues for multiple runs of a problem with a moderate number of DoFs with system noise, where the seed for random numbers is different for multiple runs. Here, rank $r$ is set to be 10.}
	\label{fig:error_eigen_medium_sonoise}
\end{figure}

\begin{figure}
	\centering
	\subfigure[$\sigma_w^2=0.0001$.]{\includegraphics[width=5cm]{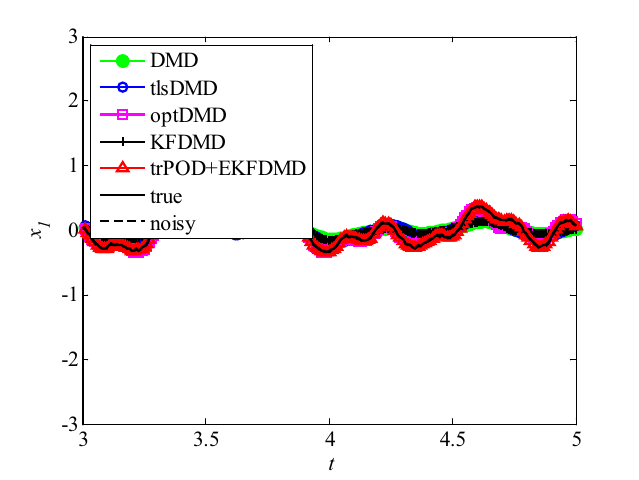}}
	\subfigure[$\sigma_w^2=0.001$. ]{\includegraphics[width=5cm]{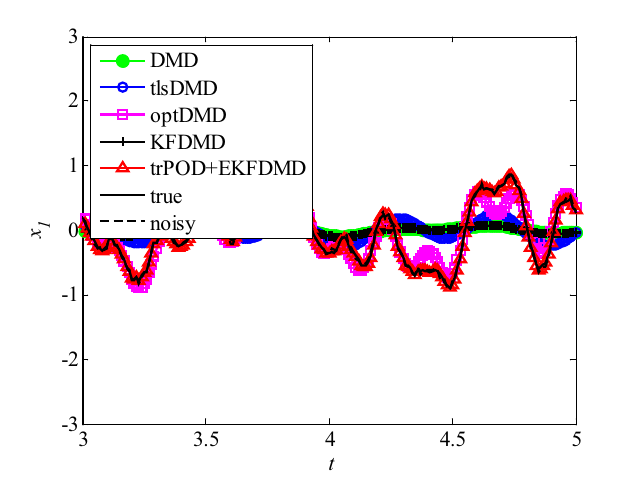}}\\
	\subfigure[$\sigma_w^2=0.01$.  ]{\includegraphics[width=5cm]{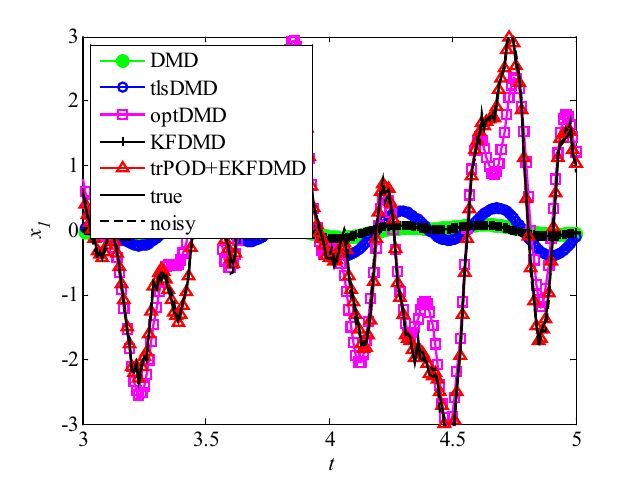}}
	\subfigure[$\sigma_w^2=0.1$.   ]{\includegraphics[width=5cm]{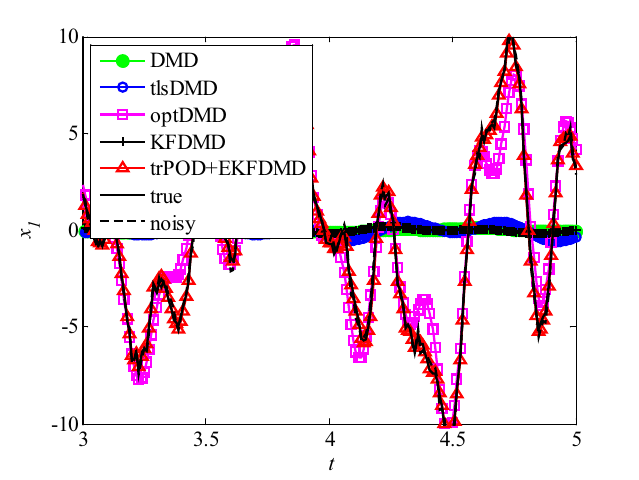}}\\		
	\subfigure[$\sigma_w^2=0.01$, tlsDMD.]{\includegraphics[width=5cm]{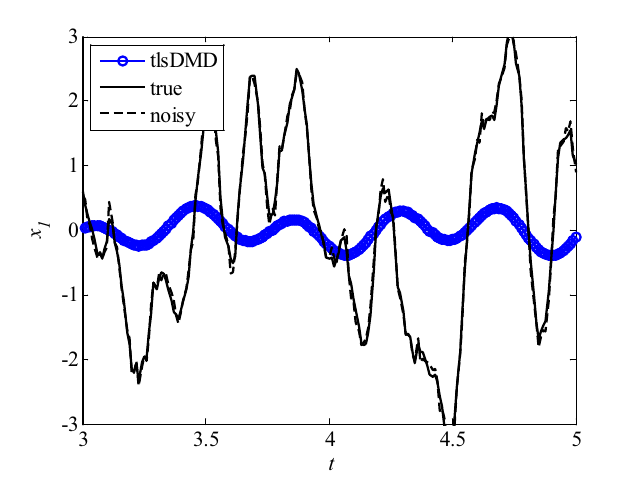}}
	\subfigure[$\sigma_w^2=0.01$, optDMD.]{\includegraphics[width=5cm]{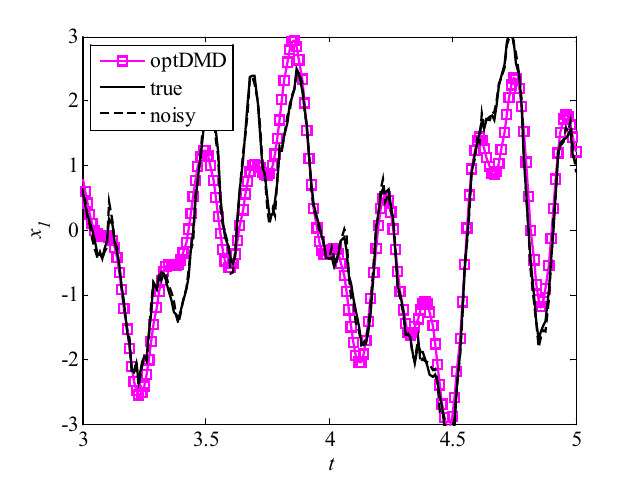}}
	\subfigure[$\sigma_w^2=0.01$, trPOD+EKFDMD.]{\includegraphics[width=5cm]{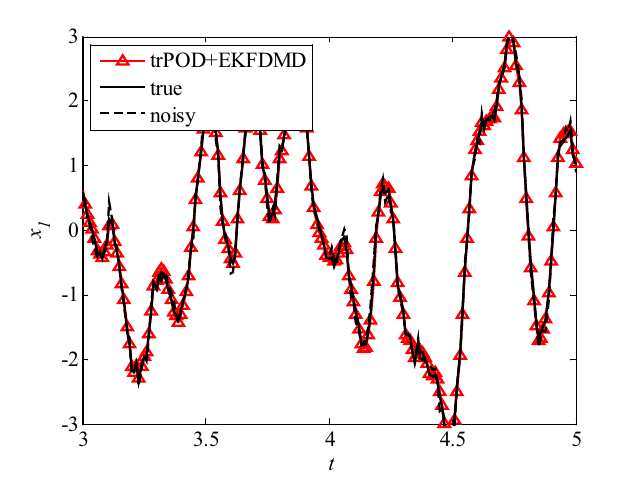}}\\
	\subfigure[$\sigma_w^2=0.1$,  tlsDMD.]{\includegraphics[width=5cm]{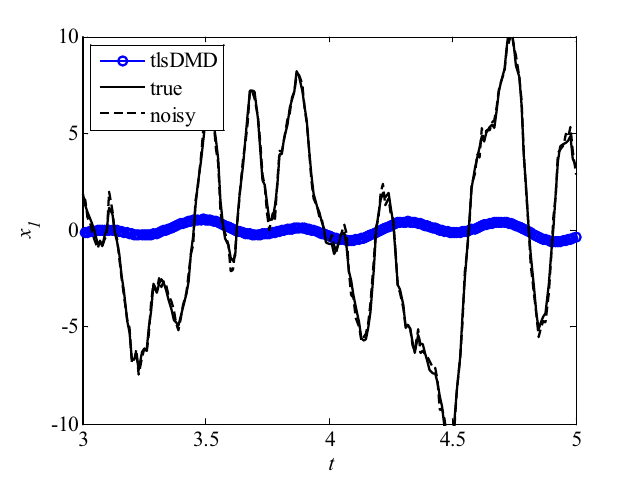}}
	\subfigure[$\sigma_w^2=0.1$,  optDMD.]{\includegraphics[width=5cm]{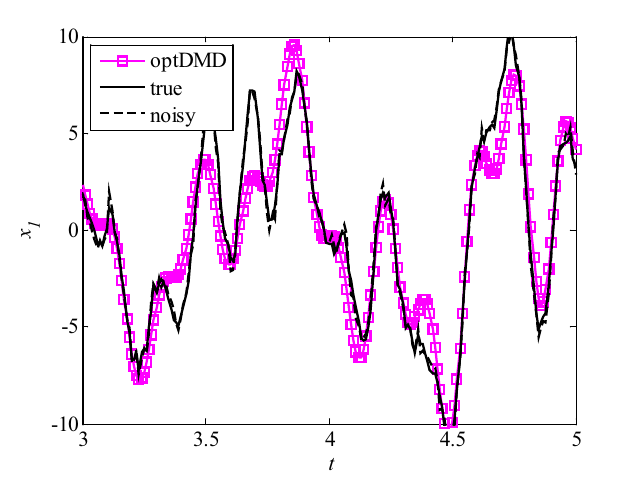}}
	\subfigure[$\sigma_w^2=0.1$,  trPOD+EKFDMD.]{\includegraphics[width=5cm]{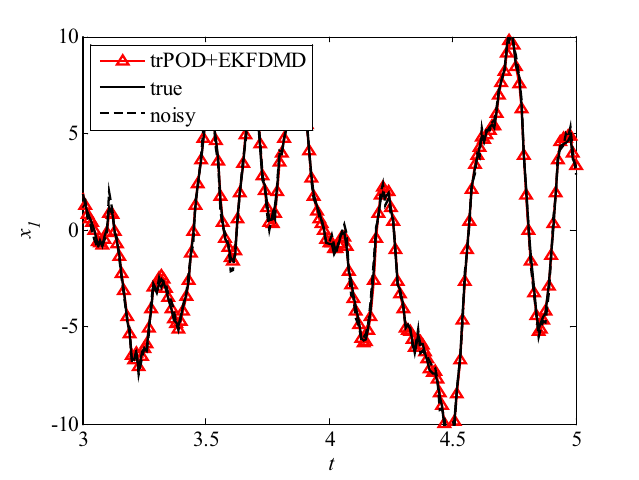}}
	\vspace{-0.2cm}
	\caption{Reconstructed data of the first node for a problem with a moderate number of DoFs with system noise. Here, rank $r$ is set to be 10.}
	\label{fig:history_medium_sonoise}
\end{figure}

\begin{figure}
	\centering
	\includegraphics[width=5cm]{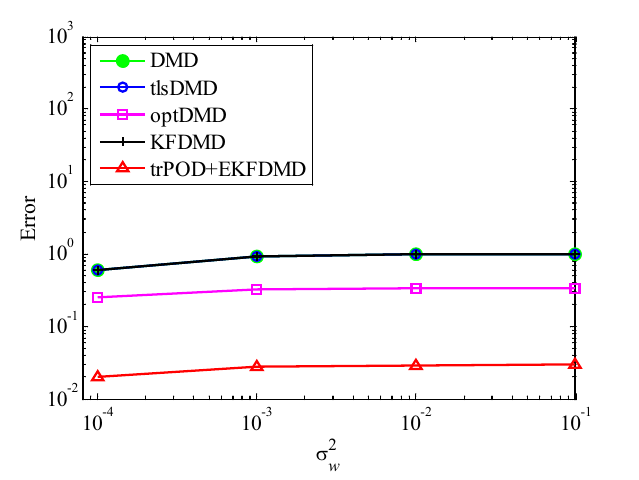}
	\caption{Errors in the reconstructed data for multiple runs of a problem with a moderate number of DoFs with system noise. Here, rank $r$ is set to be 10.}
	\label{fig:error_history_medium_sonoise}
\end{figure}

The eigenvalues and their errors for this problem are shown in Figs.~\ref{fig:eigen_medium_sonoise}, \ref{fig:eigen_medium_sonoise_multi}, and \ref{fig:error_eigen_medium_sonoise}. Interestingly, all the algorithm work similarly each other in this condition. The degradation in performance for trPOD+EKFDMD is not found in this case, together with the results later shown herein. Then, the reconstructed data are shown in Fig. \ref{fig:history_medium_sonoise}. Figure \ref{fig:history_medium_sonoise} illustrates that DMD, KFDMD, and tlsDMD fail to capture the behavior of original data while optDMD works reasonably but sometimes fails to capture the behaviour around peaks. Even if we apply the POD decomposition, the data reconstructed by trPOD+EKFDMD agree the best with original data.  The error in reconstructed data is shown in Fig. \ref{fig:error_history_medium_sonoise}. As shown earlier, the error of EKFDMD is smallest in the algorithm investigated, similar to the small DoFs problem. Thus, trPOD+EKFDMD works well to reconstruct the data especially for the case in which system noise is present, even in the moderate number of DoFs problem. 
\color{black}


\subsubsection{Effects of POD truncation}
\begin{figure}
	\centering
	\subfigure[$\lambda_1$.]{\includegraphics[width=5cm]{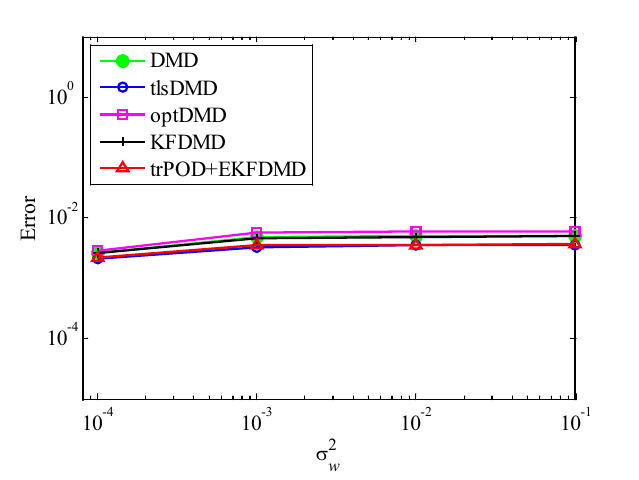}}
	\subfigure[$\lambda_2$.]{\includegraphics[width=5cm]{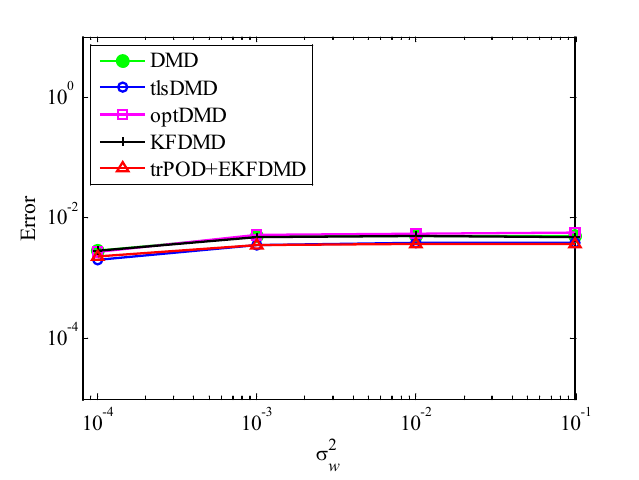}}
	\subfigure[$\lambda_3$.]{\includegraphics[width=5cm]{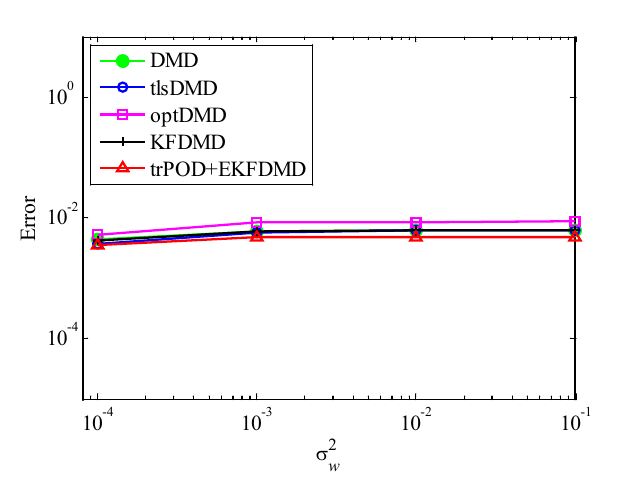}}\\
	\caption{Errors in the eigenvalues for multiple runs of a problem with a moderate number of DoFs with system noise for the case in which rank $r$ is set to be 6. The algorithms are almost identical.}
    \label{fig:error_eigen_medium_sonoise_r6}
\end{figure}

\begin{figure}
	\centering
	{\includegraphics[width=5cm]{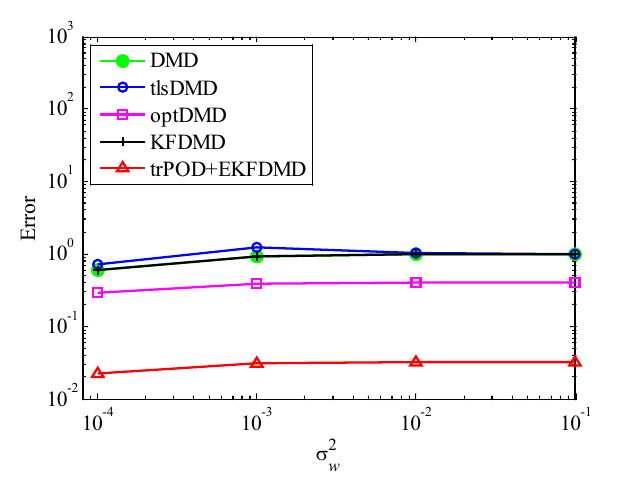}}
	\caption{Errors in the reconstructed data for multiple runs of a problem with a moderate number of DoFs without system noise for the case in which rank $r$ is set to be 6, where the seed for random numbers is different for multiple runs. }
    \label{fig:error_history_medium_sonoise_r6}
\end{figure}

\begin{figure}
	\centering
	\subfigure[$\lambda_1$.]{\includegraphics[width=5cm]{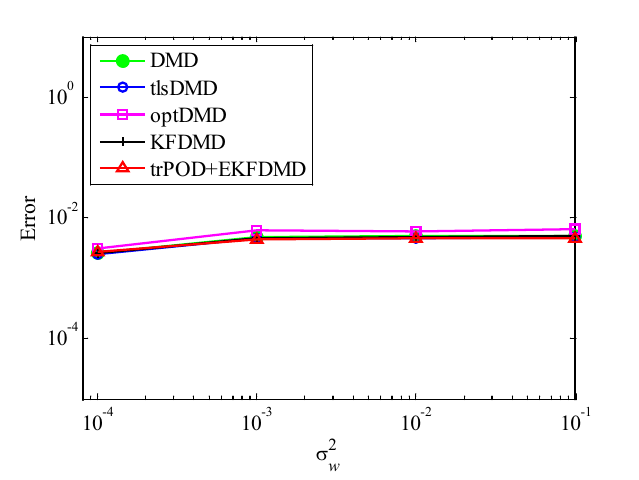}}
	\subfigure[$\lambda_2$.]{\includegraphics[width=5cm]{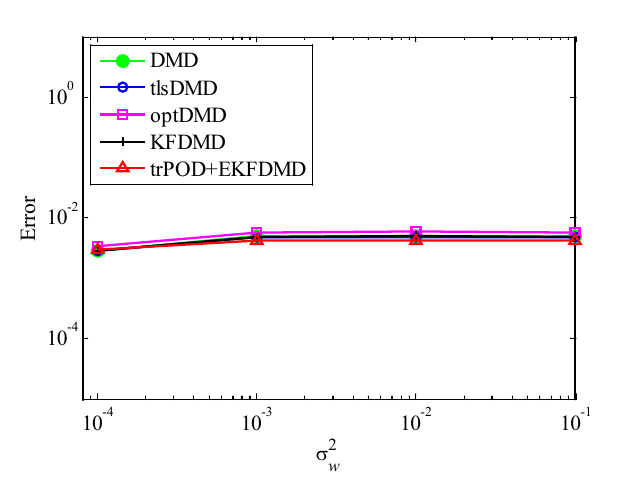}}
	\subfigure[$\lambda_3$.]{\includegraphics[width=5cm]{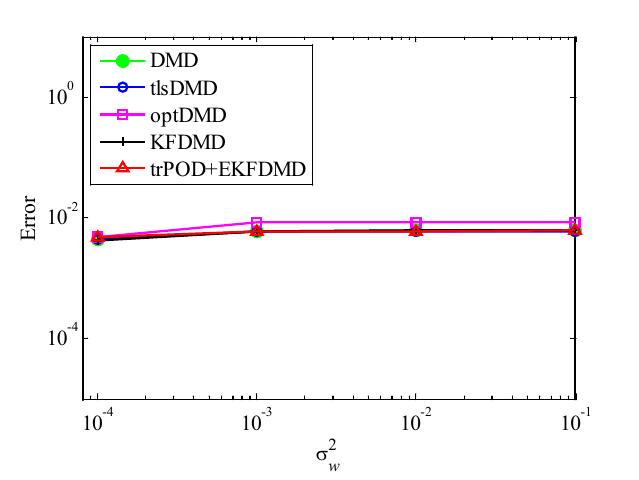}}\\
	\caption{Errors in the eigenvalues for multiple runs of a problem with a moderate number of DoFs with system noise for the case in which rank $r$ is set to be 20, where the seed for random numbers is different for multiple runs. }
\label{fig:error_eigen_medium_sonoise_r20}
\end{figure}

\begin{figure}
	\centering
	{\includegraphics[width=5cm]{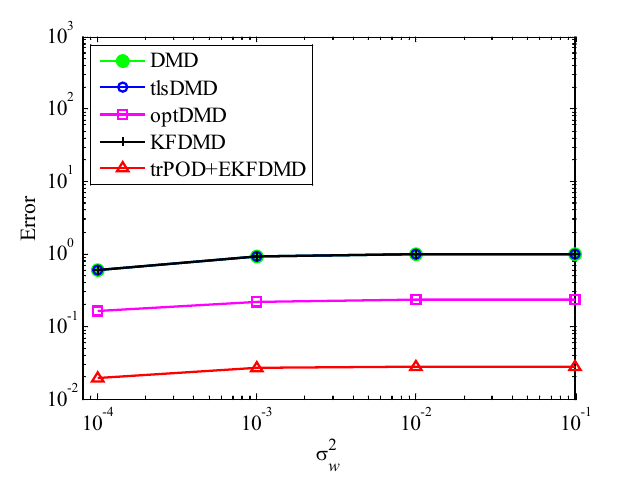}}
	\caption{Errors in the reconstructed data for multiple runs of a problem with a moderate number of DoFs with system noise for the case in which rank $r$ is set to be 20, where the seed for random numbers is different for multiple runs. }
\label{fig:error_history_medium_sonoise_r20}
\end{figure}

Similar to the cases without system noise, the effects of the rank number chosen by the user are investigated. Here, $r=6$ and $r=20$ are investigated, where the previous standard cases are computed with $r=10$, as noted earlier. The errors in the eigenvalues estimated with a truncated PODs of $r=6$ and $r=20$ and the errors in the reconstructed data with a truncated POD of $r=6$ and  $r=20$ are shown in Figs.~\ref{fig:error_eigen_medium_sonoise_r6}, \ref{fig:error_history_medium_sonoise_r6}, \ref{fig:error_eigen_medium_sonoise_r20}, and \ref{fig:error_history_medium_sonoise_r20}. These plots are similar to those with a truncated POD of $r=10$, which indicates that the rank for the POD truncation does not significantly affect the results for the case in which system noise is present. 
\color{black}

\subsection{Application to a fluid problem}
The simulation of a two-dimensional flow around a cylinder is conducted. The Mach number of the freestream velocity is set to be 0.3, and the Reynolds number based on the freestream velocity and the cylinder diameter is set to be 300. For the analysis, LANS3D,\cite{Fujii1990a} which is an in-house compressible fluid solver, is adopted. A cylindrical computational mesh is used, with the numbers of the radial- and azimuthal-direction grid points being 250 and 111, respectively. A compact difference scheme\cite{Lele1992} of the sixth order of accuracy is used for spatial derivatives and a second-order backward differencing scheme converged by an alternative-directional-implicit symmetric-Gauss-Seidel method\cite{Fujii1999,Nishida2009} is used for time integration. See Reference \cite{Sato2015b} for further details. The origin point is set to be the center of the cylinder, and a resolved region (where the mesh density is finer) is set to be inside 10$d$ far from the origin point. Here, $d$ is the diameter of the cylinder. For any DMD analyses, the quasi-steady flow data at $x=[0,10d], y=[-5d,5d]$, which is in the wake region, are used. The data are mapped to an equally distributed 100$\times$100 mesh. The DMD analyses processed 500 samples of five flow-through data with or without adding observation noise of $\mathcal{N}\left(0,\sigma_w^2\right)$, whereas the variance ($\sigma_w^2$) is set to be $0.02$. In the EFKDMD algorithm, the diagonal parts of the covariance matrix are initially set to be $10^3$, similar to previous problems. The diagonal elements of $Q$ and $R$ are set to be $0$ and $0.02$, respectively, while nondiagonal elements of $Q$ and $R$ are set to be 0. 

\begin{figure}
\centering
\includegraphics[width=5cm]{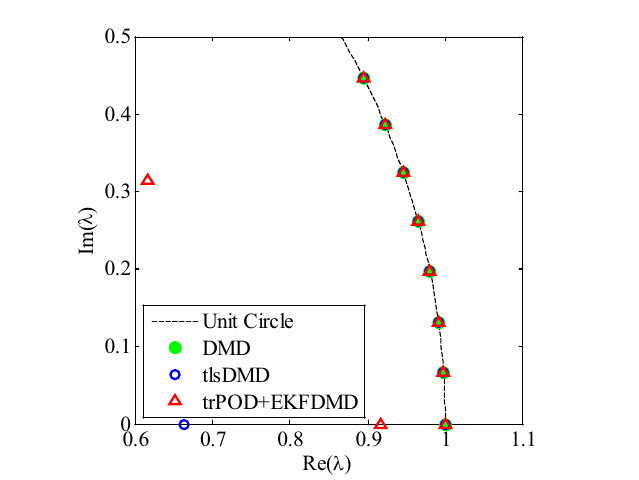}
\vspace{-0.2cm}
\caption{Eigenvalues for a flow problem without noise.}
\label{fig:eigen_flow_nonoise}
\end{figure}

First, the results without noise are processed by DMD, tlsDMD and KFDMD, where KFDMD adopts the truncated POD (Eq.~ \ref{eq:trun}) as a preconditioner. The eigenvalues computed by the DMD, tlsDMD and trPOD+KFDMD methods are shown in Fig.~\ref{fig:eigen_flow_nonoise}. The eigenvalues computed by KFDMD agree well with those of the standard DMD. The lowest frequencies computed by DMD and KFDMD correspond to the Strouhal number $St=fd/u_\infty \sim 0.2$, which is a well-known characteristic frequency for the K\'{a}rm\'{a}n vortex street of a cylinder wake, where $f$ and $u_\infty$ are the frequency and the freestream velocity, respectively. 

\begin{figure}
\centering
\includegraphics[width=5cm]{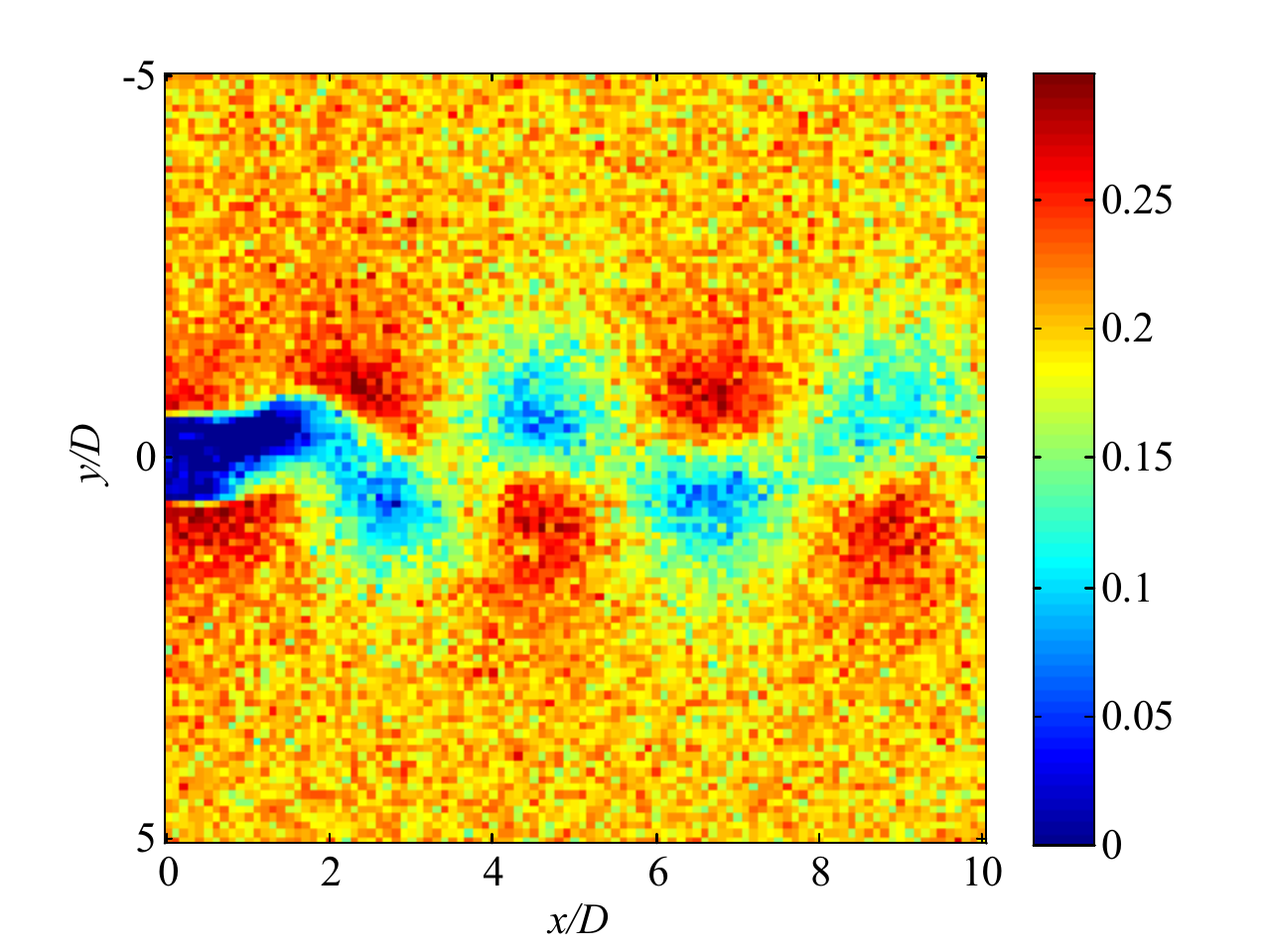}
\vspace{-0.2cm}
\caption{Noisy flow field data processed by several DMD methods. The $x$-direction velocity is visualized, where the freestream velocity is set to be 0.3.}
\label{fig:orgflow}
\end{figure}

\begin{figure}
\centering
\includegraphics[width=5cm]{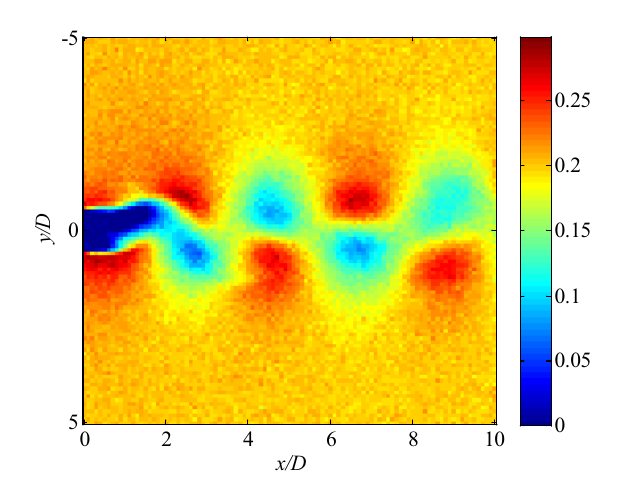}
\vspace{-0.2cm}
\caption{trPOD 30-mode reconstruction of flow fields. The $x$-direction velocity is visualized, where the freestream velocity is set to be 0.3.}
\label{fig:svdflow}
\end{figure}

Then, the data with noise are processed. 
The snapshot data of the instantaneous flow field are shown in Fig.~\ref{fig:orgflow}.
Flow fields filtered using only trPOD are shown in Fig.~\ref{fig:svdflow}. The noise can be reduced using trPOD. These 30-DoF data are used for KFDMD analyses. 

\begin{figure}
\centering
\includegraphics[width=5cm]{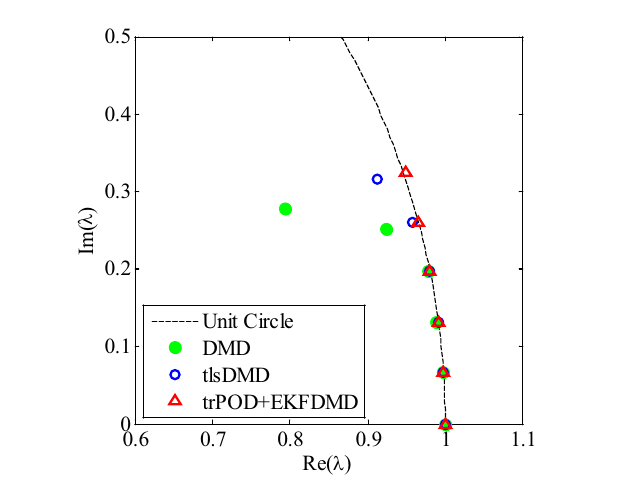}
\vspace{-0.2cm}
\caption{Eigenvalues for a flow problem with noise.}
\label{fig:eigen_flow_noise}
\end{figure}

\begin{figure}
\subfigure[Mode 2.]{\includegraphics[width=5cm]{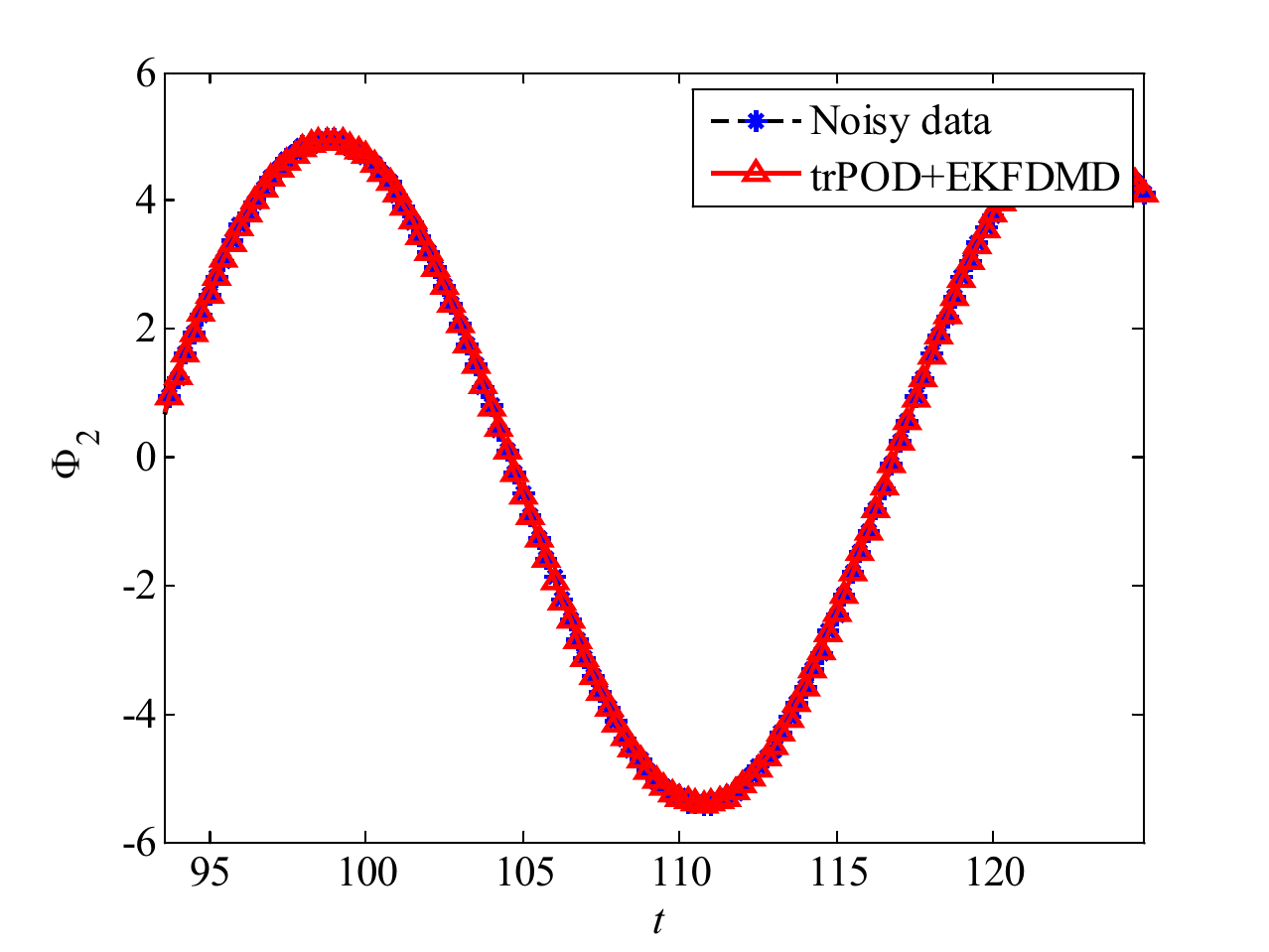}}
\subfigure[Mode 4.]{\includegraphics[width=5cm]{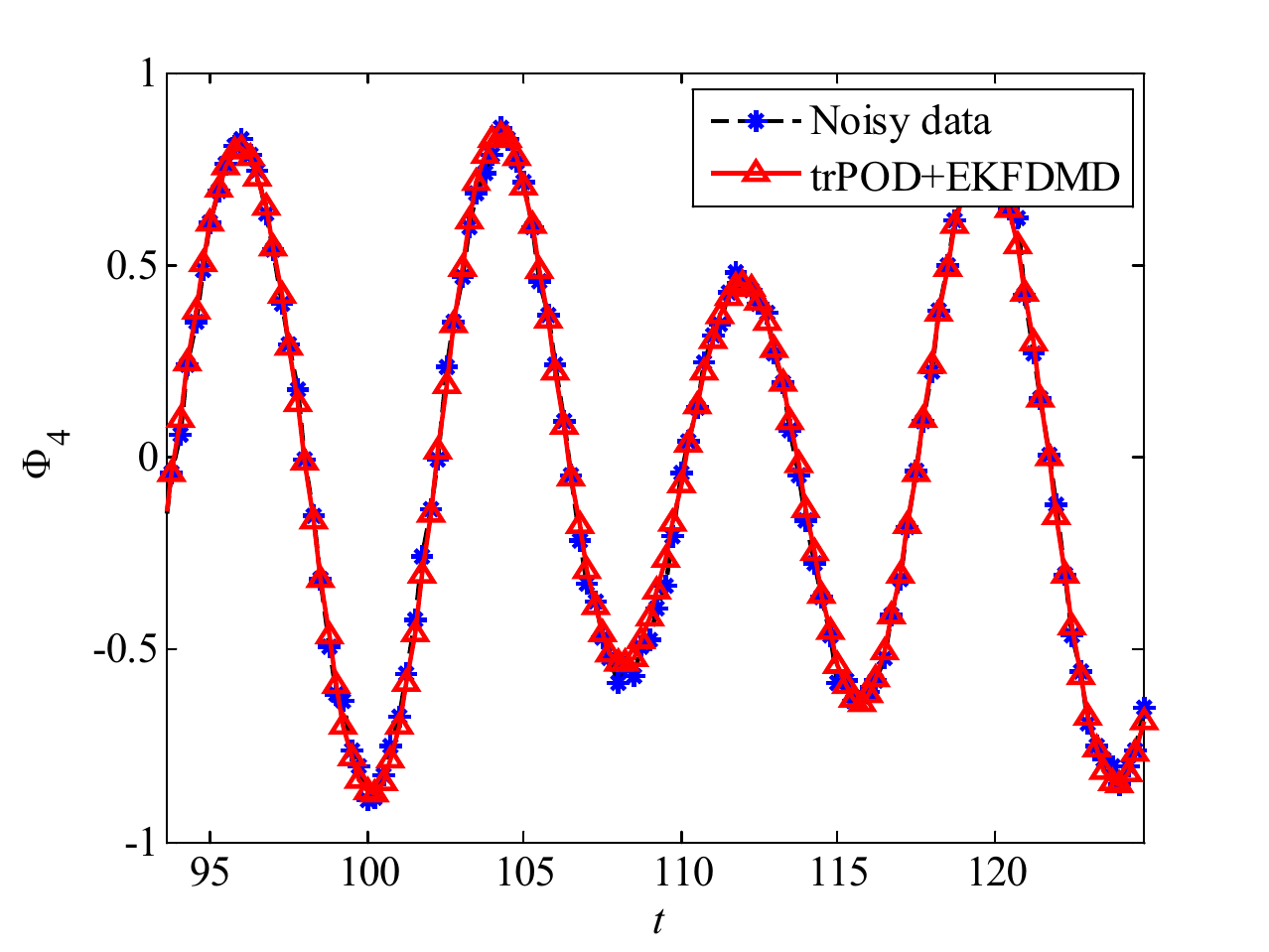}}\\
\subfigure[Mode 6.]{\includegraphics[width=5cm]{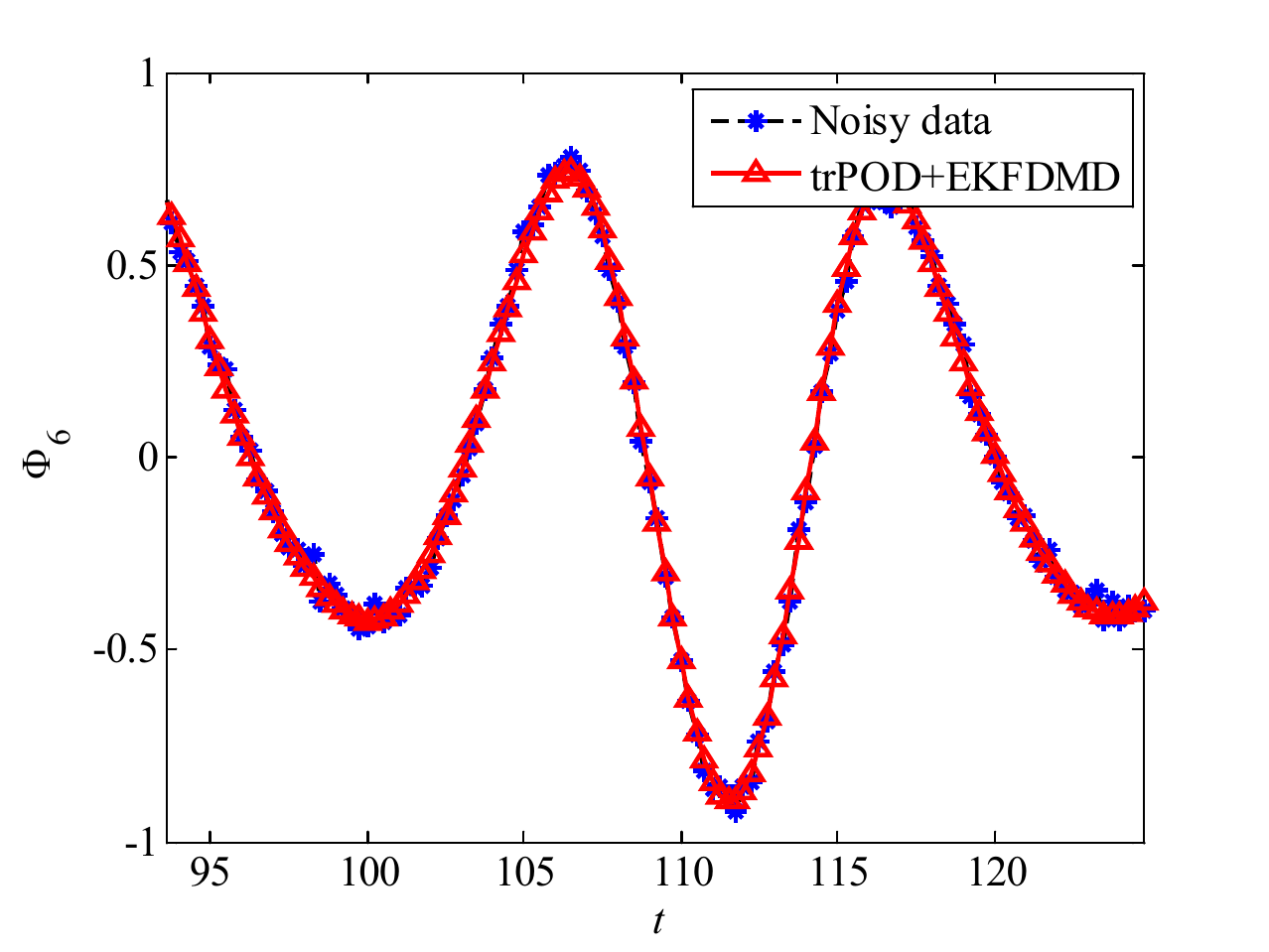}}
\subfigure[Mode 8.]{\includegraphics[width=5cm]{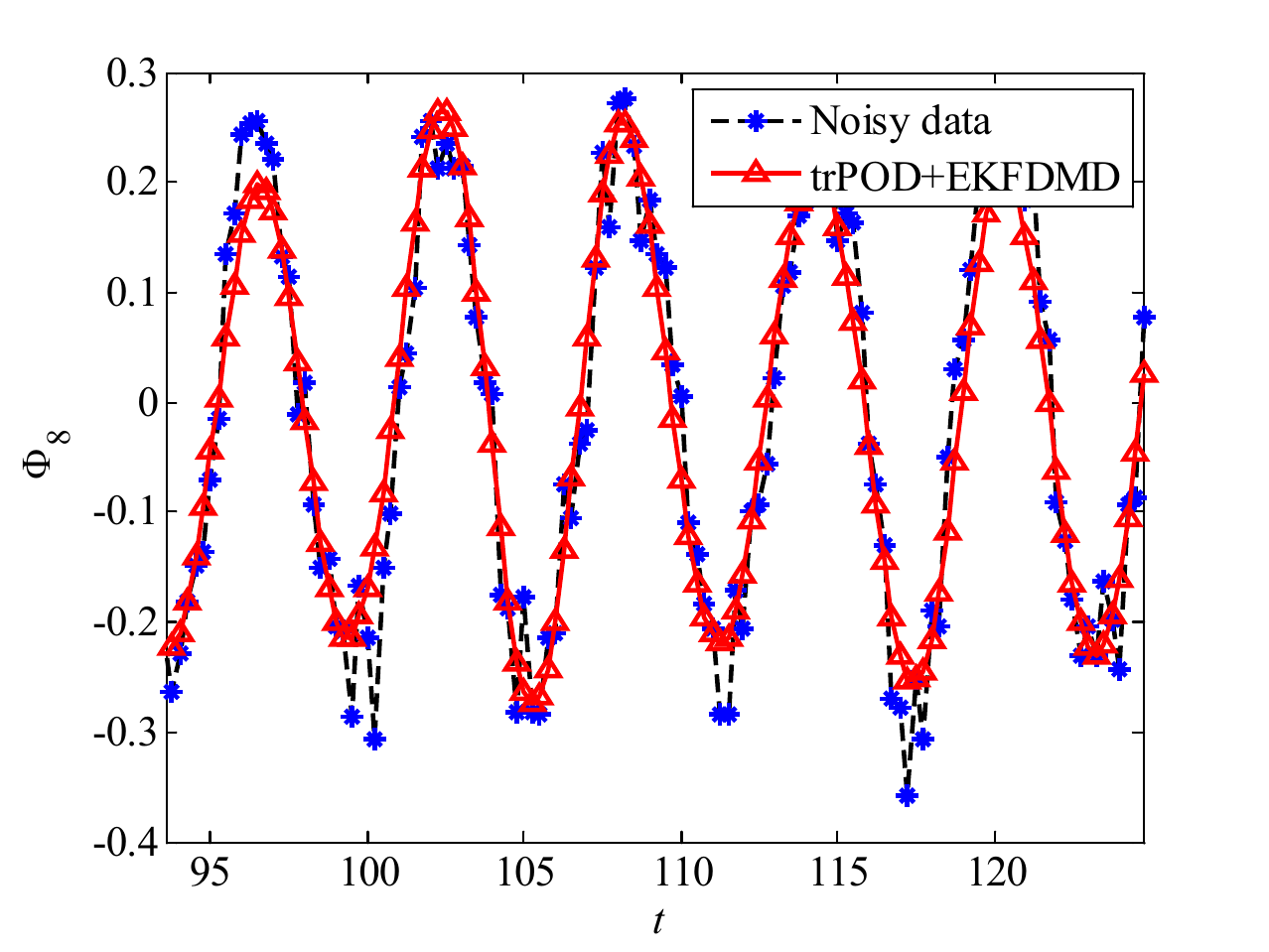}}
\vspace{-0.2cm}
\caption{Time histories of POD modes 2, 4, 6, and 8 of the data of the flow problem.}
\label{fig:history_mode_flow}
\end{figure}

\begin{figure}
\centering
\includegraphics[width=5cm]{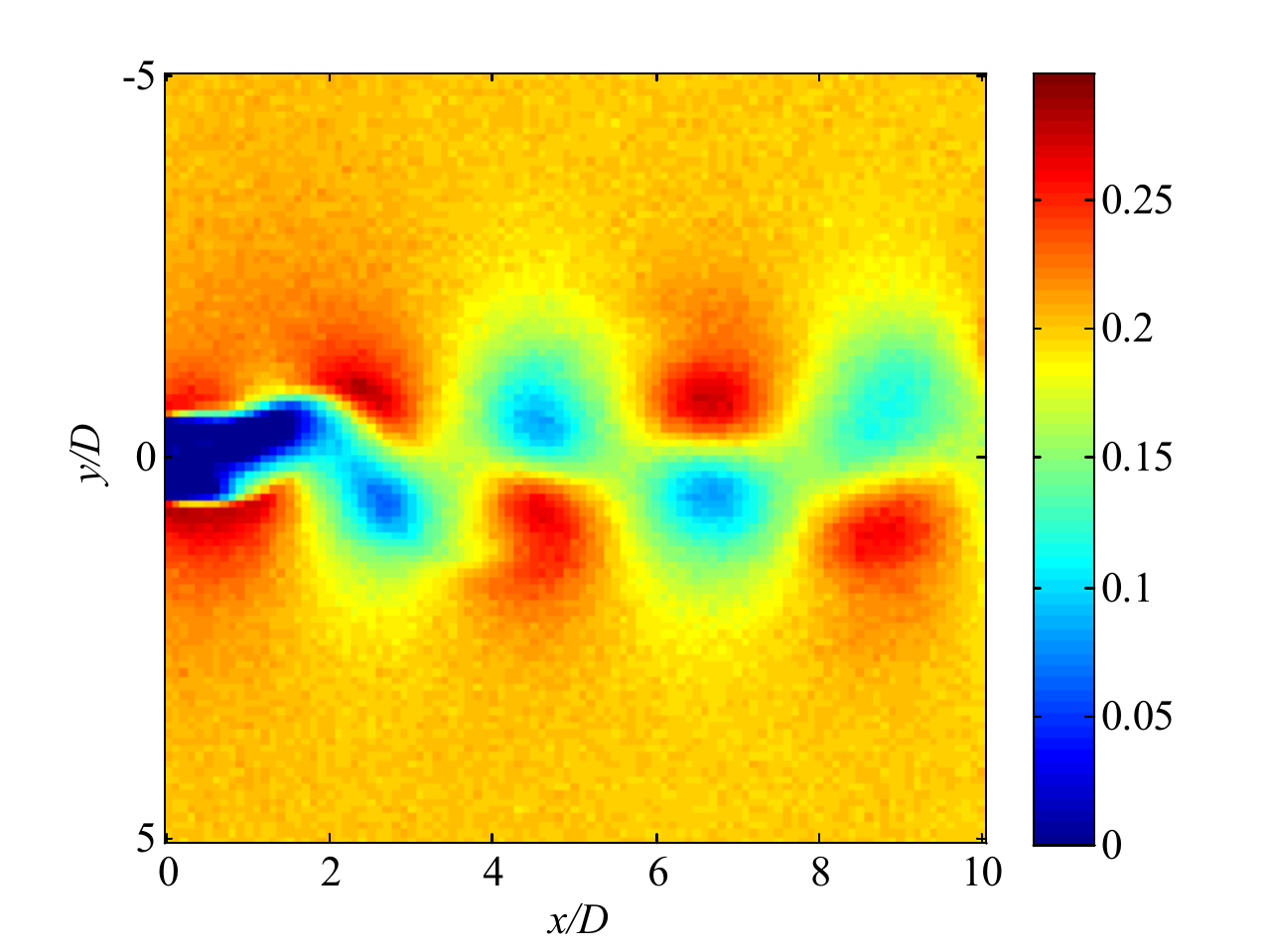}
\vspace{-0.2cm}
\caption{EKFDMD-filtered flow fields. The $x$-direction velocity is visualized, where the freestream velocity is set to be 0.3.}
\label{fig:kfdmdflow}
\end{figure}

Figure \ref{fig:eigen_flow_noise}, which illustrates the eigenvalues of DMD, tlsDMD, and EKFDMD, shows that the EKFDMD results are better than the results of the standard DMD and tlsDMD. Here, trPOD+EKFDMD accurately predicts from the steady flow mode (eigenvalue of unity) up to the fourth oscillataion mode, which corresponds to nine points on the unit circle. In addition, it should be noted that the strength of EKFDMD is that the data are denoised online. Figure \ref{fig:history_mode_flow} shows the mode histories of trPOD modes 2, 4, 6, and 8. The histories of modes 2 and 4 are approximately the same for noisy data and EKFDMD combined with the trPOD preconditioner, because these modes are strong enough compared with the noise level. On the other hand, the histories of modes 6 and 8 are cleaned up well. Finally, the flow fields of denoised data (in this case, the temporal coefficients of the trPOD modes are filtered) are shown in Fig.~\ref{fig:kfdmdflow}, and the data are slightly further cleaned up compared to the results obtained only with trPOD, as shown in Fig.~\ref{fig:svdflow}.


\section{Complexity and Computational Cost}
\label{sec:Comp}
\begin{figure}[htb]
	\centering
	\includegraphics[width=5cm]{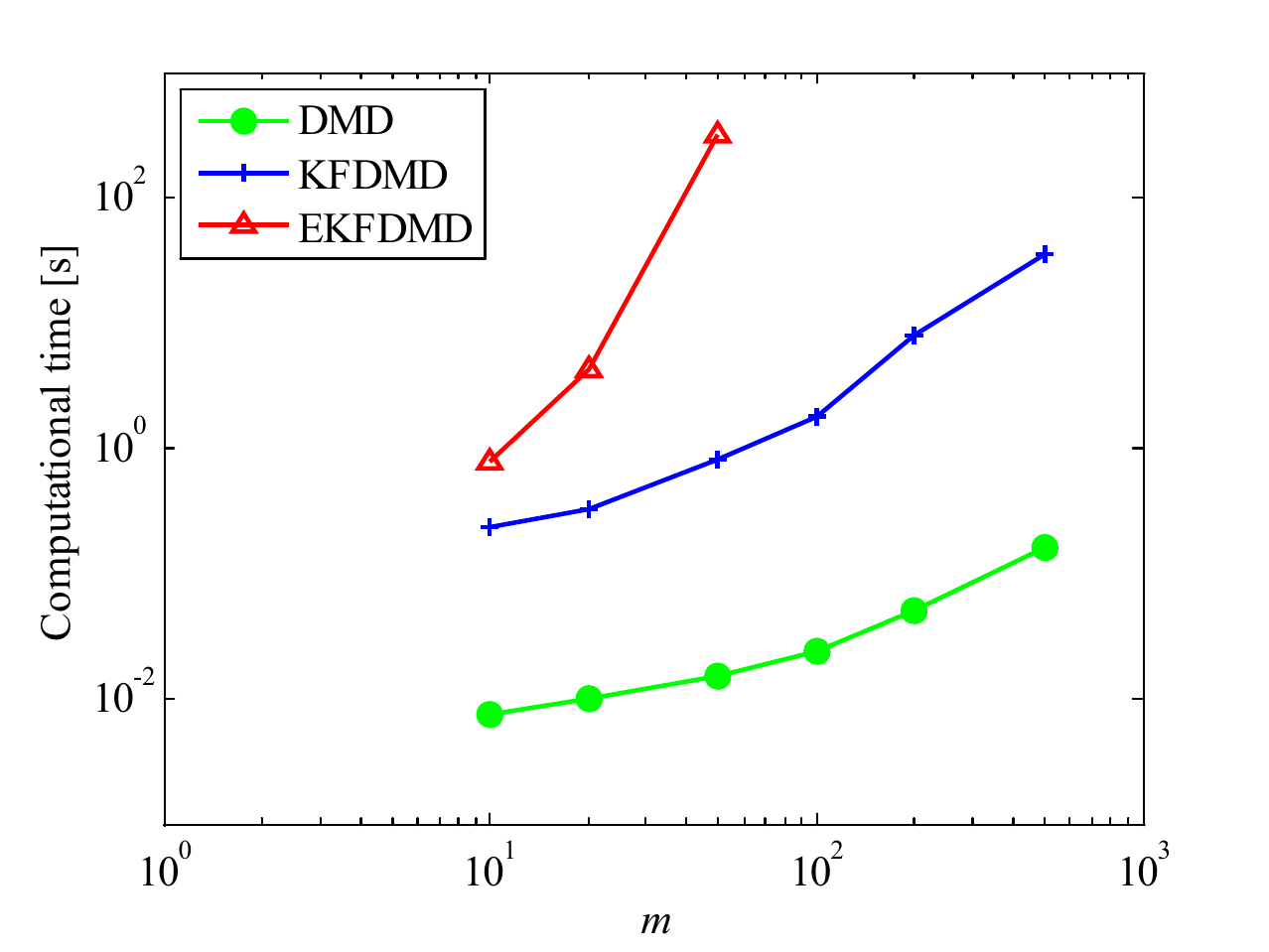}
	\vspace{-0.2cm}
	\caption{Computational time for DMD, KFDMD, and EKFDMD}
	\label{fig:comp_time}
\end{figure}

In this section, the complexity and computational cost of EKFDMD are discussed. Here, multiplication for single elements is assumed to have a complexity of $O(1)$, and the multiplication of matrices of size of $l\times m$ and $m\times n$ is estimated to be $O(lmn)$ under the dense matrix computation.
In the EKFDMD procedure, except when using trPOD as a preconditioner, the main computational cost comes from Eqs. \ref{eq:predicttheta} and \ref{eq:predictP} for the prediction step and from Eqs. \ref{eq:updates}, \ref{eq:updatekg}, \ref{eq:deltatheta}, and \ref{eq:updateP} for the updating step. For each step, the computational complexity is summarized in Table 
\ref{tab:complexity}. In total, the most significant complexity is considered to be $O(n^6)$ for one step. Therefore, if we have $m$ samples, then the computational complexity for $m$-time steps becomes $O(mn^6)$. The complexity and the required memory of EKFDMD are compared with those of the other algorithms in Table \ref{tab:complexitycomp}, where estimation of the complexities of DMD and online DMD in the previous study\cite{Zhang2017} are adopted, and the complexity of KFDMD is estimated in the present study. In addition, Fig.~\ref{fig:comp_time} shows the computational time for 500 samples with different DoF problems. The Matlab software is used with Intel Xeon E5620 2.4GHz processor.  The computational time is averaged over 20 runs for the small size of $m<50$, while it is not for the large size but the repeatability is confirmed. Both Table \ref{tab:complexitycomp} and Fig.~\ref{fig:comp_time} show that EKFDMD requires significant computational cost, and applying trPOD as a preconditioner is strongly recommended for the practical use of EKFDMD. In practical use, matrices $F$ and $H$ for EKFDMD are sparse and the corresponding computational cost and memory of EKFDMD can be decreased by using implementations of routines for the sparse matrix in the software utilized as we did. However, the complexity of EKFDMD is still higher with the routines for the sparse matrix than the other algorithms as shown in Fig.~\ref{fig:comp_time}.
\color{black}

\begin{table}[htb]
	\caption{Computational time for each procedure in KFDMD.}
	\label{tab:complexity}	
	\begin{tabular}{ccc} \hline		
		procedure & equation & complexity \\ \hline 
		predicting step & Eq. \ref{eq:predicttheta} & $O(n^2)$ \\		
		& Eq. \ref{eq:predictP}     & $O((n+n^2)^3)$ \\
		updating step   & Eq. \ref{eq:updates}      & $2O((n+n^2)^2n)$ \\		
		& Eq. \ref{eq:updatekg}     & $O(n^3)+O(n^2(n+n^2))+O((n+n^2)^2n)$ \\
		& Eq. \ref{eq:deltatheta} &$2O(n(n+n^2))$\\
		& Eq. \ref{eq:updateP}     &$2O(n(n+n^2)^2)+O((n+n^2)^3)$
		\\\hline 		       
	\end{tabular}
\end{table}

\begin{table}[htb]
	\caption{Comparison of complexity and memory for $m$-sample computation for the estimation in the final time step once.}
	\label{tab:complexitycomp}
	\begin{tabular}{ccc} \hline		
	    algorithm                             &  computational time & memory\\ \hline 
	    DMD                                   &  $O(mn^2)$          & $mn$ \\
	    online DMD                            &  $O(mn^2) $          & $2n^2$ \\
        KFDMD without trPOD (fast algorithm)  &  $O(mn^2)$           & $2n^2$ \\
	    EKFDMD without trPOD                  &  $O(mn^6)$           & $(n+n^2)(n+n^2+1)$  \\ \hline
	\end{tabular}
\end{table}

\section{Conclusions}
A dynamic mode decomposition method based on the extended Kalman filter (EKFDMD) was proposed for simultaneous online parameter estimation and denoising. The numerical experiments of the present study reveal that the proposed method can estimate the eigenstructure of matrix $A$ better than or as well as  existing algorithms, whereas EKFDMD denoises the data in its online procedure for a problem with a small number of DoFs. In particular, the EKFDMD works better  for data reconstruction in the case in which the system noise is present than existing algorithms, despite being an online procedure.  
However, this algorithm has the drawback of computational cost. This drawback is addressed by preconditioning of truncated POD (trPOD), and EKFDMD with trPOD is applied to a problem with a moderate number of DoFs and a fluid system. The performance of EKFDMD is slightly degraded by decreasing the rank number of trPOD in the case without system noise while the performance does not change in the case with system noise with regardless of the rank number. It should be noted that all the performance of EKFDMD is preferable in the analysis of noisy data. 

\section*{Acknowledgment}
The present study was supported in part by JST Presto (Grant Number JPMJPR1678).

\bibliography{pof}
\end{document}